\documentclass[%
 preprint,5p,times,twocolumn
]{elsarticle}

\usepackage{graphicx}% Include figure files
\usepackage{dcolumn}% Align table columns on decimal point
\usepackage{bm}% bold math
\usepackage{amsmath}
\usepackage{amssymb}
\usepackage{hyperref}

\usepackage[utf8]{inputenc} 
\usepackage{epsfig}
\usepackage{hyperref} % for highlighting refs, plots and links
\hypersetup{
    colorlinks=true,
    linkcolor=red,
    filecolor=blue,      
    urlcolor=magenta,
    citecolor=blue
    }
\usepackage[anythingbreaks]{breakurl} % to break url in references
\usepackage{orcidlink}

\journal{Computer Physics Communications}

\begin{document}

 \begin{frontmatter}
%%%%%%%%%%%%%%%%%%%%%%%%%%%%%%%%%%%%%%%%%%%%%%%%%%%%%%%%%%%%%%%%%%%%%%%%%
% Title. Author. Date
\title{DEMPgen: Physics event generator for Deep Exclusive Meson Production\\
at Jefferson Lab and the EIC}

\author[uregina]{Z.~Ahmed}

\author[uregina]{R.~S.~Evans}

\author[sbu]{I.~Goel
\orcidlink{0000-0002-2553-4100}}

\author[uregina]{G.~M.~Huber
\orcidlink{0000-0002-5658-1065}}
\cortext[author] {Corresponding author.\\\textit{Email address:} huberg@uregina.ca}

\author[uregina,york]{S.~J.~D.~Kay\orcidlink{0000-0002-8855-3034}}

\author[sbu,cfns,wm]{W.~B.~Li\orcidlink{0000-0002-8108-8045}}

\author[uregina]{L.~Preet\orcidlink{0009-0004-7667-077X}}

\author[uregina]{A.~Usman}

\address[uregina]{University of Regina, Regina, SK S4S 0A2, Canada}

\address[sbu]{Stony Brook University, Stony Brook NY 11794, USA}

\address[york]{University of York, Heslington, York, Y010 5DD, UK}

\address[cfns]{Center for Frontiers in Nuclear Science, Stony Brook University, Stony Brook NY 11794, USA}

\address[wm]{College of William and Mary, Williamsburg, VA 23187, USA}

%%%%%%%%%%%%%%%%%%%%%%%%%%%%%%%%%%%%%%%%%%%%%%%%%%%%%%%%%%%%%%%%%%%%%%%%%
% Abstract (in new format)
\begin{abstract}

%\begin{description} 
%\item[Background] 
There is increasing interest in deep exclusive meson production (DEMP) reactions, as they provide access to Generalized Parton Distributions over a broad kinematic range, and are the only means of measuring pion and kaon charged electric form factors at high $Q^2$. Such investigations are a particularly useful tool in the study of hadronic structure in QCD's transition regime from long-distance interactions described in terms of meson-nucleon degrees of freedom, to short-distance interactions governed by hard quark-gluon degrees of freedom.
%\item[Purpose] 
To assist the planning of future experimental investigations of DEMP reactions in this transition regime, such as at Jefferson Lab and the Electron-Ion Collider (EIC), we have written a special purpose event generator, DEMPgen. Currently, DEMPgen can generate the following reactions: $t$-channel $p(e,e^{\prime}\pi^+)n$, $p(e,e^{\prime}K^+)\Lambda[\Sigma^0]$, and $\vec{n}(e,e^{\prime}\pi^-)p$ from a polarized $^3$He target. DEMPgen is modular in form, so that additional reactions can be added over time.

%\item[Method]
The generator produces kinematically-complete reaction events which are absolutely-normalized, so that projected event rates can be predicted, and detector resolution requirements studied. The event normalization is based on parameterizations of theoretical models, appropriate to the kinematic regime under study. Both fixed target modes and collider beam modes are supported.
%\item[Results]
This paper presents the structure of the generator, the model parameterizations used for absolute event weighting, the kinematic distributions of the generated particles, some initial results using the generator, and instructions for its use.
%\item[Conclusion] 
%\end{description}

\end{abstract}

%%%%%%%%%%%%%%%%%%%%%%%%%%%%%%%%%%%%%%%%%%%%%%%%%%%%%%%%%%%%%%%%%%%%%%%%%
% Title

\end{frontmatter}

%\begin{linenumbers}
    
%%%%%%%%%%%%%%%%%%%%%%%%%%%%%%%%%%%%%%%%%%%%%%%%%%%%%%%%%%%%%%%%%%%%%%%%%
% Sec 1
\section{Introduction}

% 04/10/22 - SJDK - Should follow ordering we set out here in the opening throughout the paper

We have written a \textbf{D}eep \textbf{E}xclusive \textbf{M}eson \textbf{P}roduction (DEMP) event generator (DEMPgen) \cite{url:DEMPgen}, which is modular in form, so that the variety of reactions it simulates can be expanded over time. The motivation for the writing of the event generator is to evaluate the feasibility of hadron structure studies with polarized targets at Jefferson Lab (JLab), and with colliding beams at the \textbf{E}lectron-\textbf{I}on \textbf{C}ollider (EIC).

\begin{figure}[!hbt]
  \centering
  \includegraphics[width=0.9\linewidth]{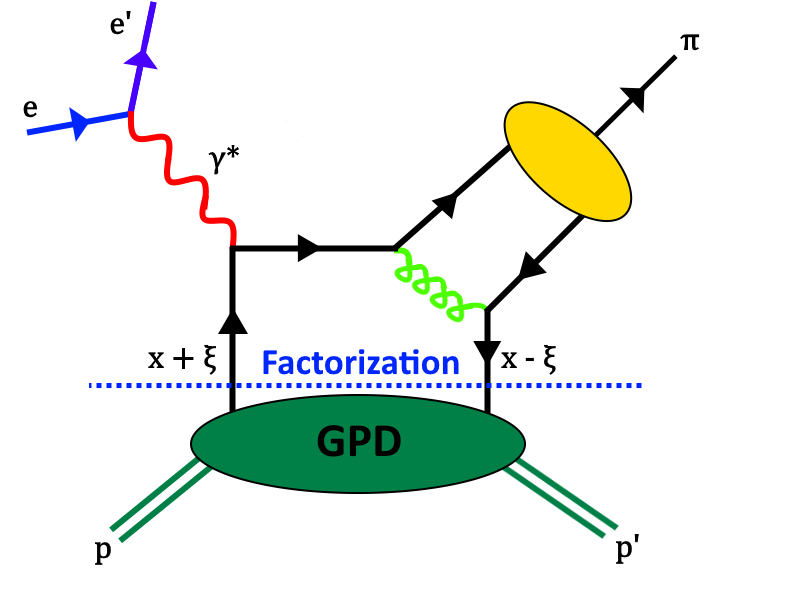}
  \caption{DEMP handbag diagram for the colinear factorization regime at large $Q^2$. The portion above the dashed line represents the hard scattering with a parton of momentum fraction $x+\xi$, which can be treated perturbatively. The portion below the line contains the soft contributions, where the Generalized Parton Distributions (GPD) encode the response of the nucleon to the exchange of a parton $x+\xi$ with one with fraction $x-\xi$.}
  \label{fig:fig_demp}
\end{figure}

%SK - This second sentence is specific to the EIC studies, should be generalised to all (e.g. the SoLID study is actually electron neutron scattering - SK%
The process of interest is deep inelastic scattering of an electron and nucleon.
%At the EIC we have electron beam of 5 GeV and proton beam of 100 GeV. Four
%momentum transfer $Q^2$ can go up to 30 GeV$^2$. 
The value of $Q^2$ is high enough to probe the parton structure via deep inelastic scattering. % SK - Edited grammar and sentence structure slightly - SK 14/12/20
DEMP is a kind of inelastic scattering in which a target nucleon is split into a meson and recoil hadron, where either all three outgoing particles are detected, or two are detected with sufficient resolution to construct the missing mass and momentum of the third particle. Generically, a DEMP reaction involves an electron ($e$) interacting with a nucleon ($N$) yielding a scattered electron ($e^{\prime}$), an ejectile ($X_{Ej}$, the produced meson) and a recoil hadron ($X_{Rec}$) which can be written as 
\begin{gather}
    N\left(e,e^{\prime} X_{Ej} X_{Rec}\right).
\end{gather}
An example $p(e,e^{\prime}\pi^+ n)$ DEMP reaction is shown in Fig.~\ref{fig:fig_demp}. At moderate $Q^2$, DEMP reactions have significant higher twist contributions at the amplitude level, but these contributions are expected to largely cancel in some asymmetries \cite{frank}.

At present, two modules are available in DEMPgen:
\begin{enumerate}

    \item A colliding beam kinematics module for the EIC. In this module, the ejectile is emitted at small $-t$, at forward angles in the center of mass frame. In this case, the recoil hadron takes most of the incident nucleon (protons typically) beam energy and is scattered at very small angles ($<1^{\circ}$). In this module, the following reactions are currently available:
    \begin{itemize}
        \item $p(e,e^{\prime}\pi^{+}n)$,
        \item $p(e,e^{\prime}K^{+}\Lambda)$,
        \item $p(e,e^{\prime}K^{+}\Sigma^{0})$.
    \end{itemize}

    \item A fixed target kinematics module for polarized targets at Jefferson Lab. This module includes optional corrections for Fermi momentum and other nuclear effects. This module is primarily designed to generate exclusive $^3He(e,e^{\prime}\pi^{-}p)(pp)_{sp}$ events from a transversely polarized $^{3}He$ target.

\end{enumerate}

% SK - Minor sentence adjusment SK 14/12/20 
This paper is divided into sections as follows. In Sec.~\ref{sec:motivation}, we will briefly describe the scientific motivation for our studies, so the structure and kinematic ranges of applicability of the generator can be better understood. In Sec.~\ref{sec:generator}, we summarize the coding structure of the two modules of the generator. The cross-section parameterizations of the currently implemented physics processes are also discussed. In Sec.~\ref{sec:results}, we present some results obtained with the generator, to display some of the ways in which it can be used. Sec.~\ref{sec:summary} presents a summary of our work to date, and an outlook of some extensions to the generator that are being considered.

%%%%%%%%%%%%%%%%%%%%%%%%%%%%%%%%%%%%%%%%%%%%%%%%%%%%%%%%%%%%%%%%%%%%%%%%%
% Chapter 2
\section{Scientific Motivation for our Studies
\label{sec:motivation}}

%------------------------------------------------------------------------
\subsection{Motivation for EIC $\pi^+$ and $K^+$ studies}

The Electron-Ion Collider, EIC, is a next generation collider to be constructed at Brookhaven National Laboratory to study the structure of nucleons in detail. According to the National Academies of Science assessment \cite{NAP25171}, the EIC will address three major questions.

\begin{enumerate}
    \item \textit{How does the mass of the nucleon arise?} The problem is that while gluons have no mass, and $u$, $d$ quarks are nearly massless, the nucleons that contain them are heavy; the total mass of a nucleon is significantly larger than the mass of the valence quarks it contains. The largest contribution to the mass of the proton originates from the gluon field energy. In this sense, the source of the vast majority of the visible mass in the universe is not the Higgs field, but the gluon field. Measurements of deep exclusive $\pi^+$ production from the nucleon at the EIC allow the quark-gluon energy contributions to the nucleon mass budget to be studied.
    \item \textit{How does the spin of the nucleon arise?} How the angular momentum, both intrinsic as well as orbital, of the internal quarks and gluons gives rise to the known nucleon spin is not well understood. The quark polarization contribution to the nucleon spin is only about 30\%. The remainder of the spin must reside in the orbital angular momenta of quarks and gluons or gluon polarization. A central goal of the EIC program is to provide a determination of the gluon spin and orbital angular momentum contributions.
    \item \textit{What are the emergent properties of dense systems of gluons?} The nature of gluons in matter, i.e.~their arrangements or states, and the details of how they hold matter together are not well known. Gluons in matter are somewhat like dark matter in the universe, unseen but playing a crucial role. The EIC would be able to study the gluons that bind quarks and antiquarks into nucleons and nuclei with unprecedented precision. A central goal of such studies is to explore the limit of low parton momentum fraction $x$, where the number of gluons in the target is very large. The EIC would also be able to explore modifications of the quark distributions in nuclei.
\end{enumerate}

Answers to these questions are essential for understanding the nature of visible matter in the universe.

As for our specific investigation, the pion and kaon are two of the simplest systems available to study the structure of hadrons. The elastic electromagnetic form factors of the charged pion and kaon, $F_{\pi}(Q^2)$ and $F_K(Q^2)$, are a rich source of insights into basic features of hadron structure. For example, $F_{\pi}(Q^2)$ and $F_K(Q^2)$ can provide insight on the roles played by confinement and Dynamical Chiral Symmetry Breaking (DCSB) in fixing the hadron’s size, determining its mass, and defining the transition from the strong- to perturbative-QCD domains \cite{Horn_2016}. 

The experimental determination of the $\pi^+$ electric form factor ($F_{\pi}$) is challenging. The best way to determine $F_{\pi}$ would be electron-pion elastic scattering. However, the lifetime of the $\pi^+$ is only 26.0~ns. Since $\pi^+$ targets are not possible, and $\pi^+$ beams with the required properties are not yet available, one of the experimentally feasible approaches is high-energy exclusive electroproduction of a $\pi^+$ from a proton at low Mandelstam four-momentum transfer to the target proton, $-t=-(p_p-p_n)^2$. 
This is best described as quasi-elastic ($t$-channel) scattering of the electron from the virtual $\pi^+$ cloud of the proton. Scattering from the $\pi^+$ cloud dominates the longitudinal photon cross section ($d\sigma_L/dt$), when $|t|\ll M_p^2$ \cite{Carlson:1990zn}. To reduce background contributions, normally one separates the components of the cross-section due to longitudinal (L) and transverse (T) virtual photons (and the LT, TT interference contributions), via a Rosenbluth separation (Eqn.~\ref{equation:cross-3}). The value of $F_{\pi}(Q^2)$ is determined by comparing the measured $d\sigma_L/dt$ values at small $-t$ to the best available electroproduction model. The obtained $F_{\pi}$ values are in principle dependent upon the model used, but one anticipates this dependence to be reduced at sufficiently small $-t$. 

Using this approach, the charged pion form factor, $F_{\pi}(Q^2)$, has been measured in Jefferson Lab (JLab) Hall C via $\pi^+$ electroproduction up to $Q^2 = -(p_e-p_{e^{\prime}})^2 = 2.45$ GeV$^2$ with high precision \cite{gmhuber}. This result generated confidence in the reliability of $\pi^+$ electroproduction as a tool for pion form factor extraction. JLab experiment E12-19-006 \cite{fpi-12gev}, one of the flagships of the 12 GeV upgrade, will extend the high precision studies to $Q^2=6.0$ GeV$^2$, and with lower precision to $Q^2=8.5$ GeV$^2$. This experiment is expected to deliver pion form factor data bridging the region where QCD transitions from the strong (color confinement, long-distance) to perturbative (asymptotic freedom, short-distance) domains.
The measurement of ${F}_{\pi}(Q^2)$ at the EIC is a continuation of the study of the pion form factor at higher $Q^2$ kinematics. The comparison of the pion and kaon form factors in this regime would provide vital information for the understanding of the role of DCSB in the generation of hadronic mass.

The reliability of the electroproduction method to determine the $K^+$ form factor, $F_{K}(Q^{2}),$ is not yet established. JLab E12-09-011 \cite{kaon-lt} has acquired data for the $p(e,e^{\prime} K^+)\Lambda$, $p(e,e^{\prime} K^+)\Sigma^0$ reactions at hadronic invariant mass $W=\sqrt{(p_{K}+p_{\Lambda,\Sigma})^2}>2.5$ GeV, to search for evidence of scattering from the proton's ``kaon cloud''. The data are still being analyzed, with L/T-separated cross-sections expected in the near future. If they confirm that the scattering from the virtual $K^+$ in the nucleon contributes significantly to $d\sigma_L/dt$ at low four-momentum transfer to the target $|t|\ll M_p^2$, the experiment will yield the world's first quality data for $F_K$ above $Q^2>0.2$ GeV$^2$. This would then open up the possibility of using the same exclusive reactions to determine the kaon form factor over a wide range of $Q^2$ at the EIC. 
%This is a further possible module that could be added into the event generator, this potential extension is discussed further in section \ref{sec:summary}.
DEMPgen includes two modules of the $p(e,e^{\prime}\pi^+n)$ and $p(e,e^{\prime}K^+\Lambda[\Sigma^0])$ reactions in colliding beam mode, to enable feasibility studies for these measurements at the EIC. These are discussed further in Secs. \ref{Sec:EIC_PiPlus} and \ref{Sec:EIC_KPlus}.

%------------------------------------------------------------------------
\subsection{Motivation for Jefferson Lab $\vec{n}(e,e'\pi^-)p$ studies}

The development of the Generalized Parton Distribution (GPD) formalism in the last 20 years is a notable advance in our understanding of the structure of the nucleon. GPDs unify the concepts of parton distributions and hadronic form factors, and are ``universal objects'' that provide a comprehensive framework for describing the quark and gluon structure of the nucleon. GPDs are probed through Deep Exclusive reactions, and their knowledge would allow a tomographic 3D understanding of the nucleon to be built up \cite{Belitsky_2005,Diehl_2003,Goeke_2001}. A special kinematic regime is probed in DEMP, where the initial hadron emits a quark-antiquark or gluon pair. This has no counterpart in the usual parton distributions and carries information about $q\bar{q}$ and $gg$-components in the hadron wavefunction.

%The purpose of our studies is to obtain an improved picture of hadrons with the help of Generalized Parton Distributions. 

%\subsection{Properties of GPD}
The four lowest-order GPDs are parameterized in terms of quark chirality. At leading twist, there are four GPDs: $H(x, \xi, t)$, $E(x, \xi, t)$, $\tilde{H}(x, \xi, t)$ and $\tilde{E}(x, \xi, t)$ associated with each quark flavor as well as gluons. $H$ and $\tilde{H}$ GPDs conserve helicity, while the $E$ and $\tilde{E}$ GPDs are associated with a helicity flip of the nucleon. Each GPD depends upon three variables, the four-momentum transferred, $Q^{2}$, the average longitudinal momentum of the struck quark $x$ (in the high $Q^{2}$ regime $x=x_{B}$) and the skewness, $\xi=(p_1^+-p_2^+)/(p_1^++p_2^+)$, where $p_1$, $p_2$ refer to the light-cone plus components of the initial and final nucleon momenta in Fig.~\ref{fig:fig_demp}.

GPDs also describe the correlation between partons in a nucleon. By utilizing a Fourier transform, one can access the longitudinal momentum fraction of quarks and their position in the transverse plane simultaneously \cite{burkardt}. The first moments of GPDs are related to the elastic form factors of the nucleon. The GPDs integrals over $x$ give \cite{ji_1}:
\begin{alignat}{4}
  \label{gpd_ffh}
  \int^{1}_{-1} dx \quad H^{q}(x,\xi,t) = F^{q}_{1}(t),\\
%\end{equation}
%\begin{equation}
%  \label{gpd_ffe}
  \int^{1}_{-1} dx \quad E^{q}(x,\xi,t) = F^{q}_{2}(t),\\
%\end{equation}
%\begin{equation}
%  \label{gpd_ffht}
  \int^{1}_{-1} dx \quad \tilde{H}^{q}(x,\xi,t) = g^{q}_{A}(t),\\
%\end{equation}
%\begin{equation}
%  \label{gpd_ffet}
  \int^{1}_{-1} dx \quad \tilde{E}^{q}(x,\xi,t) = h^{q}_{A}(t),
\end{alignat}
where $F_{1}$, $F_{2}$, $g_{A}$ and $h_{A}$ are the Dirac, Pauli, pseudoscalar and axialvector form factors respectively. 

GPDs provide information about the nucleon in a manner that is independent of the probing reaction. One way to determine GPDs is via DEMP reactions. Because quark helicity is conserved in the hard scattering regime, the produced meson acts as a helicity filter \cite{Goeke_2001}. In particular, leading order perturbative QCD predicts that longitudinally polarized vector meson production (e.g. $\rho^{0,\pm}_{L},\omega_{L}$) is sensitive only to the unpolarized GPDs ($H$ and $E$), whereas pseudoscalar mesons (e.g. $\pi$, $\eta$) produced via longitudinally polarized virtual photons are sensitive only to the polarized GPDs ($\tilde{H}$ and $\tilde{E}$). Thus, DEMP reactions are complementary to the Deeply Virtual Compton Scattering (DVCS) process, as they provide the additional data needed to disentangle the different GPDs.

The GPD $\tilde{E}$ is particularly poorly known \cite{cuic}. It is related to the pseudoscalar nucleon form factor $G_P(t)$, which is itself highly uncertain because it is negligible at the momentum transfer of nucleon $\beta$-decay. $\tilde{E}$ is believed to contain an important pion pole contribution and hence is optimally studied in DEMP. $\tilde{E}$ cannot be related to already known parton distributions, and so experimental information about it can provide new information on nucleon structure, which is unlikely to be available from any other source. Furthermore, this observable has been noted as being important for the reliable extraction of $F_{\pi}$ from pion electroproduction \cite{Fr00}, due to the significant $\pi$ pole contribution.

\begin{figure}[hbt]
\begin{center}
\includegraphics[width=\linewidth]{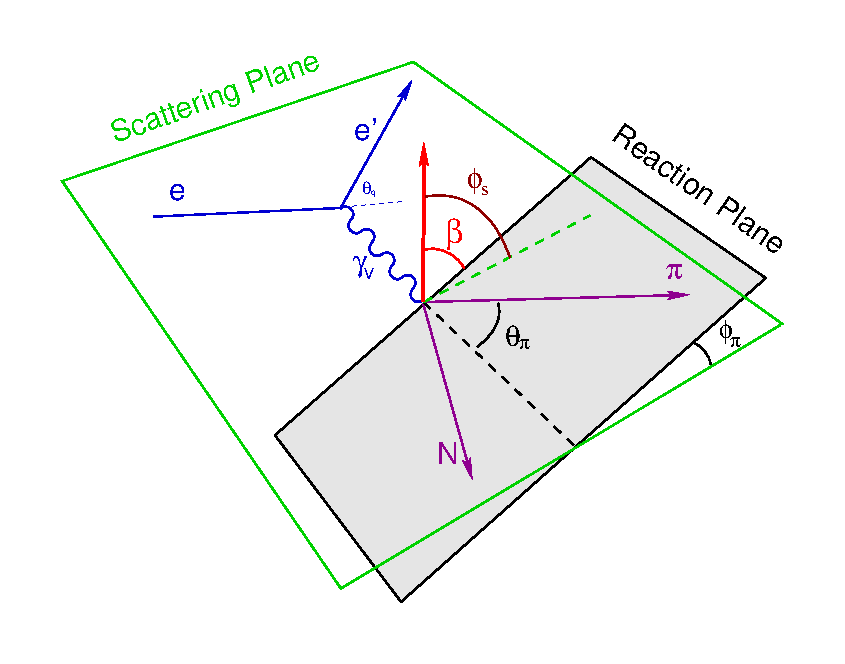}
\end{center}
\caption{\label{fig:planes}
\footnotesize{
Scattering and hadronic reaction planes for exclusive the $\vec{N}(e,e^{\prime}\pi)N'$ reaction. $\theta_q$ is the scattering angle of the virtual photon with respect to the incident electron beam, and $\theta_{\pi}$ is the scattering angle of the pion with respect to the $q$-vector. $\phi_{\pi}$ is the azimuthal angle between the hadronic reaction plane and the electron scattering plane. $\phi_S$ is the azimuthal angle between the target nucleon polarization and the scattering plane, and $\beta=(\phi-\phi_s)$ is the angle between the target nucleon
polarization vector and the reaction plane.}}
\end{figure}

The most sensitive observable to probe $\tilde{E}$ is the transverse single spin asymmetry in exclusive $\pi^{\pm}$ production:
\begin{equation}
  A^{\perp}_{L} = \frac{ \int^{\pi}_{0}d\beta \frac{ d\sigma^{\pi}_{L}}{ d\beta } - 
                       \int^{2\pi}_{\pi}d\beta \frac{ d\sigma^{\pi}_{L}}{ d\beta } }
                     { \int^{2\pi}_{0}d\beta \frac{ d\sigma^{\pi}_{L}}{ d\beta } },
  \label{equation:transasym}
\end{equation}
where d$\sigma^{L}_{\pi}$ is the exclusive $\pi$ cross-section for longitudinal virtual photons and $\beta$ is the angle between the transversely polarized target vector and the reaction plane (Fig.~\ref{fig:planes}). Frankfurt et al. have shown that $A^{\perp}_{L}$ vanishes if $\tilde{E}$ is zero \cite{frank}. If $\tilde{E} \neq 0$, the asymmetry will produce a $\sin(\beta)$ dependence. Refs. \cite{frank,Belitsky2004} note that ``precocious scaling'' is likely to set in at moderate $Q^2\sim 2-4$ GeV$^2$ for this observable, as opposed to the absolute cross section, where scaling is not expected until $Q^2>10$ GeV$^2$.

So that all final state particles are charged, and hence relatively easily detectable experimentally, the fixed target DEMPgen module simulates the reaction $^3He(e,e^{\prime}\pi^-p)(pp_{sp})$ from a transversely polarized target, including Fermi momentum and final state interaction (FSI) effects. This is described in more detail in Sec.~\ref{Para:fermi}.

%%%%%%%%%%%%%%%%%%%%%%%%%%%%%%%%%%%%%%%%%%%%%%%%%%%%%%%%%%%%%%%%%%%%%%%%%%%%%%%%%%%%%%%%%%%%
% Chapter 3
\section{DEMP Event Generator}
\label{sec:generator}

This section describes the structure of the event generator, as well as the model parameterization, scattering cross-section and kinematic ranges for each of the DEMP processes. DEMPgen has two distinct modules, one for colliding beam event generation and one for beam-on-fixed-target event generation. Both modules use a common .json input file, the options for which are detailed in \ref{sec:Appendix_json}.

The colliding beams module has initially been developed to enable the simulation of various DEMP channels at the upcoming EIC. As such, this is referred to as the EIC module. 

The fixed target module is primarily focused on the generation of $^3He(e,e^{\prime}\pi^-p)(pp)_{sp}$ events from a polarized $^{3}He$ target at the upcoming \textbf{So}lenoidal \textbf{L}arge \textbf{I}ntensity \textbf{D}evice (SoLID) experiment at JLab \cite{arrington2023solenoidal}. Consequently, this module is referred to as the SoLID module.

The EIC and SoLID modules are detailed in the following subsections, including the structure of each module and the relevant event kinematics. The physics models utilized in the generator are described in Sec.~\ref{SubSec:PhysicsModels}. Results from studies using both modules are presented in Sec.~\ref{sec:results}. In the following discussion, reaction kinematics will be referred to in generic terms as much as possible using the nomenclature of scattered electron, ejectile (the produced meson), and recoil hadron.

\subsection{EIC Module}
\label{SubSec:EIC_Module}

The EIC module reads in the input .json file and calculates events in colliding beam kinematics. Input parameters specified in this file include the incoming beam energies, random number generator seed, number of events to generate, the output file type, generation ranges, and the reaction to generate. Details on the names of the input parameters, and where relevant their default values, are included in \ref{sec:Appendix_json}. 

Currently, DEMPgen is capable of generating events for the following reactions:
\begin{enumerate}
    \item $p(e,e^{\prime}\pi^+ n)$
    \item $p(e,e^{\prime}K^+\Lambda)$
    \item $p(e,e^{\prime}K^+\Sigma^{0})$.
\end{enumerate}
In the future, DEMPgen will be extended to handle deep exclusive meson production from $t$-channel $e$-A and $u$-channel $e$-$p$ collisions.

The EIC module considers electrons and nucleons as the incoming particles at different beam energies. For electron-proton collisions, four of the proposed beam energy combinations are $5 (e) \times 41 (p)$, $5(e) \times 100(p)$, $10(e) \times 100(p)$, and $18(e) \times 275(p)$. DEMPgen can produce events with any arbitrary electron-proton beam energies, however, only a few ``standard'' beam energy combinations (including those defined above) have projected luminosity values. The luminosity is required for event weight calculations, see Sec. \ref{SubSubSec:EventWeight} for further details on this, and \ref{sec:Appendix_EIC_Lumi} for details on the luminosity values. DEMPgen generates three outgoing particles in the inertial frame of the EIC detector (collider frame) based on the pion or kaon electroproduction reactions. 

\subsubsection{Event Generation}
\label{SubSubSec:PiModel_EventGen}

%If an event fails a cut, no further processing is carried out for this event. Another event is generated, so long as $N_{Gen}$ < $ N_{Requested}$, where $N_{Gen}$ is the number of events generated so far and $N_{Requested$  

\begin{figure*}
	\centering
	\includegraphics[width=1.0\textwidth]{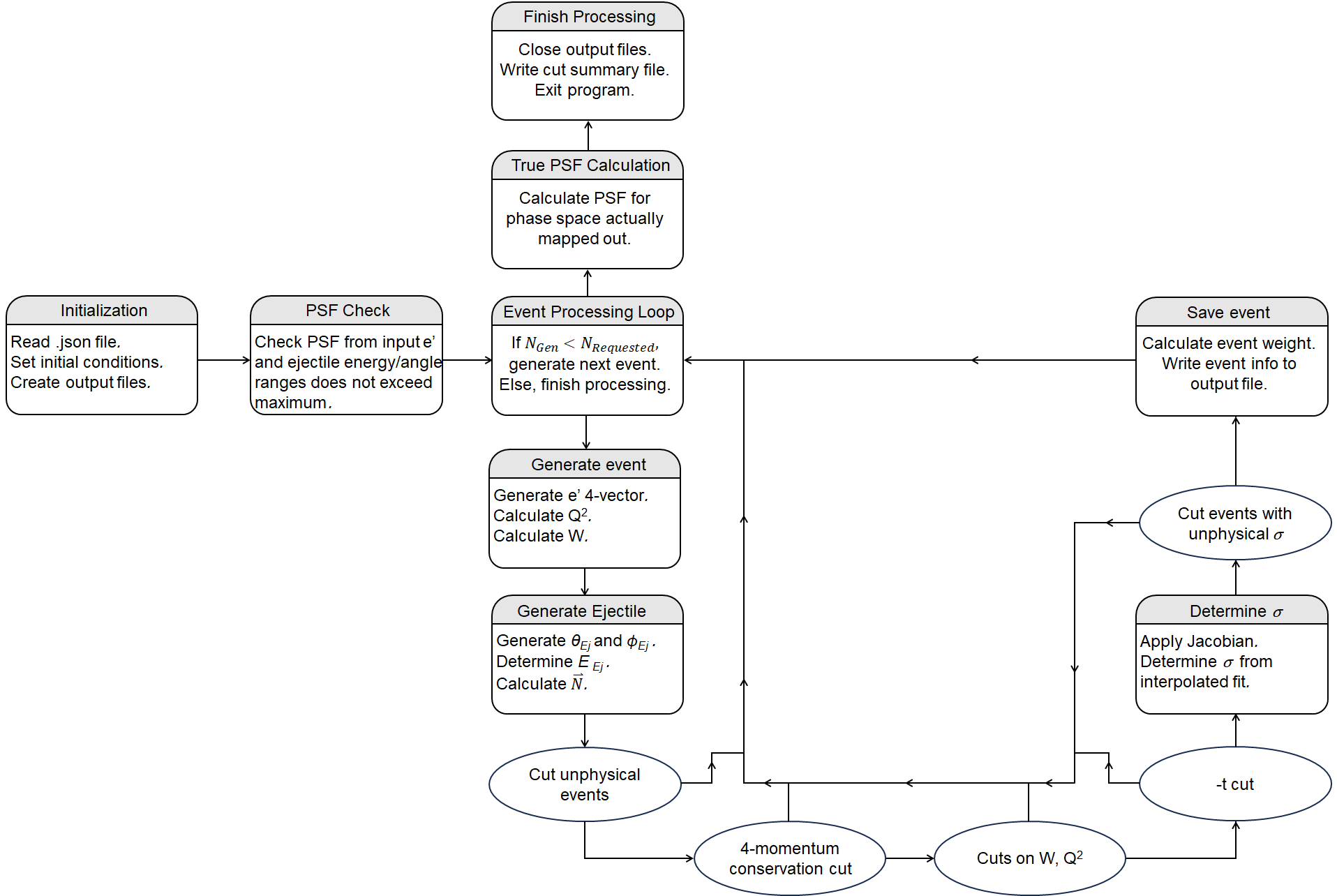} 
	\caption{Flowchart of the basic operation of the EIC module in simplified steps. Key process and steps are described in the rounded boxes. Cuts on the generated events are represented by the elliptical boxes. Beginning from the initialisation box, the chart then flows counter-clockwise. Events passing cuts continue on this counter-clockwise loop, those failing go back to the beginning of the loop as shown. See \ref{sec:Appendix_PSF} for further details on the two PSF calculation process boxes. Many of the event selection cuts are on kinematic quantities that are detailed further in Sec.~\ref{SubSubSec:PiModel_KinDef}. The cut values vary depending upon the requested reaction, cut values for the $p(e,e^{\prime}\pi^+ n)$ reaction are shown in Table~\ref{tab:EIC_CutVals}. This table also indicates which cuts are reaction dependent.}
	\label{fig:EIC_Flowchart}
\end{figure*}

After reading in the .json input file, DEMPgen initialises several parameters and begins generating events. The main event processing loop generates $N_{Requested}$ events (specified in the input .json file). Each pass through the main event processing loop increments $N_{Gen}$ by 1, effectively this is the number of ``tries'' so far. Generally, the majority of the generated events are discarded by various event selection cuts. If an event fails a cut, no further processing is carried out for this event. Another event is generated, so long as $N_{Gen}$ $<$ $ N_{Requested}$. The general sequence and flow of events through the EIC module of DEMPgen are summarized in a flowchart, Fig.~\ref{fig:EIC_Flowchart}. As shown in this figure, cuts are conducted in a specific order and conducted sequentially to reject events as soon as possible and prevent any redundant calculations being conducted. Cut values for $p(e,e^{\prime}\pi^+ n)$ reactions are shown in Tab.~\ref{tab:EIC_CutVals}. As an example, an event generated with $Q^{2} = 0.1~GeV^{2}$ would be rejected and no further processing would occur for this event. 

\begin{table}[!htb]
    \centering
\begin{tabular}{|>{\centering\arraybackslash}m{2cm} | >{\centering\arraybackslash}m{3cm}| >{\centering\arraybackslash}m{2cm}|} 
  \hline
 {\bf Cut} & {\bf Typical Values} & {\bf Reaction Dependent?} \\
  \hline
    Unphysical & $E_{X}$ returns NaN & No\\
  \hline
  4-Momentum Conservation & $0.00001$~GeV & No\\
  \hline
  $W^{2}$ & $W^{2} < 0$~GeV$^{2}$ & No\\
  \hline
  $Q^{2}$ & $Q^{2} < 3~$GeV$^{2}$ \& $Q^{2} > 35$~GeV$^{2}$ & Yes\\
  \hline
  $W$ & $W < 2$~GeV \& $W > 10.2$~GeV & Yes\\
  \hline
  $-t$ & $-t > 1.3$~GeV$^{2}$ & Yes \\
  \hline
  Unphysical $\sigma$ & $\sigma < 0$ or $\sigma$ NaN & No \\
  \hline
  \end{tabular}
    \caption{Table of event selection cuts and their values in the EIC module. 
     Some cut ranges depend upon the reaction specified, as indicated in the table, the values shown are the values for $p(e,e^{\prime}\pi^+ n)$ reactions. Note that for the 4-momentum conservation cut, the difference between the initial and final 4-momenta sum is calculated. All components of this resulting 4-vector are checked against this value. The cut will fail upon any component exceeding the specified value.} 
    \label{tab:EIC_CutVals}
\end{table}

The exclusive nature of DEMP reactions pose some constraints on event generation. Specifically, the relevant differential cross section is 5-fold ($\theta_{e},\phi_{e},E_{e},\theta_{Ej},\phi_{Ej}$), rather than 6-fold ($\theta_{e},\phi_{e},E_{e},\theta_{Ej},\phi_{Ej},E_{Ej}$), as the final states are discrete for the considered reactions. To calculate the cross section, the outgoing ejectile momentum magnitude is uniquely determined from four-momentum conservation at the photon-ejectile vertex once the scattered electron energy and angles, and the outgoing ejectile $(\theta,\phi)$ angles are chosen. 
%The particles are defined in the inertial frame of the EIC detector (collider frame). The energy of the scattered electron is selected from a uniform random distribution in a configurable range. The direction of the scattered electron and the produced ejectile is selected using sphere point picking \cite{spherepp}. The energy of the ejectile is left to be solved for. The angular ranges over which the scattered electron and ejectile are distributed over is a configurable parameter in DEMPgen that is set the .json input card. Similarly, the energy range over which the scattered electron is produced is an input parameter. Default values and further details are included in \ref{sec:Appendix_json}.%
 The energy of the scattered electron is selected from a uniform random distribution in a configurable range specified in the DEMPgen .json input card, given in \ref{sec:Appendix_json}. The direction of the scattered electron and the produced ejectile is selected using sphere point picking \cite{spherepp}. The angular range over which the scattered electron and ejectile are distributed is also a configurable parameter in the DEMPgen .json input card. The energy of the ejectile is left to be solved for.

These variables provide all the information necessary to uniquely solve for all remaining kinematic variables. Applying conservation of energy and momentum yields the following equation:
\begin{equation}
  \label{eqn:kinsolve}
  \nu + E_{N} -\sqrt{m_{Ej}^2+|\vec{X_{Ej}}|^2 } -
  \sqrt{m_{Rec}^2+|\vec{q}+\vec{N}-\vec{X_{Ej}}|^2 }=0,
\end{equation}
where $\nu$ and $E_{N}$ represent the energy of the virtual photon and the nucleon, respectively. The vectors represent the three-momenta of the respective particles. The only unknown in this equation is the momentum vector of the ejectile, $\vec{X_{Ej}}$. Since the direction of the ejectile has already been specified, Eqn.~\ref{eqn:kinsolve} can be further reduced to a single-valued unknown: the magnitude of the ejectile momentum.

The energy of the ejectile is determined analytically. Eqn.~\ref{eqn:kinsolve} is modified in terms of the energy of the ejectile in the collider frame and then solved to get a quadratic equation. Finally, the coefficients of the quadratic equation are defined and calculated,
\begin{equation}
  \label{eqn:kinsolve2}
  a [E^{2}_{Ej}] + b [E_{Ej}] + c=0.
\end{equation}
 Here, $a$, $b$, and $c$ depend on the known quantities, such as the four-momenta of other particles and direction of ejectile and are specified in ~\ref{sec:Appendix_quad}. Using the quadratic formula, the solutions of Eqn.~\ref{eqn:kinsolve2} are determined. The direction and momentum of the recoil hadron are generated by applying conservation of energy and momentum at the physics reaction vertex. %Some cuts are applied on the kinematic variables to select the required events as given in Table~\ref{tab:kin}.

%{\bf The below text is for the solve method for generating the particle 4-vectors.}
%The left hand side of the equation is defined in the event generator as a function of $|\vec{\pi}|$ in a ROOT TF1 (1-Dim function) object.  The TF1 object incorporates a root finding algorithm, using Brent's method \cite{ROOT}, which is used to find values for $|\vec{\pi}|$ which solve Equation \ref{eqn:kinsolve}.  If more than one solution is found, one is picked at random.

 Kinematic quantities are determined as soon as the relevant information is available to calculate them. E.g., as soon as the ejectile 4-vector is determined, $t$ and $u$ can be calculated. Selection cuts are applied to the events, as shown in Fig~\ref{fig:EIC_Flowchart}, and detailed in Tab.~\ref{tab:EIC_CutVals}. As a consequence of these cuts, the actual number of successfully generated events will be different (or smaller) than the number of events tried. If an event passes all selection cuts, a cross section value is determined from a parameterized model as described in Sec.~\ref{sec:pi-model}. As a validation check of this determination, events that return negative cross section values or NaN are removed. Such cases are very rare, this check is only present as a final verification of the generated events. This cross section is then used to determine a weight for the event, as described in the next section, and the event information is saved to the output file. 

\subsubsection{Event Weighting}
\label{SubSubSec:EventWeight}

DEMPgen produces events with variable weight, corresponding to the rate of the given reaction at the input luminosity. Following the generation of events as presented in Sec.~\ref{SubSubSec:PiModel_EventGen}, every event that passes all of the selection cuts is assigned a weight value as follows, 
\begin{equation}
    \label{eqn:Weight}
    \mathrm{Weight} = \frac{\sigma \times PSF \times CF \times \mathcal{L}}{N_{Requested}},
\end{equation}
where $\sigma$ is the 5-fold differential cross-section in the collider frame, $PSF$ is the phase space factor (see \ref{sec:Appendix_PSF} for more details), $CF$ is a conversion factor to convert $\mu$b to cm$^{2}$, $\mathcal{L}$ is the luminosity (in units of cm$^{-2}$s$^{-1}$), and $N_{Requested}$ is the total number of events that the generator tried to produce. $N_{Requested}$ includes events that were discarded due to either falling outside of acceptable parameters, or having no valid solutions in the kinematics solver. 
If desired, the correction factor to convert the denominator in Eqn.~\ref{eqn:Weight} to $N_{physical}$, where unphysical events are excluded, is given in the cut summary output file.
The value of $\sigma$ is determined for the generated kinematics from a parameterized model calculation, further details on the models used and the parameterization for each physics process are provided in Sec.~\ref{SubSec:PhysicsModels}.

The resulting weight value is in units of Hz. The output of the generator is a CERN-ROOT file and a LUND format file for the SoLID module. For the EIC module, only a text file of events is produced, this can be in LUND, Pythia6 or HEPMC3 format depending upon what the user specifies in the control card (see \ref{sec:Appendix_json} for more info).

%------------------------------------------------------------------------------
%== from Rory's thesis
\subsection{SoLID Module}
\label{SubSec:SoLID_Module}
% 04/10/22 - SJDK - Go through section A along with the flowchart and check that this is still valid/accurate

% SK - Some sentence tweaks - SK 16/12/20
\begin{figure*}
	\centering
	\includegraphics[width=.95\textwidth]{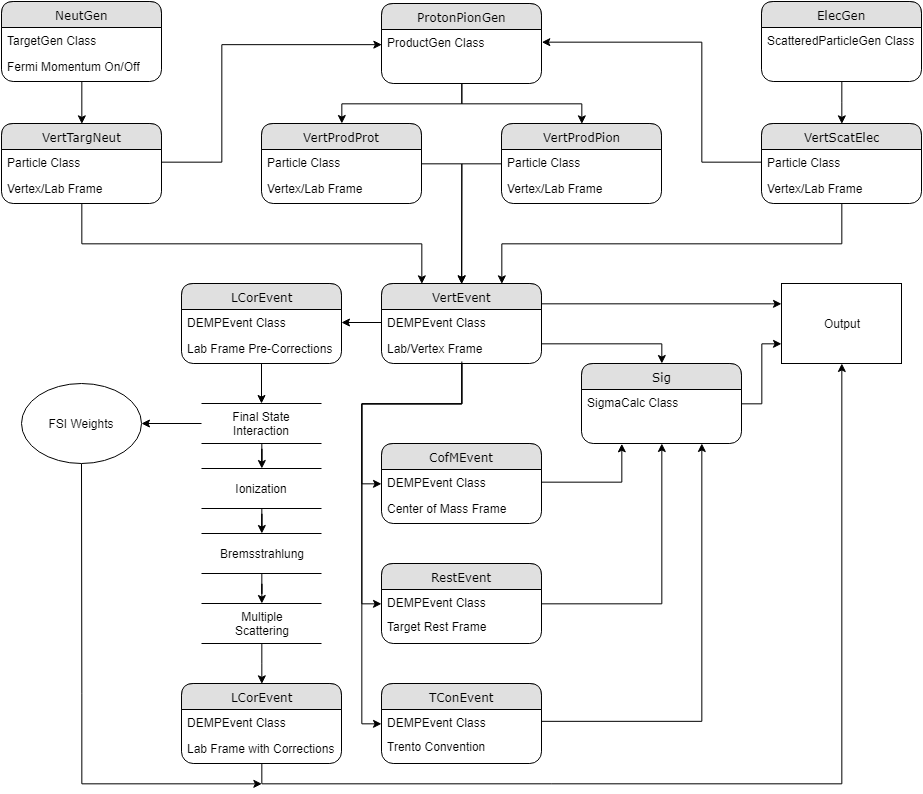} 
	\caption{Flowchart describing the flow of data through the program, and the structure of the SoLID module of the event generator. The rounded boxes signify the most important named variables that appear in the main file of the program. Their values are pointers to an instance of the given class. Arrows indicate how data is moved between these objects. The four open boxes indicate static functions. The ``FSI Weights'' node represents a simple collection of doubles. The ``Output'' box represents the destination for data to be saved into the output CERN-ROOT tree.}
	\label{fig:flowchart}
\end{figure*}

Fig.~\ref{fig:flowchart} demonstrates the flow of data in the event generator for the SoLID module, starting from the random number generators, and leading to the output file. The rounded boxes in this chart each indicate the main named variables whose values are pointers to instances of the indicated class. For example, ``VertTargNeut'' is a pointer to an object of class ``Particle''. These variables, the classes, and their place in the structure of the event generator, are discussed in the following paragraphs.

\subsubsection{Particle Class}\label{SubSubSec:particleclass}

Instances of the particle class contain all of the pieces of information about a single particle, in a single frame of reference, that are relevant to the event generator. This includes the four-momentum, rest mass, charge, etc. The particle inherits from CERN-ROOT's TLorentzVector class, which allows for the creation and manipulation of general four-vectors. The TLorentzVector class includes methods to calculate components, angles, and magnitudes of a four-momentum, as well as perform Lorentz boosts and rotations. It also defines algebraic operators for four-vectors. Implementation of this class significantly simplifies calculations within the event generator, and eliminates a large number of messy algorithms that would otherwise need to be present.

\subsubsection{DEMPEvent class}\label{SubSubSec:dempevent}

Instances of the DEMPEvent class represent the event viewed from a single reference frame. The class stores seven particle objects: the incident electron, target nucleon, virtual photon, scattered electron, produced meson, and recoiled nucleon. The class has methods to calculate Mandelstam variables $s$, $t$, $u$, and functions to perform coordinate transformations on the event. There are five DEMPEvent objects initialized in the event generator, as seen in Fig.~\ref{fig:flowchart}. VertEvent contains the particles as viewed at the vertex of the interaction, in the laboratory rest frame. Once the kinematics calculations have been completed, this object is no longer modified. All other DEMPEvent objects are calculated from this object by copying and then transforming them. CofMEvent is the event viewed at the vertex in the center of the momentum reference frame. RestEvent is the event viewed at the vertex in the rest frame of the target neutron. TConEvent is the event viewed at the vertex in the coordinate system defined by the Trento Conventions \cite{Trento}.

\subsection{Physics Models}
\label{SubSec:PhysicsModels}

As discussed in Sec.~\ref{SubSubSec:EventWeight}, generated events are assigned a weight calculated from various quantities. One of these quantities is the 5-fold differential cross-section, $d^5\sigma/dE{e'}d\Omega_{e'}d\Omega^*_{Ej}$. This cross section is determined based on the kinematics of the generated event using parameterized physics models. The utilized models and their parameterizations are described in this section.

\subsubsection{Kinematic definitions}
\label{SubSubSec:PiModel_KinDef}

%------------------------------------------------------------------------------

In general, the following Lorentz invariants are used to define the kinematics of $p\left(e,e^{\prime}X_{Ej}X_{Rec}\right)$ events:
\begin{gather}
  -Q^2 = (p_e-p_{e^{\prime}})^2\\
  \label{equation:equation-qsq}
%\end{equation}
%\begin{equation}
  W^2 = (p_{\gamma^*}+p_{p})^2 = (p_{Ej}+p_{Rec})^2\\
%  \label{equation:equation-w}
%\end{equation}
%\begin{equation}
  t = (p_{\gamma^*}-p_{Ej})^2 = (p_{p}-p_{Rec})^2\\
%  \label{equation:equation-t}
  u = (p_{\gamma^*}-p_{Rec})^2 = (p_{p}-p_{Ej})^2,
\end{gather}
where for the EIC module, $p_{e}$, $p_{e^{\prime}}$, $p_{\gamma^*}$, $p_{p}$, $p_{Ej}$, and $p_{Rec}$ represent four vectors for the electron beam, scattered electron, virtual photon, proton beam, ejectile, and recoil hadron, respectively. For the SoLID module with a quasi-free $^3$He target configuration, substitute the neutron four momentum ($p_{n}$) for the proton four momentum ($p_{p}$) in the above equations.

In the fixed target frame, $t$ can be expressed as 
\begin{equation}
    t = (E_{Ej}-\nu)^2 - |\vec{p_{Ej}}|^2 - |\vec{q}|^2 + 2|\vec{p_{Ej}}||\vec{q}|\cos \theta_{Ej q},
\end{equation}
where $\nu, \vec{q}$ are the energy and three-momentum of the virtual photon, and 
$\theta_{Ej q}$ is the angle between the ejectile and the $q$-vector shown in Fig.~\ref{fig:planes} (as $\theta_{\pi}$).
For spacelike DEMP reactions, $t$ is always negative, so the variable $-t$ is used throughout the paper. The minimal value of $-t$ (or $-t_{min}$) is obtained when $\theta_{Ej q}=0$, i.e. when the ejectile is emitted in the direction of the virtual photon (referred to as parallel kinematics). $-t_{max}$ is the maximum value of $-t$, corresponding to anti-parallel kinematics, $\theta_{Ej q}=\pi$. $u$ is not always negative; it takes its most negative value $-u_{max}$ at $\theta_{Ej q}=0$, passes through zero just before anti-parallel kinematics, and ends slightly positive at $\theta_{Ej q}=\pi$ (taken as $-u_{min}$). For convenience, $t'=t-t_{min}$ and $u'=u-u_{min}$ are defined.

We define the missing mass and momentum in DEMP as follows:
\begin{align}
\centering
	\vec{p}_{miss} &= \vec{p}_{e} + \vec{p}_{N}- \vec{p}_{e^{\prime}} - \vec{p}_{Ej} 
\label{eqn:pm}
 \\
	E_{miss} &= E_{e}+E_{N}-E_{e^{\prime}}-E_{Ej} \\
	m_{miss}^2 &= E_{miss}^2 - p_{miss}^2 ,
    \label{eqn:mm}
\end{align}
where $N$ refers to the quasi-free target nucleon (proton or neutron, e.g. from $^{3}$He). For DEMP reactions, $m_{miss}$ must equal the mass of the recoil particle.

\subsubsection{Cross Section Formalism}
\label{Sec:Cross_Sec_Form}

For the event weighting in Eqn. \ref{eqn:Weight}, the five-fold differential cross section in the relevant detector frame is required.
In the one-photon exchange approximation, the reduced five-fold differential cross section for DEMP in terms of virtual photon flux factor, $\Gamma_{V}$, and a virtual photon cross-section, $\frac{d{^2} \sigma}{d\Omega^{cm}_{Ej}}$ is given by Eqn.~\ref{equation:cross-1}, with incoming and outgoing particles described by plane waves
\begin{equation}
  \frac{d^{5} \sigma}{d E_{e^{\prime}} d\Omega_{e^{\prime}} d\Omega_{Ej}} =
  \Gamma_{V} \biggl( \frac{d{^2} \sigma}{d\Omega^{cm}_{Ej}} \biggr) \biggl( \frac{d\Omega^{cm}_{Ej}}{d\Omega_{Ej}} \biggr) ,
  \label{equation:cross-1}
\end{equation}
where $E_{e^{\prime}}$, $\Omega_{e^{\prime}}$ are the scattered electron's energy and solid angle in the detector frame, respectively, and $\Omega^{cm}_{Ej}$, $\Omega_{Ej}$ is the ejectile solid angle in the center-of-mass and detector frames. 

The virtual photon flux factor, $\Gamma_{V}$, can be written as
\begin{equation}
  \Gamma_{V}=\frac{\alpha}{2\pi^2} \frac{E_{e^{\prime}}}{E_e} \frac{K}{Q^2}\frac{1}{1-\epsilon},
  \label{equation:photon-flux-1}
\end{equation}
where $\alpha$ is the fine structure constant, the factor $K=(W^2-M_{N}^2)/(2 M_{N})$ is the equivalent real photon energy \cite{hand}, i.e., the laboratory energy required by a real photon to excite a target of mass, $M_{N}$, and create a system with invariant mass equal to $W$. 

$\epsilon$, the polarization of the virtual photon, is given in the fixed target frame as 
\begin{equation}
  \epsilon=\left(1+\frac{2 |\mathbf{q}|^2}{Q^2} \tan^2\frac{\theta_{e}}{2}
  \right)^{-1}.
  \label{equation:photon-flux-3}
\end{equation}
For colliding beams, $\epsilon$ can be expressed in terms of the fractional energy loss of the collision, $y$, where 
\begin{gather}
    y = \frac{Q^{2}}{x_B(s_{tot}-M_{N}^{2})},
    \label{eqn:y_frac_ELoss}
\end{gather}
and
\begin{gather}
    \epsilon = \frac{2(1-y)}{1+(1-y)^{2}},
\end{gather}
where $x_B$ is the Bjorken scaling variable and $s_{tot}$ is the square of the center of mass energy of the system.

The two-fold differential cross-section in Eqn.~\ref{equation:cross-1} can be expressed in terms of the invariant cross-section as
\begin{equation}
  \frac{d^2 \sigma}{d\Omega^{cm}_{Ej}}= J_A \frac{d^2 \sigma}{dt d\phi},\\
  \label{equation:cross-2}
\end{equation}
where $J_A$ is the Jacobian factor to transform $t, \phi$ to $\Omega^{cm}_{Ej}$.

The general two-fold differential cross section in Eqn.~\ref{equation:cross-2} can be expressed in terms of four structure functions:
\begin{equation}
\begin{split}
  2\pi \frac{d^2 \sigma}{dt d\phi} &= \epsilon  \frac{d\sigma_{\mathrm{L}}}{dt}
  + \frac{d\sigma_{\mathrm{T}}}{dt} \\
  &+ \sqrt{2\epsilon (\epsilon +1)} \frac{d\sigma_{\mathrm{LT}}}{dt} \cos{\phi} +
  \epsilon \frac{d\sigma_{\mathrm{TT}}}{dt} \cos{2 \phi},
  \end{split}
  \label{equation:cross-3}
\end{equation}
%------------------------------------------------------------------------------
where the subscripts $L$ and $T$ represent the longitudinal and transverse polarizations of the virtual photon. For brevity, we refer to $d\sigma_L/dt$ as $\sigma_L$, and so on.
To study DEMP at the EIC, the cross terms $\sigma_{LT}$ and $\sigma_{TT}$, which arise from longitudinal transverse and transverse transverse interference states of the virtual photon, are ignored as they are expected to be small, and even more highly uncertain than $\sigma_L$ and $\sigma_T$.

%------------------------------------------------------------------------------
% SJDK - 09/12/22 - These equations are defined on lines 468-482 of PiPlusProd.cc, slowly making sense of them
Finally, for the EIC module, the five-fold cross-section is transformed into the collider frame with the help of the following Jacobians:
\begin{gather}
    \label{eqn:EIC_Jacobian_Line1} J = J_A \times J^{cm}_{col},\\
        \label{eqn:EIC_Jacobian_Line2} J^{cm}_{col} = \frac{d\Omega^{cm}_{Ej}}{d\Omega^{col}_{Ej}}=\frac{ |\vec{p}^{col}_{Ej}|^{2}}{\gamma^{col}_{cm} |\vec{p}^{cm}_{Ej}| \left( 
    |\vec{p}^{col}_{Ej}| - \beta^{col}_{cm} E^{col}_{Ej} \cos\theta^{col}_{Ej} \right) },\\
       \label{eqn:EIC_Jacobian_Line3} J_A = \frac{dt}{d\cos\theta^{cm}_{Ej}}=
       2 | \vec{p}^{cm}_{\gamma^*}| |\vec{p}^{cm}_{Ej}|,\\
%  \label{eqn:EIC_Jacobian_Line3} A = J_{cm} \frac{ |\vec{p}^{cm}_{Ej}| }{ \pi },\\
   \label{eqn:EIC_Jacobian_Line4}\mathrm{where}\ |\vec{p}^{cm}_{\gamma^*}| =   
   \gamma^{rf}_{cm} \left(
   p^{rf}_{\gamma^*} - \beta^{rf}_{cm} E^{rf}_{\gamma^*} \right) .
\end{gather}
In Eqns.~\ref{eqn:EIC_Jacobian_Line2}-\ref{eqn:EIC_Jacobian_Line4}:
$\vec{p}^{col}_{\gamma^*}$, $\vec{p}^{rf}_{\gamma^*}$ are the three momentum vectors of the virtual photon in the collider and proton's rest frames,
$E^{rf}_{\gamma^*}$ is the energy of the virtual photon in the proton's rest frame, 
$\vec{p}^{col}_{Ej}$, $\vec{p}^{cm}_{Ej}$ are the three-momentum vectors of the ejectile in the collider and center-of-mass frames, 
$E^{col}_{Ej}$ is the ejectile's energy in the collider frame,  
and $\theta^{col}_{Ej}$ is the ejectile angle in the collider frame. 
Additionally,
$\beta^{col}_{cm}$, $\gamma^{col}_{cm}$ refer to the speed of the center of mass frame in the collider frame, and $\beta^{rf}_{cm}$, $\gamma^{rf}_{cm}$  are of the center of mass frame relative to proton's rest frame, respectively.  An additional factor of $1/2\pi$ comes from $d\phi$ when converting from $d\cos\theta^{cm}_{Ej}$ to $d\Omega^{cm}_{Ej}$.

\begin{figure*}[!hbtp]
    \centering
    \includegraphics[width=0.48\linewidth]{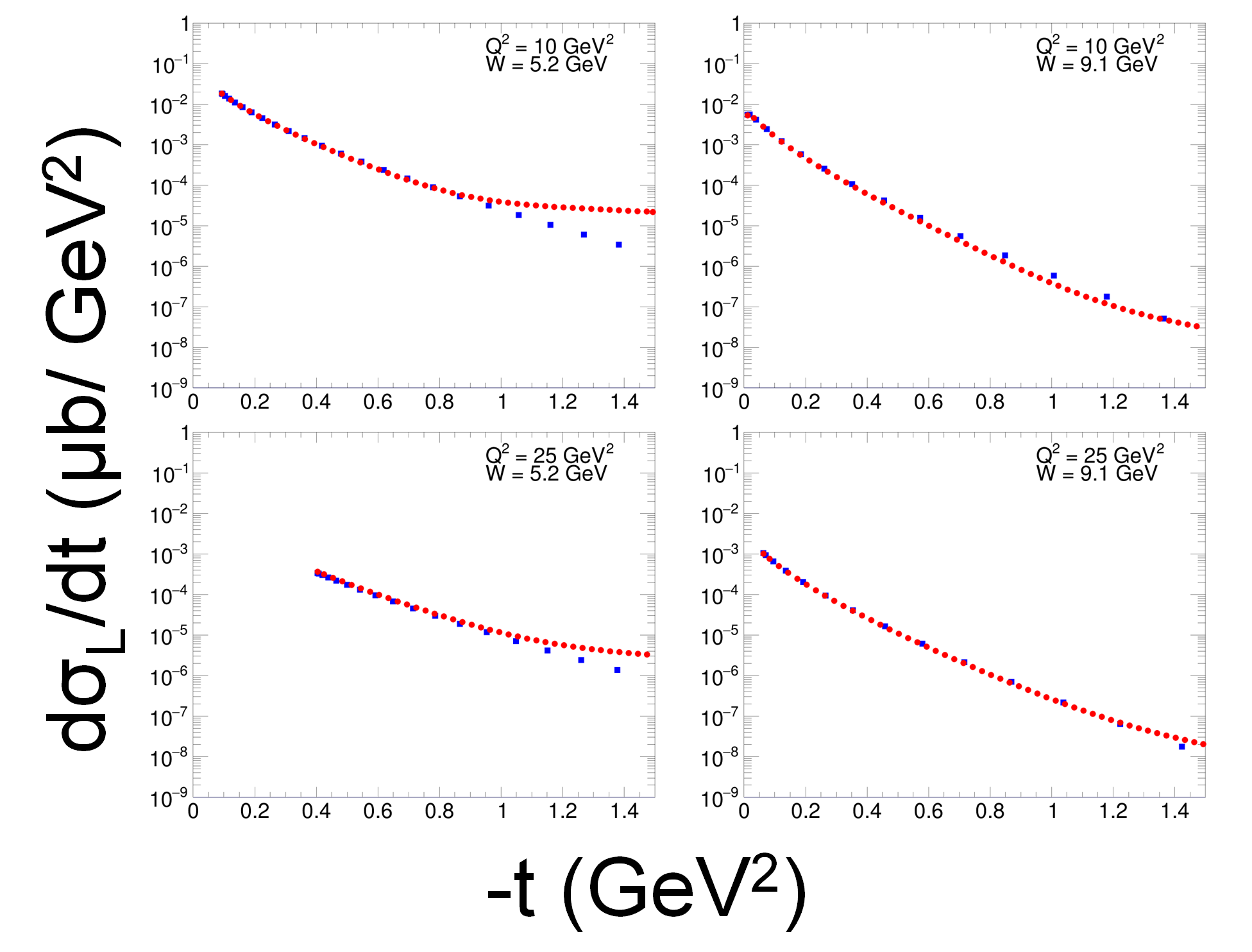}
    \includegraphics[width=0.48\linewidth]{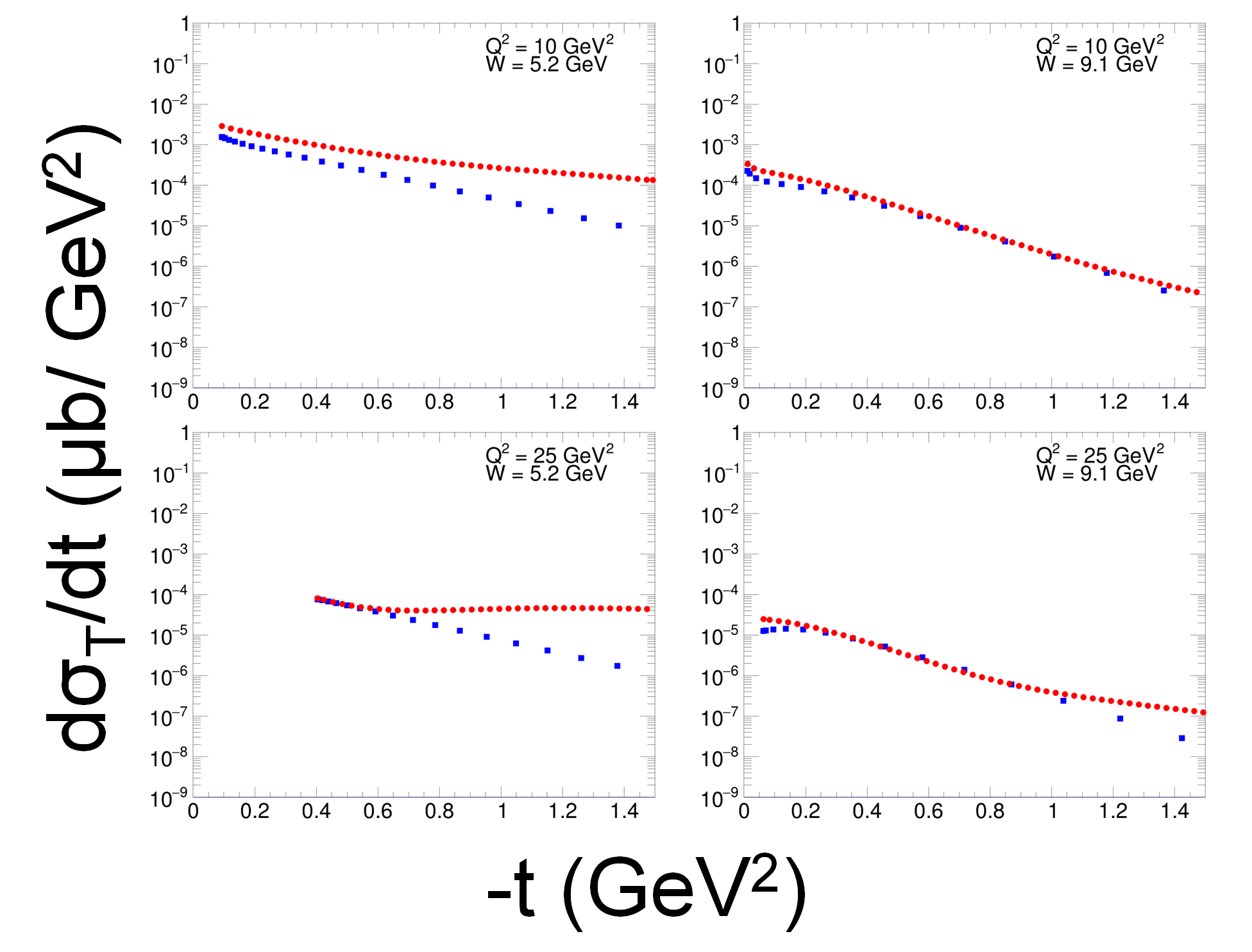}
        \caption{Comparison of the VR (red circles) and CKY (blue squares) models for $\sigma_{L}$ (left) and $\sigma_{T}$ (right) of the EIC pion module for selected kinematics.}
    \label{fig:eic_sig}
\end{figure*}

\begin{figure*}[!hbtp]
  \centering
\mbox{
    \includegraphics[width=0.95\linewidth]{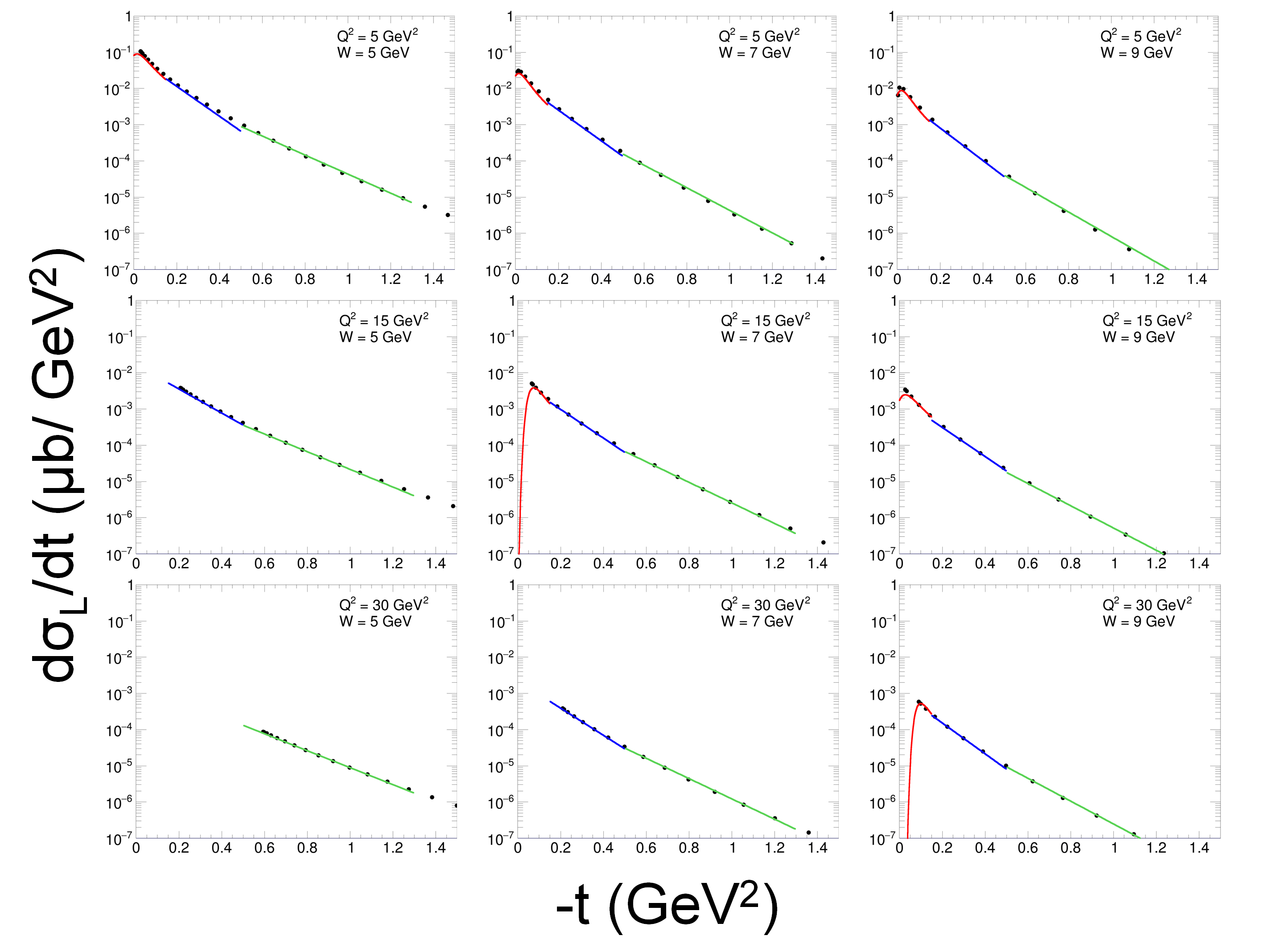}
}
    \caption{Parameterization of CKY $\sigma_{L}$ for the EIC pion module with CKY model values (black circles) plotted along with the landau (red line) and exponential fits (blue and green lines) in Eqn. \ref{sigl_para}.}
    \label{fig:eic_sigl}
%  \end{minipage}
\end{figure*}

\begin{figure*}[!hbtp]
  \centering
\mbox{
    \includegraphics[width=0.95\linewidth]{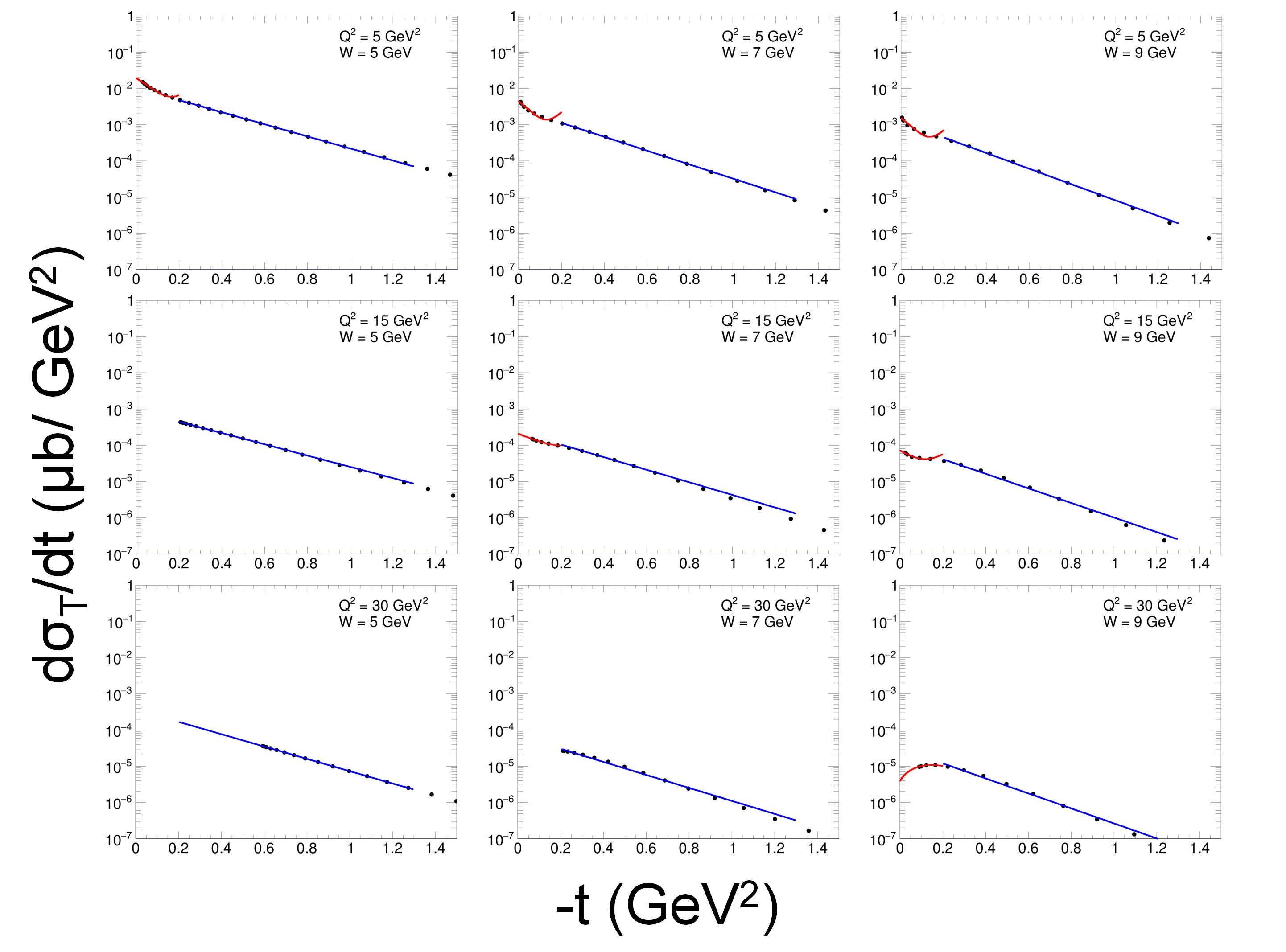}
}
    \caption{Parameterization of CKY $\sigma_{T}$ for the EIC pion module with CKY model values (black circles) plotted along with the second order polynomial (red line) and exponential fits (blue line) in Eqn. \ref{sigt_para1}.}
   \label{fig:eic_sigt}
%  \end{minipage}
\end{figure*}

\subsubsection{Cross Section Parameterization}

The following sections provide details of selecting the appropriate theoretical model, parameterizing various cross section components using the model, and implementing the parameterizations into DEMPgen to determine the total differential cross-section, along with assigning the suitable weights to the events to enable the study of different reactions.

\subsubsection{Exclusive $p(e,e^{\prime}\pi^+ n)$ Physics Model
\label{sec:pi-model}}

\label{SubSubSec:PiModel_Parameter}
\paragraph{VR and CKY models}

The VR model \cite{VRPaper} by Tom Vrancx and Jan Ryckebusch introduces a strong hadronic form factor in the Reggeized background amplitudes to improve the description of $\sigma_T$, while retaining good agreement with $\sigma_L$ data, at low $-t$ and $W>2$ GeV. The VR model for pion electroproduction is fine-tuned with L/T-separated data up to $-t \lesssim 0.5$ GeV$^2$ and $0.7<Q^2<4.35$ GeV$^2$. 

The CKY \cite{Choi:2015yia} model by Tae Keun Choi, Kook Jin Kong and Byung Geel Yu is also a Regge-based model. The CKY model accounts for the importance of the roles of the pion and proton form factors in DEMP to provide a good description of separated $W>2$ GeV DESY and JLab data for $-t<0.7$ GeV$^2$, and unseparated data for $-t< 5$ GeV$^2$, both up to $Q^2\approx5$ GeV$^2$ \cite{Choi:2015yia,Basnet:2019cpg}. 

%------------------------------------------------------------------------------
\paragraph{Comparison of two models}

For EIC kinematics, a detailed comparison of the VR and CKY models was undertaken for $\sigma_{L}$ and $\sigma_{T}$. A typical graph of each $\sigma_{L,T}$ is shown in Fig.~\ref{fig:eic_sig}. As shown in Fig.~\ref{fig:eic_sig}, the VR and CKY models are in generally good agreement with each other at low $-t$. However, at higher $-t$, the two models differ from each other in some cases and it was observed that the CKY model behaves more consistently than the VR model.

Therefore, DEMPgen utilizes the CKY model to determine the cross section and assign each event an appropriate weight. The cutoff mass parameter values for $\pi$, $\rho$,  and $p$ (proton) trajectories were chosen to be $\Lambda_{\pi}$ = 0.65 GeV, $\Lambda_{\rho}$ = 0.782 GeV, and $\Lambda_{p}$ = 1.55 GeV to fit high the $Q^{2}$ region.  We also have performed simulations demonstrating the feasibility of pion electric form factor, $F_{\pi}$, measurements at the EIC using this event generator, as presented in Sec.~\ref{Sec:EIC_PiPlus}.

%------------------------------------------------------------------------------ 

%SK - Sentence corrections - grammar/formatting- SK 
\paragraph{Model Implementation in DEMPgen}

For the cross-section parameterization of the pion module, the following ranges are chosen, $W$ from 2 to 10.2 GeV, $Q^{2}$ from 3 to 35 GeV$^{2}$, and $-t$ up to 1.3 GeV$^{2}$. There are 22 bins of $W$, each of 0.2 GeV width. For each $W$ bin, there are 33 $Q^{2}$ bins, each of 1 GeV$^{2}$ width. For each unique bin of $W$ and $Q^{2}$, $\sigma_{L}$ and $\sigma_{T}$ are parameterized against $-t$ from 0 GeV$^{2}$ to 1.3 GeV$^{2}$, as shown in Figs.~\ref{fig:eic_sigl},~\ref{fig:eic_sigt}. 

In order to make the event generator more efficient and save CPU time, some hard cuts are applied, as discussed in Sec.~\ref{SubSubSec:PiModel_EventGen}. Events with $Q^{2} < 5$~GeV$^{2}$, $W < 3$~GeV, $W > 10.6$~GeV are ignored. If the FF (Form Factor) generator option is used, then $-t > 0.6$~GeV$^{2}$ events are removed, or if the TSSA (Transverse Single Spin Asymmetry) generator option is used, then $-t>1.3$~GeV$^{2}$ events are removed. 

When an event is generated with specific values of $Q^{2}$ and $W$, the generator looks for the closest parameterization combination bin and uses the relevant function to determine the cross-section value.

%------------------------------------------------------------------------------
\paragraph{Parameterization of $\sigma_{L}$}

$\sigma_{L}$ is parameterized with a Landau function, $\mathcal{L}_{Landau}$, and two exponential functions as described below:
\begin{equation}
  \label{sigl_para}
  \sigma_{L}(Q^{2}_{bin},-t,W_{bin}) = \left\{
  \begin{array}{l l}
    \mathcal{L}_{Landau}, \quad 0 \leq -t < 0.15 \\
    exp(c_{1}+c_{2}|-t|), \quad 0.15 \leq -t < 0.5 \\
    exp(c_{3}+c_{4}|-t|), \quad 0.5 \leq -t < 1.3 
  \end{array} \right.
\end{equation}
For an analytic expression of $\mathcal{L}_{Landau}$, please see the relevant CERN-ROOT documentation \cite{ROOT}.

%------------------------------------------------------------------------------ 
\paragraph{Parameterization of $\sigma_{T}$}

%$\sigma_{T}$ is parameterized with a second order polynomial function of $-t$ up to 0.2 GeV$^{2}$ and then with an exponential function of $-t$ for $0.2 < -t <1.3$ for each bin of $Q^{2}$ and $W$.

$\sigma_{T}$ is parameterized with a polynomial and an exponential function as described by Eqn.~\ref{sigt_para1}
\begin{equation}
  \label{sigt_para1}
  \sigma_{T}(Q^{2}_{bin},-t,W_{bin}) = \left\{
  \begin{array}{l l}
    c_{0}+c_{1}|-t|+c_{2}|-t|^{2}, \quad 0 \leq -t < 0.2 \\
    exp(c_{3}+c_{4}|-t|), \quad 0.2 \leq -t < 1.3 
  \end{array} \right.
\end{equation}

%--------------------------------------------------------------------------
\begin{figure*}[!hbtp]
  \centering
\mbox{
    \includegraphics[width=0.95\linewidth]{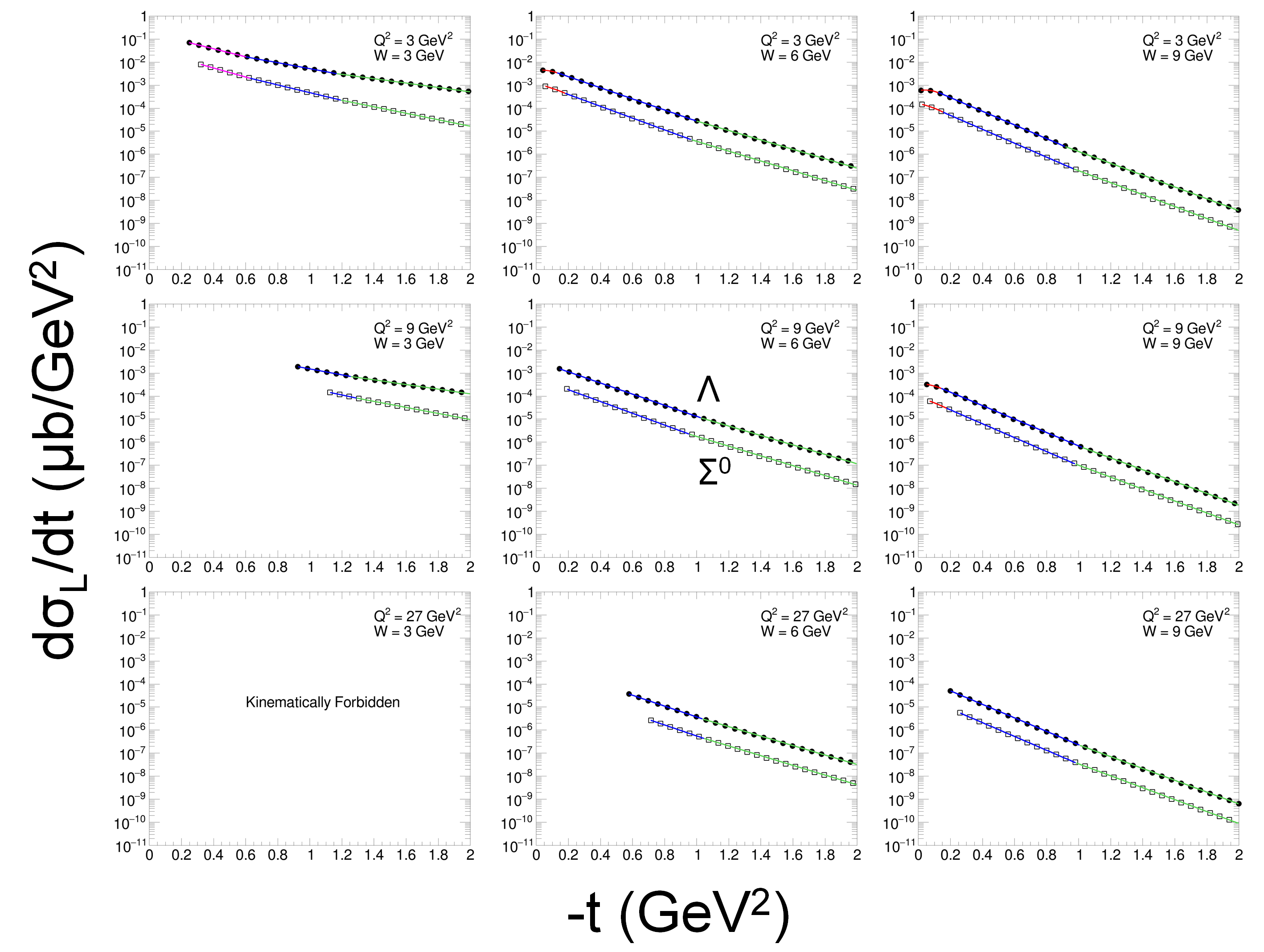}
}
    \caption{Parameterization of VGL $\sigma_{L}$ for the EIC kaon module with $\Lambda$ channel values (black circles) and $\Sigma^{0}$ channel values (black squares) plotted along with the polynomial (red line) and exponential fits (magenta, blue, and green lines) in Eqns. \ref{sigl1_para}, \ref{sigl2_para}.}
%  \end{minipage}
\label{fig:eic_sigl_kaon}
\end{figure*}

\begin{figure*}[!hbtp]
  \centering
\mbox{
    \includegraphics[width=0.95\linewidth]{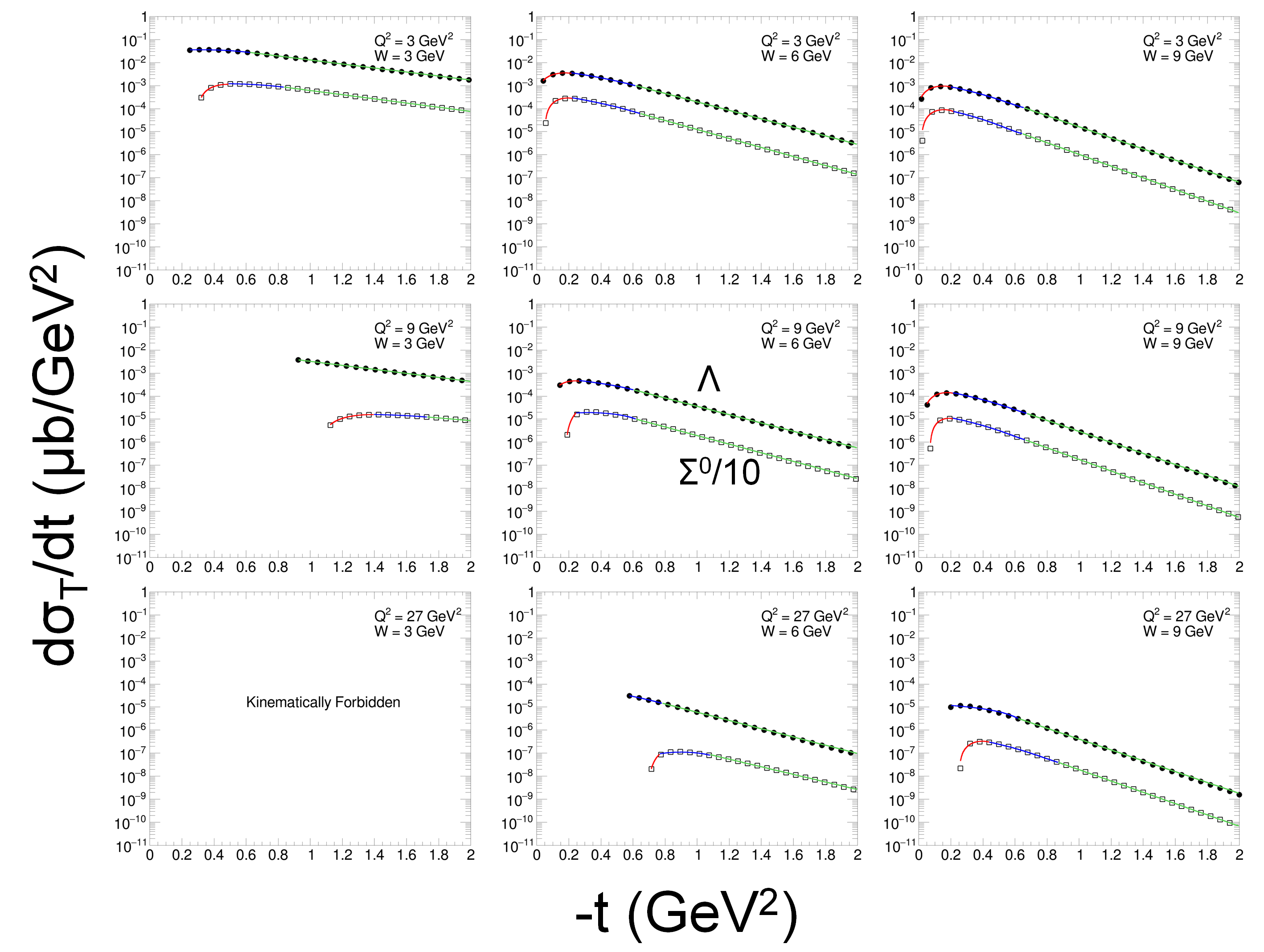}
}
    \caption{Parameterization of VGL $\sigma_{T}$ for the EIC kaon module with $\Lambda$ channel values (black circles) and $\Sigma^{0}$ channel values (black squares) plotted along with the polynomial (red and blue lines) and exponential fits (green line) in Eqn. \ref{sigt_para}. The $\Sigma^{0}$ channel is scaled down by a factor of 10 to make the two channels visible.}
    \label{fig:eic_sigt_kaon}
%  \end{minipage}
\end{figure*}

\subsubsection{Exclusive $p(e,e^{\prime}K^+\Lambda[\Sigma^{0}]$) Physics Model} 

The VGL model \cite{VGL_paper1, VGL_paper2, VGL_paper3, VGL_paper4} by Vanderhaeghen, Guidal, and Laget describes exclusive hadronic reactions above the resonance region (above $W\approx 2$ GeV) and at low four-momentum transfer ($-t<2$ GeV$^{2}$). The framework of this model is based on the exchange of one or more meson Regge trajectories ($K$ and $K^{*}$ trajectories for kaon production) in the $t$-channel. The model cutoff mass parameter values for these trajectories are taken as $\Lambda_K^{2}$ = $\Lambda_{K^{*}}^{2}$ = 1.5 GeV$^{2}$ to fit the high $Q^{2}$ behavior of $\sigma_{L}$ and $\sigma_{T}$. Similar to the pion module, a detailed comparison of the VR \cite{VRPaper2} and VGL models was undertaken for $\sigma_{L}$ and $\sigma_{T}$ over a wide range of EIC kinematics. In this case, the CKY $K^+$ model was not yet available for comparison. It was found that at higher $-t$, the VGL model behaves more consistently than the VR model, therefore, DEMPgen uses the VGL model to determine the event cross section (and eventually the weight) for the $\Lambda$ and $\Sigma^{0}$ channels of the kaon module.

\paragraph{Model Implementation in DEMPgen}

For the cross section parameterization of the $\Lambda$ and $\Sigma^{0}$ channels, the following ranges are chosen, $W$ from 2 to 10 GeV, $Q^{2}$ from 1 to 35 GeV$^{2}$, and $-t$ up to 2.0 GeV$^{2}$. There are 9 bins of $W$, each of 1 GeV width. For each $W$ bin, there are 35 $Q^{2}$ bins, each of 1 GeV$^{2}$ width. For each unique bin of $W$ and $Q^{2}$, $\sigma_{L}$ and $\sigma_{T}$ are parameterized against $-t$ from 0 GeV$^{2}$ to 2.0 GeV$^{2}$ as shown in Figs.~\ref{fig:eic_sigl_kaon},~\ref{fig:eic_sigt_kaon}.

Unlike the pion module, when an event is generated with specific values of $Q^{2}$ and $W$ in the kaon module, the generator looks for the lower and upper bound values for $Q^{2}$ and $W$ based on the parameterization ranges and gets the relevant cross-section values. After the values are computed, the generator uses a truncated Taylor series to determine the cross-section value at the desired point. The first-degree Taylor polynomial in two variables, $x$ and $y$, for a function $f(x,y)$ about the point $(a,b)$ is given by Eqn.~\ref{interpolation equation}. 
\begin{equation}
    f(x,y)=f(a,b)+f_{x}(a,b)(x-a)+f_{y}(a,b)(x-b),
    \label{interpolation equation}
\end{equation}
where the subscripts represent the respective partial derivatives.
This feature will be implemented for the EIC pion module in a future patch.

\paragraph{Parameterization of $\sigma_{L}$ }

The functional form of the $\sigma_{L}$ parameterization depends upon the $W$ range to effectively describe the VGL model points. For lower $W$ values, $2<W$ (GeV) $<3$, three exponential functions are utilized. For $4<W$(GeV)$<10$, a polynomial and two exponential functions are used for the parameterization, as in Eqns.~\ref{sigl1_para},~\ref{sigl2_para}.

For $2<W$ (GeV) $<3$, 
\begin{equation}
  \label{sigl1_para}
  \sigma_{L}(Q^{2}_{bin},-t,W_{bin}) = \left\{
  \begin{array}{l l}
    exp(c_{0}+c_{1}|-t|), & t_{1} \leq -t < t_{2} \\
    exp(c_{2}+c_{3}|-t|), & t_{2} \leq -t < t_{3} \\
    exp(c_{4}+c_{5}|-t|), & t_{3} \leq -t < 2.0 
  \end{array} \right.
\end{equation}
and for  $4<W$ (GeV) $<10$,
\begin{equation}
  \label{sigl2_para}
    \sigma_{L}(Q^{2}_{bin},-t,W_{bin}) = \left\{ 
  \begin{array}{l l}
    c_{0}+c_{1}|-t|+c_{2}|-t|^{2}, & t_{1} \leq -t < t_{2} \\
    exp(c_{3}+c_{4}|-t|), & t_{2} \leq -t < t_{3} \\
    exp(c_{5}+c_{6}|-t|), & t_{3} \leq -t < 2.0 
  \end{array} \right.
\end{equation}

\paragraph{Parameterization of $\sigma_{T}$}

Unlike $\sigma_{L}$, $\sigma_{T}$ is parameterized with a common set of functions over the full $W$ range, two polynomials and an exponential function as described by Eqn.~\ref{sigt_para},
 \begin{equation}
  \label{sigt_para}
    \sigma_{T}(Q^{2}_{bin},-t,W_{bin}) = \left\{ 
  \begin{array}{l l}
    c_{0}+c_{1}|-t|+c_{2}|-t|^{2}, & t_{1} \leq -t < t_{2} \\
    c_{3}+c_{4}|-t|+c_{5}|-t|^{2}, & t_{2} \leq -t < t_{3} \\
    exp(c_{6}+c_{7}|-t|), & t_{3} \leq -t < 2.0 
  \end{array} \right.
\end{equation}

$t_{1}$ is the minimum value of $-t$ (or $-t_{min}$) for a given set of values for $Q^2$ and $W$. The cutoff points, $t_{2}$ and $t_{3}$, between the different parameterizations are chosen according to the point where each pair of parameterizations intersect, $f_{1}(Q^{2}_{bin}, -t, W_{bin})$ = $f_{2}(Q^{2}_{bin}, -t, W_{bin})$ and $f_{2}(Q^{2}_{bin}, -t, W_{bin})$ = $f_{3}(Q^{2}_{bin}, -t, W_{bin})$~\cite{Lovepreet_thesis}. It is important to note that the $\sigma_{L}$ and $\sigma_{T}$ are parameterized similarly for the $\Lambda$ and $\Sigma^{0}$ channels.

%--------------------------------------------------------------------------
\subsubsection{Exclusive $^3He(e,e^{\prime}\pi^-p)(pp)_{sp}$ Physics Model}
\label{SubSubSec:SoLID_PhysicsModel}

Many of the elements for this module are the same as the $p(e,e^{\prime}\pi^+n)$ module detailed in Section \ref{sec:pi-model}, except that the Jacobians from target rest frame to collider frame are not needed, and the Fermi momentum of the quasi-free struck neutron in $^3$He must be included. For brevity, only the differences from the previous module are described.

\paragraph{Parameterization of $\sigma_{UU}$}\label{Para:siguu}

The unpolarized differential cross section (Eqn \ref{equation:cross-1}), shorthanded as $\sigma_{UU}$, and its components are parameterized from the phenomenological Vrancx-Ryckebusch (VR) model \cite{VRPaper} (same model as in Fig.~\ref{fig:eic_sig} except for different kinematic range). Model data are generated in the kinematic region of $Q^2$ from 4.0 to 7.5 GeV$^2$, $-t$ from $-t_{min}$ to $-1.0$ GeV$^2$, at a fixed $W=3.0$ GeV \cite{VRPaper,VRPaper2,strangecalc}, which is within the region of validity of the VR model. The $W$ dependence is then taken as $(W^2-M_p^2)^{-2}$, where $M_p$ is the proton mass \cite{Blok08}.

These data were parameterized to fit the following functions:
\begin{gather}
  \sigma_{L} = \exp{(P_1(Q^2) + |t|P^{\prime}_1(Q^2))}
  + \exp{(P_2(Q^2) + |t|P^{\prime}_2(Q^2))},\\
  \label{equation:l-fit}
%\end{equation}
%\begin{equation}
  \sigma_{T} = \frac{\exp{(P_1(Q^2) + |t|P^{\prime}_1(Q^2))}}{P_{1}(|t|)},\\
%  \label{equation:t-fit}
%\end{equation}
%\begin{equation}
  \sigma_{LT} = P_{5}(t(Q^2)),\\
%  \label{equation:lt-fit}
%\end{equation}
%\begin{equation}
  \sigma_{TT} = P_{5}(t(Q^2)).
%  \label{equation:tt-fit}
\end{gather}
The results of this parameterization are accessed by the SigmaCalc class.

\begin{figure*}
  \centering
  \includegraphics[width=0.95\linewidth]{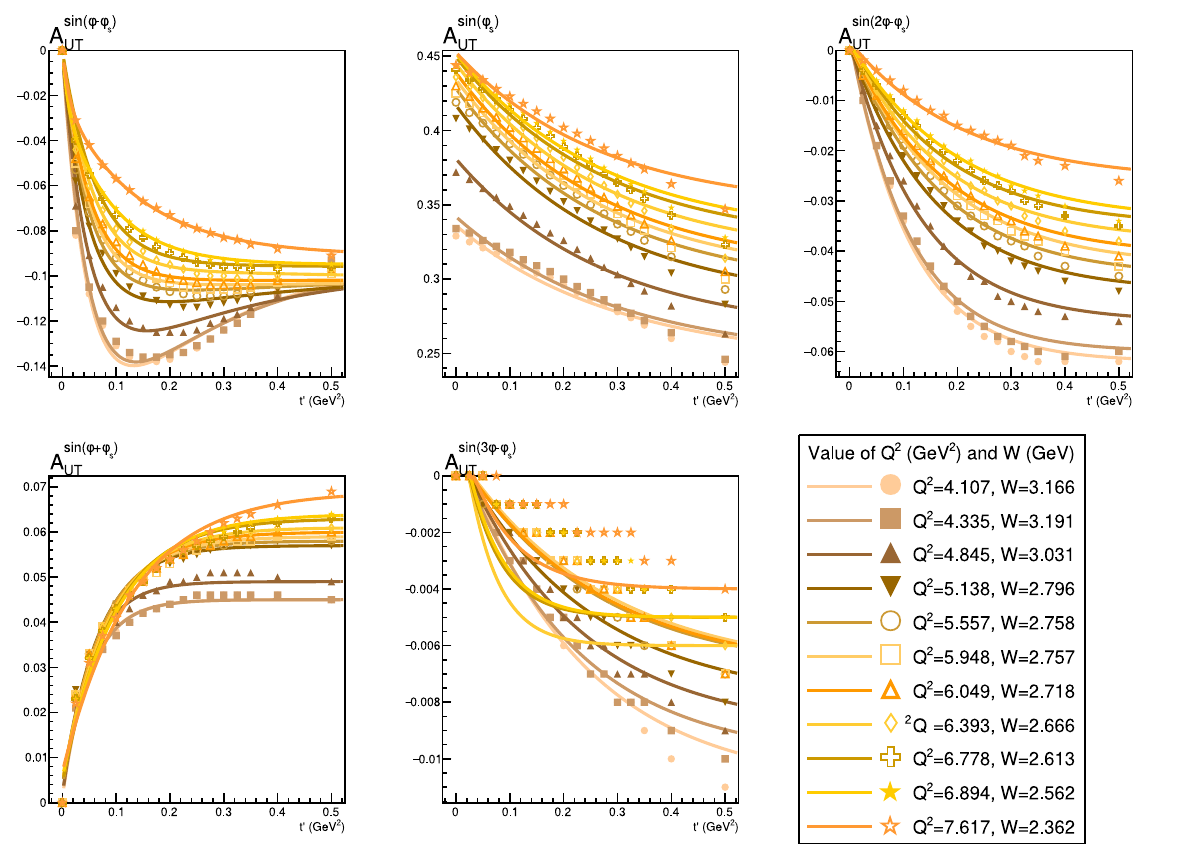}
  \caption{$\vec{n}(e,e^{\prime}\pi^-p)$ asymmetry amplitudes vs $t'$ for different values of $Q^2$ and $W$. Data points are the raw model data provided by Goloskokov and Kroll \cite{GKPriv}. The lines are the parameterized fit for each $Q^2, W$ pair (Eqn. \ref{eqn:cases}).
  %  \textcolor{red}{Made points more obvious in the legend, can change colours fairly straightforwardly if needed. A lot to change so will always have some conflict though.}
  }
  \label{fig:asymmetries}
\end{figure*}

\paragraph{Parameterization of Azimuthal Modulations}\label{Para:asympar}

The transversely polarized neutron target gives rise to a variety of azimuthal modulations,
\begin{equation}
A(\phi,\phi_s)=\frac{d\sigma_{UT}(\phi,\phi_s)}{d\sigma_{UU}(\phi)}
=-\sum_k A_{UT}^{\sin(\mu\phi+\lambda\phi_s)_k}\sin(\mu\phi+\lambda\phi_s)_k ,
\label{eqn:six_asym}
\end{equation}
where $d\sigma_{UU}$ is the unpolarized nucleon cross section in terms of the well-known L, T, LT and TT response functions described above. Six different azimuthal angular modulations contribute to $A_{UT}$ \cite{Diehl_2005}:
   \begin{eqnarray}
        A_{UT}(\phi, \phi_{S}) &=& A_{UT}^{\sin(\phi-\phi_{S})}\sin(\phi-\phi_{S}) \nonumber \\
       &+& A_{UT}^{\sin(\phi+\phi_{S})} \sin(\phi+\phi_{S}) \nonumber \\
       &+& A_{UT}^{\sin(\phi_{S})} \sin(\phi_{S}) \nonumber \\
       &+& A_{UT}^{\sin(2\phi-\phi_{S})} \sin(2\phi-\phi_{S}) \nonumber \\
       &+& A_{UT}^{\sin(3\phi-\phi_{S})} \sin(3\phi-\phi_{S}) \nonumber \\
       &+& A_{UT}^{\sin(2\phi+\phi_{S})} \sin(2\phi+\phi_{S}).
   \label{eqn:six_asym2}
   \end{eqnarray}

The main physics goal of our measurement with SoLID \cite{atpi_proposal} is to measure the $k=1$ asymmetry amplitude, $A_{UT}^{\sin(\phi-\phi_s)}$. In addition, the $k=3$ asymmetry amplitude, $A_{UT}^{\sin(\phi_s)}$, is also accessible through the SoLID experiment, and gives information on higher order transversity GPDs \cite{HERMES:2009gtv,Goloskokov:2009ia}. 

S. V. Goloskokov and P. Kroll (GK) have provided model data for the first five asymmetry amplitudes \cite{GKPriv}. These data are at discrete values of $Q^2$ from 4.107 to 7.167 GeV$^2$, $W$ from 2.362 to 3.191 GeV, and $t'=t-t_{min}$ from 0 to 0.5 GeV$^2$. The GK model data are shown in Fig.~\ref{fig:asymmetries}. The sixth asymmetry amplitude is expected to be much smaller, and is taken to be zero.

The fit functions were chosen only to closely match the shape of the GK model data, and were not based on any physical principle. They are as follows:
\begin{align}
  A_{UT}^{\sin(\mu\phi+\lambda\phi_s)_k} =
  \begin{cases}
    ae^{bt'}-(a+c)e^{dt'}+c, &\quad k=1 \\
    ae^{bt'}+c, &\quad k=2,3,4,5
  \end{cases}
  \label{eqn:cases}
\end{align}
where $a$, $b$, $c$, and $d$ are fit parameters. These fits are carried out independently for each $Q^2,W$ pair. The parameterized functions are displayed alongside the model data in Fig.~\ref{fig:asymmetries}. The $k=1$ fit function originally had an additional, independent parameter in place of $(a+c)$, but the fit did not converge reliably. As such, it was constrained to pass through the origin, justified by the requirement that all asymmetries dependent on $\phi$ must vanish at $t=t_{min}$, as $\phi$ is undefined in parallel kinematics. This also applies to all but the $k=3$ asymmetry. However, the fits were satisfactory, and the additional constraint was not deemed necessary.

%\subsubsection{Asymmetries and \texorpdfstring{$d\sigma_{UT}$}{sigut}}\label{subsec:assym}

The parameters are stored in a file and read at runtime into instances of the Asymmetry class. Each Asymmetry object corresponds to one of the five asymmetries and contains each of the parameterized functions for that asymmetry, one for each $Q^2$, $W$ pair. The Asymmetry class implements a function to retrieve the asymmetry amplitude given $Q^2$ and $t'$ as arguments. The value is calculated by selecting the two functions with the associated values of $Q^2$ closest to the input. These two functions are each evaluated at the input $t'$ value, resulting in two data points, $(Q^2_1, A_1)$ and $(Q^2_2, A_2)$. A line is then drawn between these two points to interpolate a value for the asymmetry amplitude at the input $Q^2$ value.

The asymmetries are accessed by the SigmaCalc class and used to calculate the cross section components $\sigma_{UT}$ according to Eqn. \ref{eqn:sigut}.
\begin{align}
  \Sigma_k &= d\sigma_{UU}(\phi)A_{UT}^{\sin(\mu\phi+\lambda\phi_s)_k} \\
  d\sigma_{UT}&=-\frac{P_T}{\sqrt{1-\sin^2\theta\sin^2\phi_s}}\sum_{k=1}^{6}\sin(\mu\phi+\lambda\phi_s)_k\Sigma_k.
      \label{eqn:sigut}
\end{align}

\paragraph{$\sigma$ and Event Weight}\label{Para:weight}

The cross section components $\sigma_{UU}$ and $\sigma_{UT}$ are summed to give the overall cross section $\sigma$, as shown by Eqn. \ref{eqn:dsigma},
\begin{equation}
  d\sigma = d\sigma_{UU}+d\sigma_{UT},
  \label{eqn:dsigma}
\end{equation}
where $\sigma_{UT}$ is determined from Eqns \ref{eqn:six_asym}, \ref{eqn:six_asym2}, \ref{eqn:cases}.

\begin{figure}
\centering
\includegraphics[width=.9\linewidth]{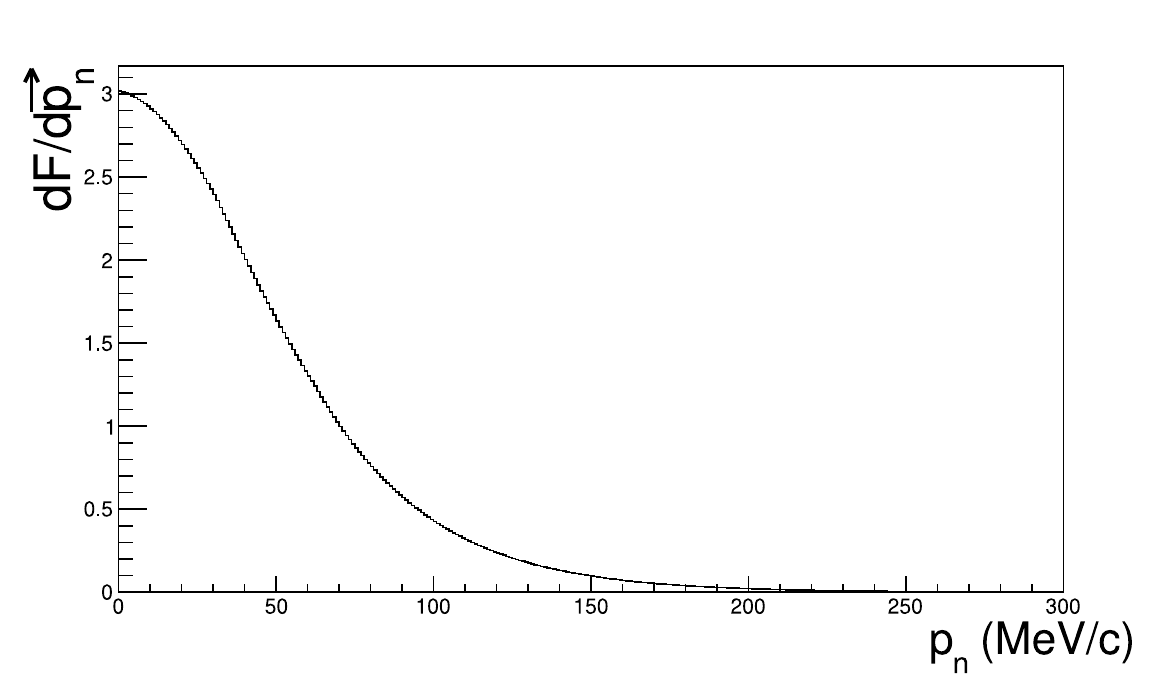}
\caption{$^3$He spectral function generated according to the Argonne Nuclear Potential \cite{SCHIAVILLA1986219}.}
\label{fig:fermidist}
\end{figure}

\begin{figure}
\centering
\includegraphics[width=.95\linewidth]{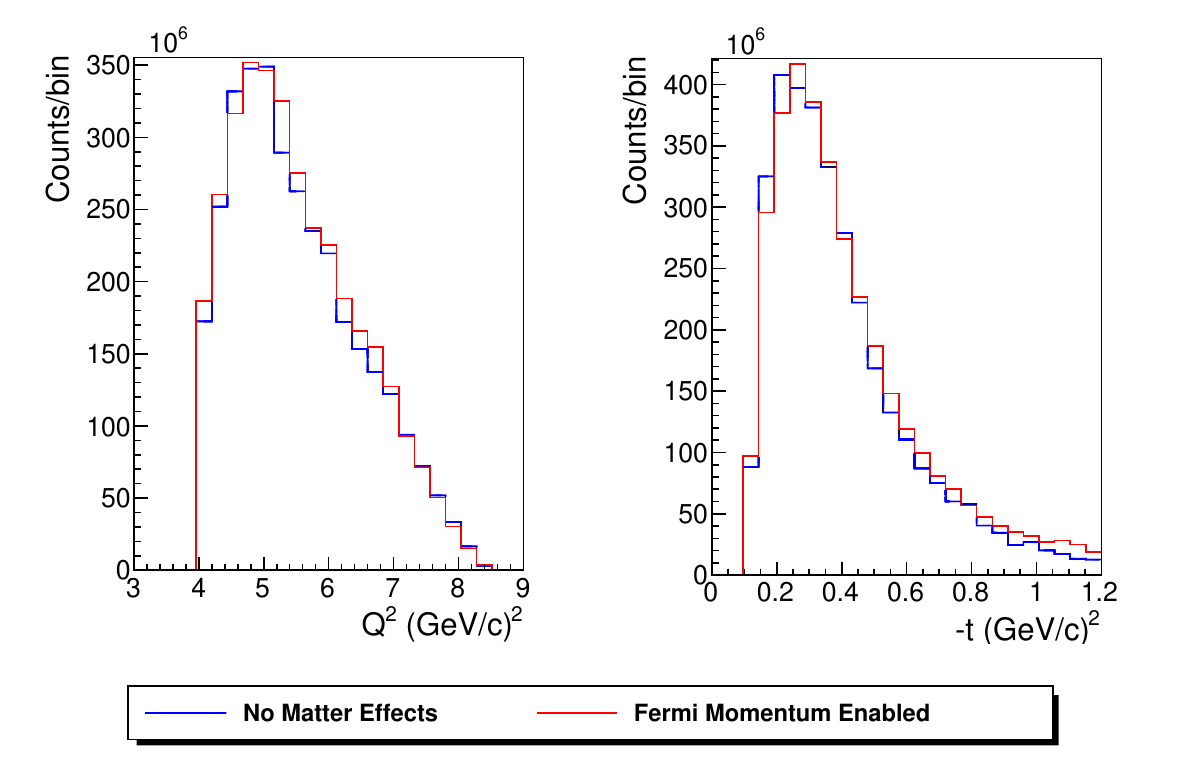}
\caption{Comparison of $Q^2$ (left) and $t$ (right) weighted distributions for the $^3$He$(e,e'\pi^-p)(pp)_{SP}$ reaction with Fermi momentum disabled (blue) and enabled (red).}
\label{fig:fermicompare}
\end{figure}

\begin{figure}
\centering
\includegraphics[width=.95\linewidth]{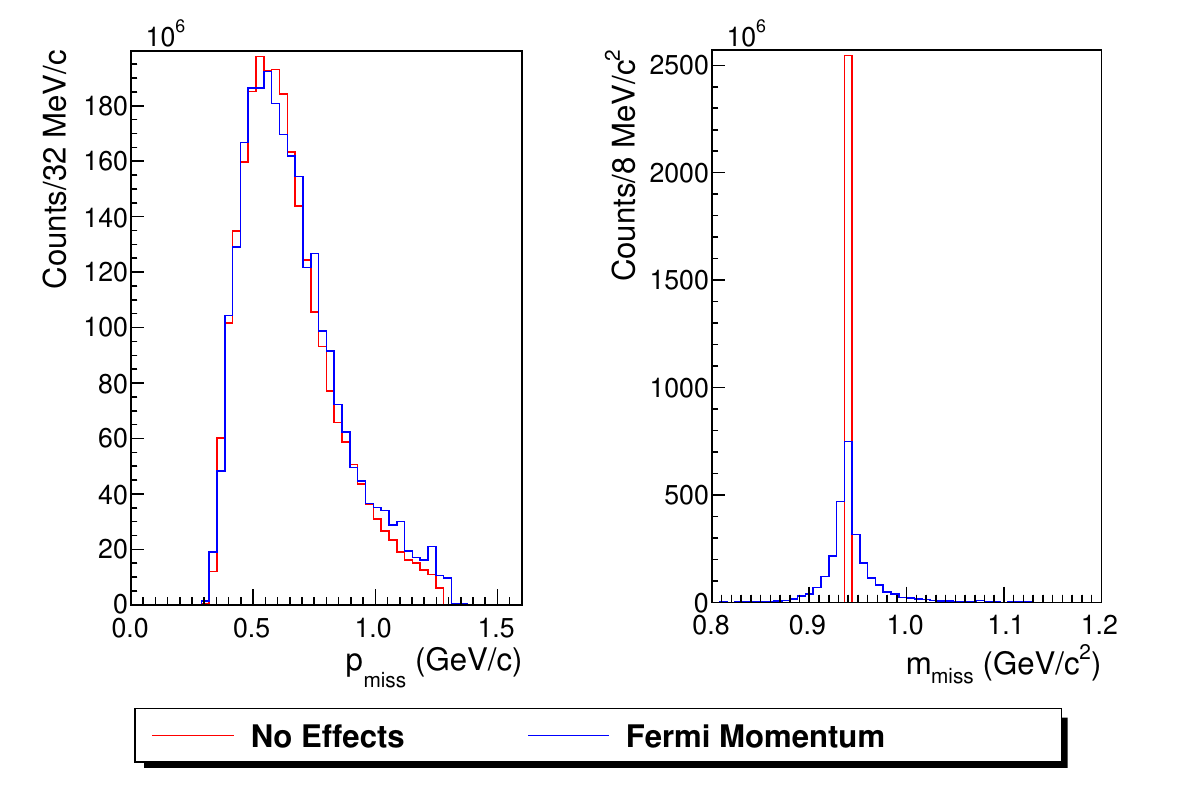}
\caption{Missing mass and momentum distributions for the $^{3}$He$(e,e^{\prime}\pi^{-})p(pp)_{SP}$ reaction with only energy loss effects enabled (red squares), and with no corrective effect enabled (blue circles).}
\label{fig:mmfermi}
\end{figure}

Explicitly, this cross section is a two-fold differential scattering cross section in the center of mass frame. In order to calculate the event weight, the five-fold differential cross section in the lab frame is needed, given by
\begin{equation}
	d^5\sigma = \frac{d^5\sigma}{de^{\prime}d\Omega_{e^{\prime}}d\Omega_\pi} = \Gamma_V J
  \frac{d^2\sigma}{dtd\phi},
\end{equation}
where $J$ is the Jacobian transformation from the center of mass frame to the lab frame, and $\Gamma_V$ is the virtual photon flux factor in Eqn. \ref{equation:photon-flux-1}.
%\begin{equation}
%	\Gamma_V = \frac{\alpha}{2\pi}\frac{e^{\prime}}{E}\frac{(W^2-M_n^2)}{2M_n Q^2}
%				\frac{1}{1-\epsilon},
%\end{equation}
%where $\alpha$ is the fine structure constant, $W$ is the invariant mass of the final state, 
%\begin{equation}
% \epsilon=\left(1+\frac{2
% |\vec{q}|^2}{Q^2} \tan^2\frac{\theta_{e}}{2} \right)^{-1}.
%  \label{eqn:epsilon}
%\end{equation}

The event weight is then given by the following expression:
\begin{equation}
    \mathrm{Weight} = \frac{\sigma \times PSF \times CF \times \mathcal{L} \times TF}{N_{Requested}},
\end{equation}
where $TF$ are target factors and $N_{Requested}$ is the number of events the generator tried to produce. The other variables given in this equation are the same as those defined in Eqn. \ref{eqn:Weight}. The assumed luminosity for the SoLID DEMP experiment is $10^{36}$ cm$^{-2}$s$^{-1}$ \cite{pcdr}. The target-factors include the $\sim$60\% target polarization and the 85.6\% effective polarized neutron 
\cite{PhysRevC.42.2310, PhysRevC.64.024004} of the Jefferson Lab polarized $^3$He target.

\paragraph{Fermi Momentum Effects} \label{Para:fermi}

The target neutron in the SoLID experiment is contained within a $^3$He nucleus. As such, the neutron has a non-zero momentum in the $^3$He target frame, known as Fermi momentum. Fermi momentum is incorporated into the event generator in the TargetGen class, which generates the target neutron's momentum before the main kinematics calculation is performed. 

The calculation of particle kinematics begins with random generation of the energy and momentum of the target neutron, if Fermi momentum is enabled in the generator. If Fermi momentum is not enabled, it is set to zero momentum, with energy equal to the neutron rest mass. The direction of the neutron's Fermi momentum is chosen uniformly using sphere point picking \cite{spherepp}. The magnitude of the Fermi momentum follows the distribution shown in Fig.~\ref{fig:fermidist}, chosen randomly according to the Argonne Nuclear Potential \cite{SCHIAVILLA1986219}.

The Fermi distribution was originally generated with the following normalization,
\begin{equation}
	4\pi \int p^2 \frac{dF}{d\vec{p}} dp\sin\theta d\theta d\phi = 2.
\end{equation}
The distribution was normalized to two in order to describe the two protons in the $^3$He nucleus. For the single neutron, the distribution needs to be normalized to one, as such the data have simply been divided by two and reused here.

The resulting momentum distribution is given by a set of 1000 discrete data points, providing the probability density from 0 to 1 GeV/c. As there is no clear, theoretically motivated, functional form for this distribution, the momentum is selected in the generator by a simple Monte Carlo procedure: A point, ($x\in[0,300],y\in[0,6.03]$), is randomly selected. If this point lies within the bounded area below the curve on Fig.~\ref{fig:fermidist}, then it is used, otherwise the procedure repeats until a point in the bounded area is found. In order to cut down on computation time, the data is truncated at 300 MeV/c, beyond which the probabilities are negligible.

Figs.~\ref{fig:fermicompare}, \ref{fig:mmfermi} demonstrate the effect of Fermi momentum on the $Q^2$ and $t$ missing momentum and mass distributions (Eqns. \ref{eqn:pm}, \ref{eqn:mm}) of the generated data. These plots indicate that the effect of Fermi momentum upon the data is minimal for SoLID kinematics.

\paragraph{Final State Interaction Effects}\label{Para:fsi}

\begin{figure*}
\centering
\includegraphics[width=.95\linewidth]{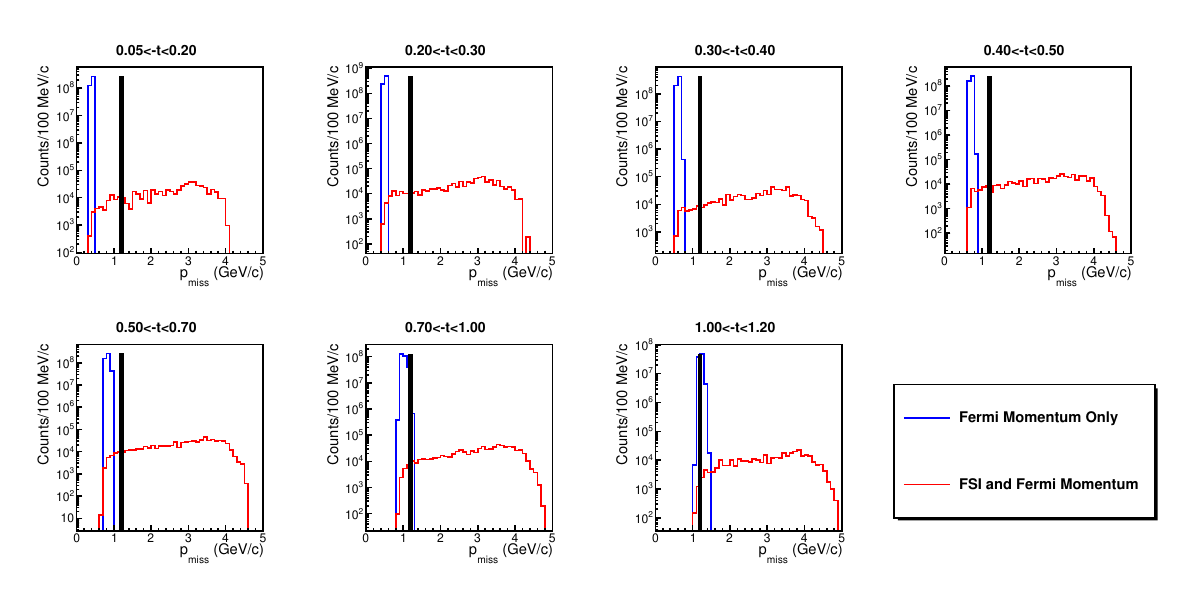}
\caption{Weighted missing momentum distribution in each $t$ bin for quasi-free $\pi^-$ production from the neutron in $^3$He, with FSI and Fermi momentum enabled, compared to the distribution with only Fermi momentum enabled, and with no effects enabled. The FSI distribution uses the Catchen weight in its weighting. The solid black line indicates the cut point $|\vec{p}_{miss}|>1.2$ GeV/c.}
\label{fig:mpfsi}
\end{figure*}

When the target nucleon emits the charged pion in the $^3$He$(e,e^{\prime}\pi^-p)(pp)_{SP}$ reaction, it is possible for the pion to scatter off of one of the spectator nucleons via the $\pi^- p \rightarrow \pi^{\prime} p^{\prime}$ process as it passes through the nuclear volume. This secondary reaction is known as a Final State Interaction (FSI). FSI effects have been estimated by calculating the kinematics using elastic scattering and the scattering cross section using phase-shift parameterizations by Rowe, Solomon and Landau \cite{phase-shifts}. Charge exchange reactions of the form $\pi^0 n\rightarrow \pi^-p $ are excluded, as the purpose of the FSI module is to study the relative effects of FSI on the beam target asymmetries, and both charge conserving and charge exchanging reactions should have similar kinematic effects.

FSI is implemented in the event generator using another instance of the TargetGen class to generate a target proton with Fermi Momentum, as described in Sec.~\ref{Para:fermi}. A random direction is selected with sphere point picking \cite{spherepp} to determine the direction of the scattered pion in the pion-nucleon center of mass frame. In the center of mass frame, the total momentum is zero, so:
\begin{align}
	|\vec{p_{\pi}}| &= |\vec{p_p}| = p \\
	|\vec{p'_{\pi}}| &= |\vec{p'_p}| = p' .
\end{align}
The conservation of energy equation then may be expressed as:
\begin{align}
	E_{\pi}+E_p &= E^{\prime}_{\pi}+E^{\prime}_{p} \\
	\sqrt{p^2+m_{\pi}^2}+\sqrt{p^2+m_p^2}&
    =\sqrt{{p'}^2+m_{\pi}^2}+\sqrt{{p'}^2+m_p^2}.
\end{align}
The only solution to this equation is $p=p'$, and so the kinematics of the outgoing particles are trivial. They are then transformed back to the lab frame.

The implementation of the FSI cross section calculation, i.e. the calculation of the $\pi^- p$ differential scattering cross section,
was written by A. Shinozaki \cite{Shinozaki}, and further modified by us. This differential cross section is given in the center of mass frame, and must be transformed via a Jacobian into a lab frame value which may be used as a correcting factor to the overall event weight. Three different formulations of the Jacobian are available.

The ``William's Weight'' uses the following Jacobian \cite{williamsweight}:
\begin{align}
	J_{Williams} &=\frac{|p^{lab}_{\pi}|^2}
        {\gamma|p^{cm}_{\pi}|(|p^{lab}_{\pi}|-\beta
          E^{lab}_{\pi}\cos\theta^{lab}_{\pi})} \\
        \gamma &= \frac{E^{lab}_{\pi}+E^{lab}_{p}}
        {|\vec{p}^{lab}_{\pi}+\vec{p}^{lab}_{p}|}\\
        \beta &= \frac{|p^{lab}_{\pi}|+|p^{lab}_{p}|}
        {E^{lab}_{\pi}+E^{lab}_{p}}
\end{align}

The ``Dedrick Weight'' uses the following Jacobian \cite{dedrickweight}:
\begin{align}
	J_{Dedrick} &=\frac{\left((g+\cos^2(\theta^{cm}_{\pi})+
	(1-\beta^2)(1-\cos^2(\theta^{cm}_{\pi}))\right)^{3/2}}
	{(1-\beta^2)|(1+g\cos(\theta^{cm}_{\pi}))|}\\
	g &=\frac{\beta E^{cm}_{\pi}}{p^{cm}_{\pi}},
\end{align}
where $\beta$ is the same as in the William's Weight.

Finally, the ``Catchen Weight'' uses the following Jacobian \cite{Catchen}:
\begin{equation}
	J_{Catchen} = \frac{|p^{lab}_{\pi}|^2 E^{cm}_{\pi}}
	{|p^{cm}_{\pi}|^2 E^{lab}_{\pi}}.
\end{equation}

The effects of FSI on the missing momentum distributions are shown in Fig.~\ref{fig:mpfsi}, using Catchen Weight in weighting the FSI-enabled data. The figure shows that a secondary interaction has a much more significant effect on the data than the other corrective effects. However, they also indicate that events which undergo FSI occur at a much smaller rate than those that do not. Furthermore, Fig.~\ref{fig:mpfsi} indicates that the majority of FSI events can be eliminated by cutting events with $|\vec{p}_{miss}|>1.2$~GeV/c. Only $~4\%$ of the FSI events remain after the cut.

%%%%%%%%%%%%%%%%%%%%%%%%%%%%%%%%%%%%%%%%%%%%%%%%%%%%%%%%%%%%%%%%%%%%%%%%%%%
% Chapter 4
%\clearpage
\section{Results}
\label{sec:results}

\begin{table}[hb]
    \centering
\begin{tabular}{|>{\centering\arraybackslash}m{1cm} | >{\centering\arraybackslash}m{7cm}|}
  \hline
  & {\bf Kinematic ranges for EIC Module} \\
  \hline
  $\theta_{e^{\prime}}$ & User specified, default is 60$^{\circ}$ to 175$^{\circ}$ \\
  \hline
  $E_{e^{\prime}}$ & User specified, expressed as fraction of electron beam energy ($E_{e}$), default is 0.5$E_{e}$ to 2.5$E_{e}$\\
  \hline
  $\phi_{e^{\prime}}$ &  0$^{\circ}$ to 360$^{\circ}$ \\
  \hline
  $\theta_{Ej}$ & User specified, default is 0$^{\circ}$ to 50$^{\circ}$ \\
  \hline
  $\phi_{Ej}$ &  0$^{\circ}$ to 360$^{\circ}$ \\
  \hline
  $Q^{2}$ &  Varies by reaction, 3~GeV$^{2}$ to 35~GeV$^{2}$ for $\pi^{+}$, 1~GeV$^{2}$ to 35~GeV$^{2}$ for $K^{+}$\\
  \hline
  $-t $ &  $\pi^+$ FF mode: up to 0.5 GeV$^{2}$, $\pi^+$ TSSA mode: up to 1.2 GeV$^2$, $K^+$ up to 2.0 GeV$^2$ \\
  \hline
  $W $ & User specified, default varies by reaction, typically $\approx$ 2 to 10~GeV\\
  \hline
  $\mathcal{L}$ &  Varies depending upon beam energy combination, see \ref{sec:Appendix_EIC_Lumi} for details, default is $1\times 10^{33}$\,cm$^{-2}$s$^{-1}$ \\
  \hline
  \end{tabular}
    \caption{Kinematic ranges for $p(e,e^{\prime}X_{Ej}X_{Rec})$ event generation for the EIC. Note that the $\pi^{+}$ TSSA mode for the EIC is currently a work in progress. For more details on user specified quantities, see \ref{sec:Appendix_json}. The chosen kinematic range will be checked and may be reduced in the ``PSF Check'' step of the generator shown in Fig.~\ref{fig:EIC_Flowchart}. See \ref{sec:Appendix_PSF} for further details.} 
    \label{tab:kin}
\end{table}

\begin{figure*}
    \centering
\includegraphics[width=0.8\linewidth]{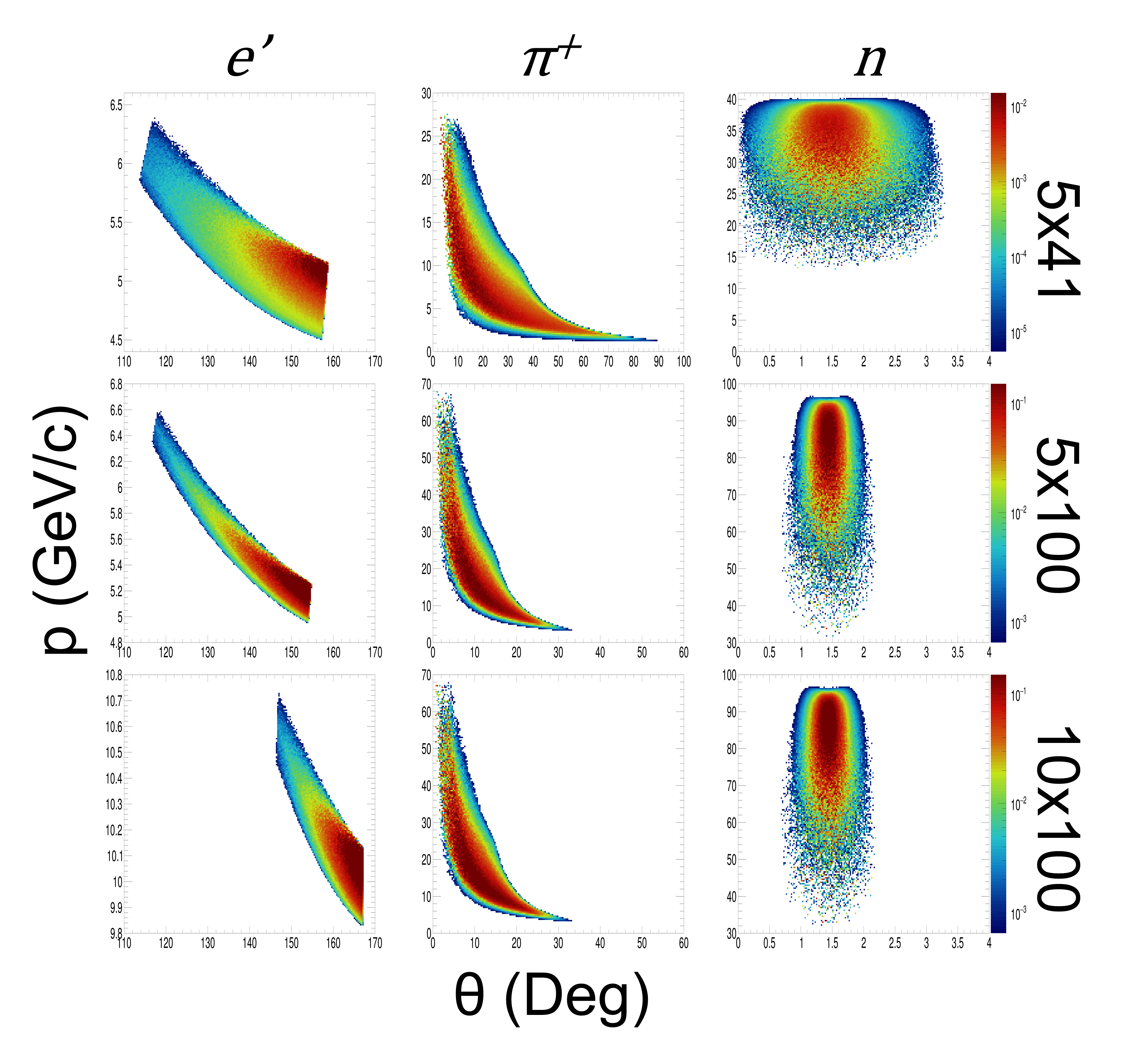}
    \caption{Exclusive $p(e,e^{\prime}\pi^+n)$ kinematic distributions for $e^{\prime}$ (left),$\pi^+$ (center), $n$ (right) at $5\times 41$ (top), $5\times 100$ (middle) and $10\times 100$ (bottom) GeV EIC beam energy combinations with $5<Q^{2}$(GeV$^{2})<35$. Due to the EIC beam crossing angle of 25~mrad, the neutron event distributions are shifted from zero -- in reality, they are centered about a line tangent to the proton beam trajectory at the interaction point. The $z$-axis (color scale) is logarithmic and shows the rate (in Hz) per bin.
      }
    \label{fig:fpi_kin}
\end{figure*}

\begin{figure}
    \centering
\includegraphics[width=\linewidth]{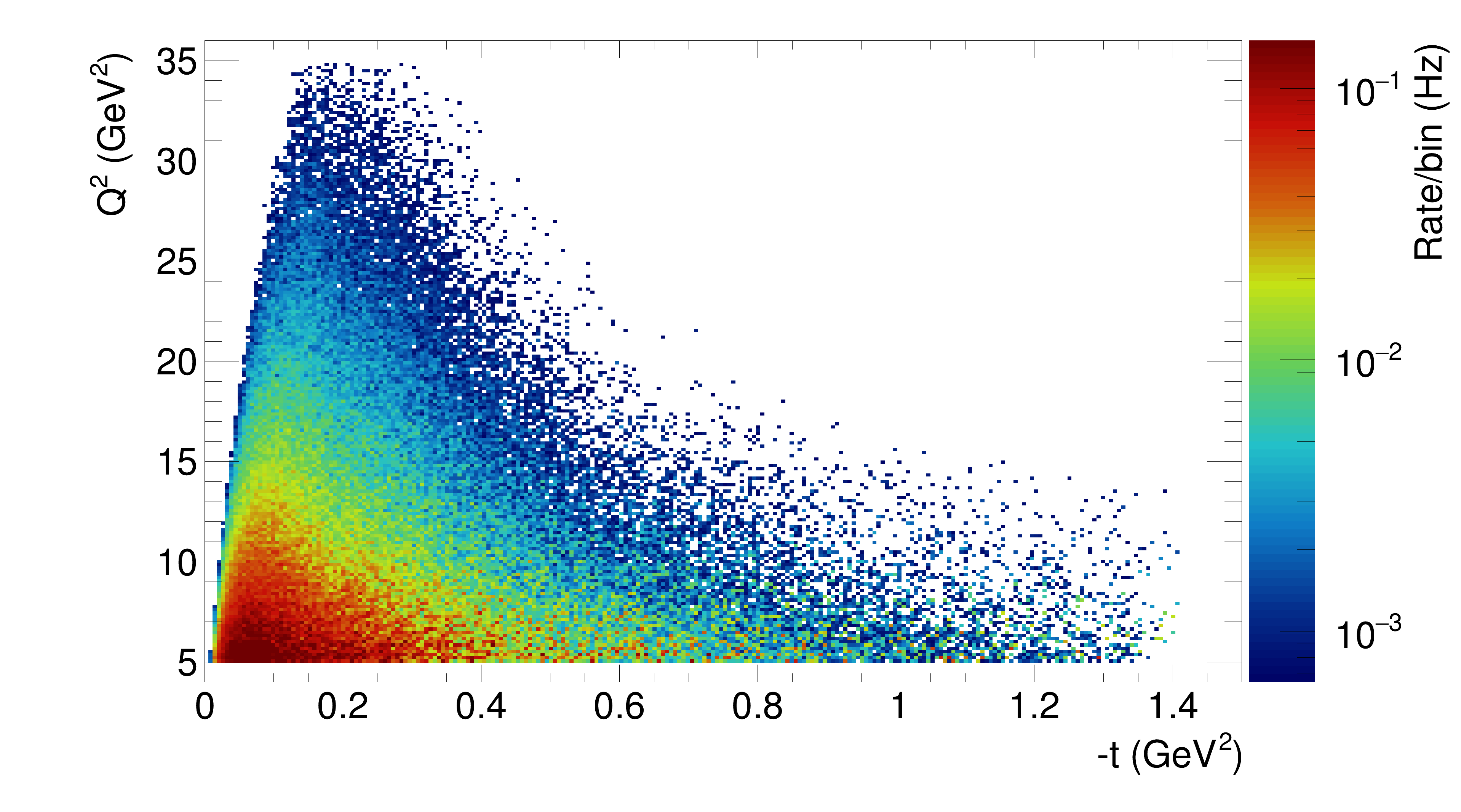}
    \caption{Weighted distribution of the $p(e,e^{\prime}\pi^+n)$ reaction produced by DEMPgen for EIC kinematics ($5\times 100$ beam energy combination). The $z$-axis (color scale) is logarithmic and shows the rate (in Hz) per bin.}
    \label{fig:fpi_Q2_t}
\end{figure}

\begin{figure*}
    \centering
        \includegraphics[width=0.8\linewidth]{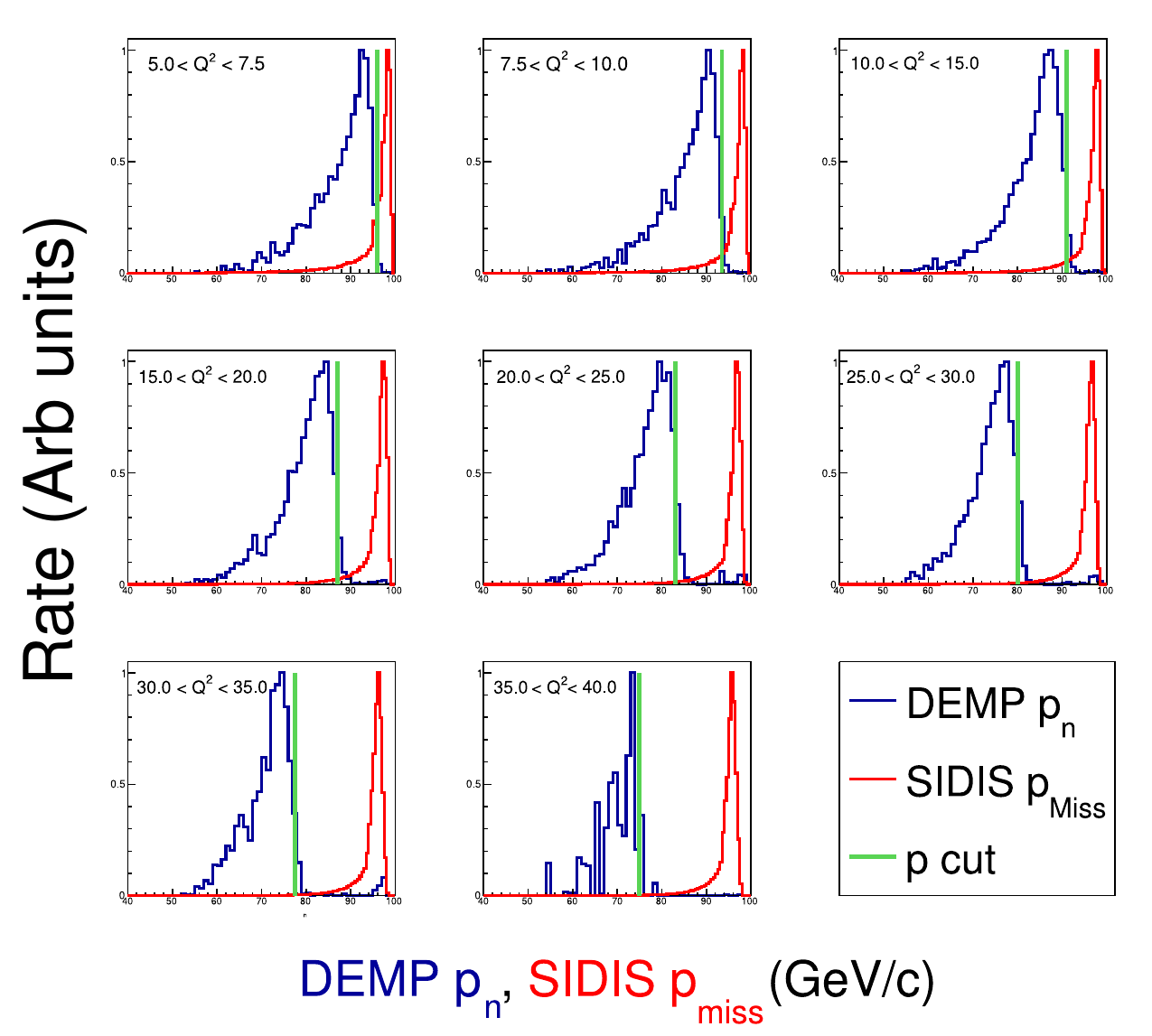}
        \caption{\label{fig:eic_DEMP_SIDIS}
Comparison of DEMP $p_{n}$ distributions and SIDIS $p_{miss}$ distributions for different bins in $Q^{2}$ from EIC simulations ($5\times 100$ beam energy combination). 
The DEMP events are generated for the kinematic ranges listed in Table \ref{tab:EIC_CutVals}, while the SIDIS events have significantly broader distributions in $W$ and $t$.
The distributions have been arbitrarily scaled to demonstrate the difference in shape. Shown in green are example cut values that could be used to distinguish between DEMP and SIDIS events~\cite{sidis}.}
\end{figure*}

\begin{figure}
    \centering
\includegraphics[width=\linewidth]{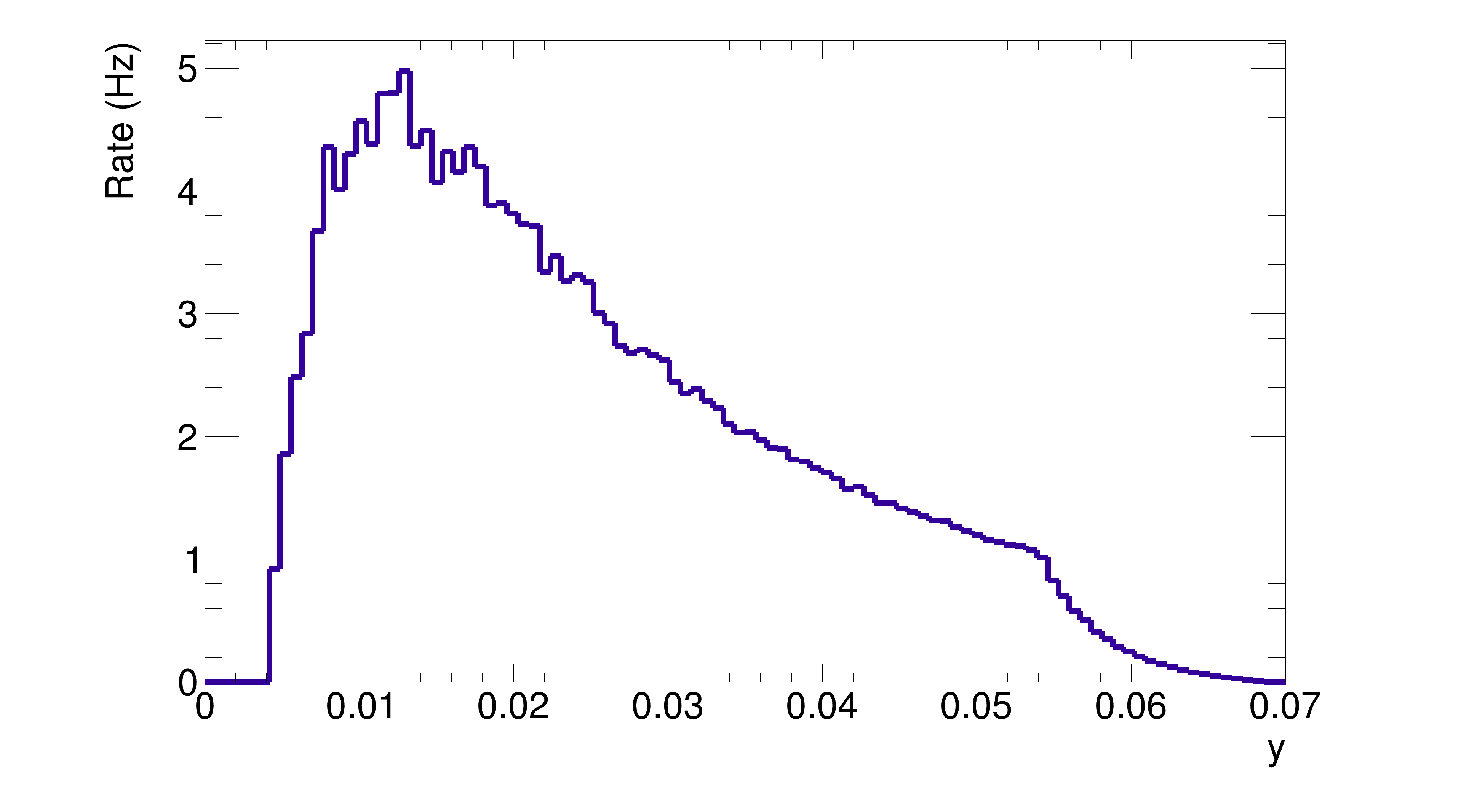}
    \caption{Weighted $y$ distribution of the $p(e,e^{\prime}\pi^+n)$ reaction produced by DEMPgen for EIC kinematics ($5\times 100$ beam energy combination) with $5<Q^{2}$ (GeV$^{2}$) $<35$ and the $W$, $t$ ranges listed in Table \ref{tab:EIC_CutVals}.}
    \label{fig:fpi_Y}
\end{figure}

\subsection{EIC Kinematic Ranges}

In DEMPgen, EIC events are generated within specific kinematic variable ranges. Many of the variable limits are user defined, and some depend upon the reaction being generated. The limits for these variables, over which events are generated for the EIC, are shown in Table \ref{tab:kin}.

% SK - Edited luminosity entry - SK 16/12/20

Note that beam energy combinations at the EIC are typically expressed as 
\begin{gather}
    E_{e} \times E_{Pr},
\end{gather}
where $E_{e}$ is the electron beam energy in GeV and $E_{Pr}$ is the proton (or ion) beam energy in GeV, i.e. $5\times 100$ represents 5~GeV electrons colliding with 100 GeV protons.

\subsection{Exclusive $p(e,e^{\prime}\pi^+ n)$ Projections for the EIC
\label{Sec:EIC_PiPlus}}

\begin{figure}[htbp]
\begin{center}
%\vspace{-0.5cm}
\includegraphics[width=0.95\linewidth]{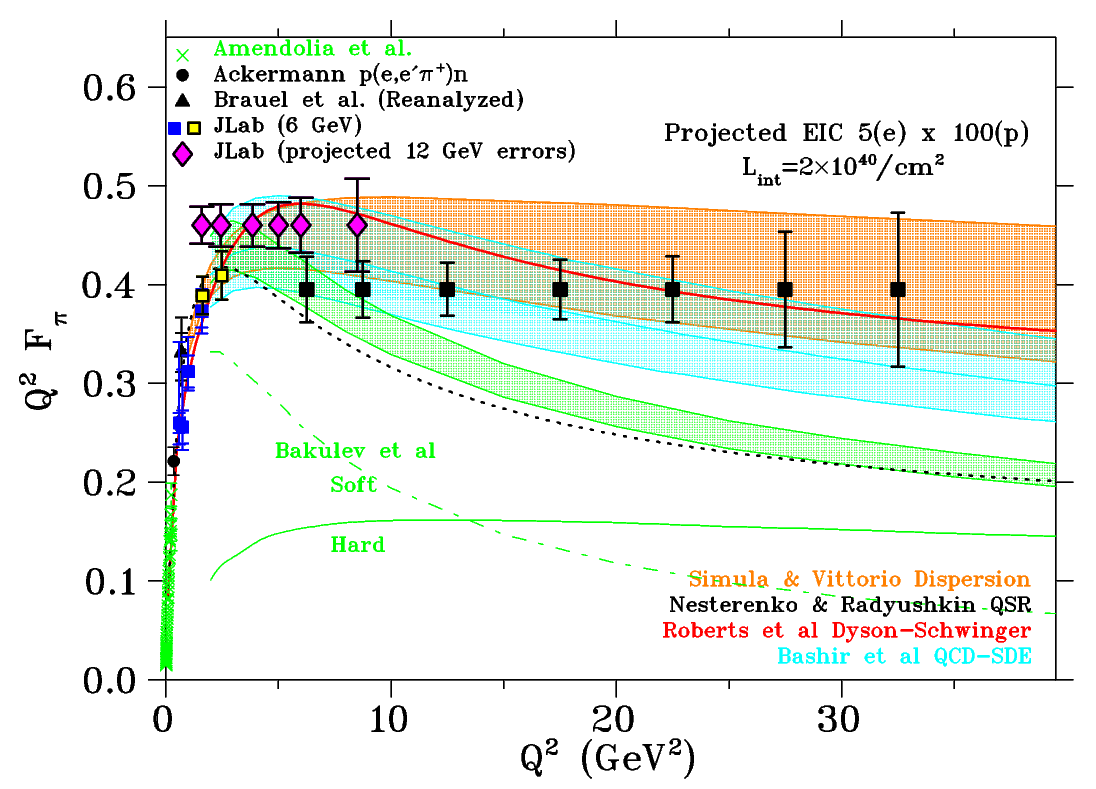}
\vspace{-0.2cm}
\caption{\label{fig:eic_fpi}
Existing data (green \cite{Amendolia:1984nz,Amendolia:1986wj};
black circles \cite{Ackermann:1977rp}; black triangles \cite{gmhuber,Brauel:1979zk}; blue and yellow \cite{gmhuber,Volmer:2000ek,Horn:2006tm}) and projected uncertainties for future $F_{\pi}$ data from JLab (violet \cite{fpi-12gev}) and the EIC (black squares), in comparison to a variety of models of charged pion structure (black dot \cite{Nesterenko:1982gc}; red solid \cite{Chang:2013nia}; orange \cite{Simula:2023ujs}; cyan \cite{Albino_2022}; and green \cite{Bakulev:2004cu}, where Hard is pQCD with analytic running coupling, and the band is Hard+Soft including non-perturbative uncertainties). The EIC projections, obtained with the use of DEMPgen, cover a wide range in $Q^2$, providing access to the emergent mass scale in QCD.}
\end{center}
\end{figure}

To illustrate the utility of DEMPgen, we briefly describe our EIC measurement feasibility studies published in Ref.~\cite{jr:ECCE_EDT}\footnote{Updated projections utilizing the latest ePIC detector design and reconstruction algorithms will be published in the forthcoming ePIC Technical Design Report (TDR).}. The EIC is projected to be capable of delivering proton beam energies of up to 275~GeV and electron beam energies up to 18~GeV \cite{jr:EIC_YR}. The beam crossing angle at IP6, the location of the ePIC detector, is planned to be 25~mrad, i.e. the proton beam will make an angle of 25~mrad with respect to the $z$ axis, with the electron beam propagating in the $-\hat{z}$ direction.

$p(e,e^{\prime}\pi^+n)$ events were generated with DEMPgen assuming an integrated luminosity of 20~fb$^{-1}$ for 5$\times$100~GeV electron/proton collisions. DEMP event kinematic distributions are shown in Fig.~\ref{fig:fpi_kin}. All three outgoing particles, the electron, pion and neutron are required to be detected to ensure exclusivity. The neutrons take nearly all of the proton beam momentum and are detected at very forward angles (in the Zero Degree Calorimeter, ZDC). The scattered electrons and pions have similar momenta, however, the electrons are distributed over a wider range of angles. For example, for 5$\times$100~GeV electron/proton collisions, the 5-6~GeV/c electrons are primarily scattered 25-45$^{\circ}$ from the electron beam, while the 5-40 GeV/c $\pi^+$ are 5-25$^{\circ}$ from the proton beam. Similarly, Fig.~\ref{fig:fpi_Q2_t} shows the $Q^{2}$ versus $-t$ distribution of the generated DEMP events.

Simulation studies demonstrated that event selection cuts were highly effective in isolating pion DEMP events from background pion SIDIS ($p(e,e^{\prime}\pi^{+})X$) events, as seen in Fig.~\ref{fig:eic_DEMP_SIDIS}. The selection cuts included a cut on $y$, the fractional energy loss (defined in Eqn.~\ref{eqn:y_frac_ELoss}). DEMP events predominantly have $y>0.01$ (Fig.~\ref{fig:fpi_Y}). Further details of the event selection cuts can be found in \cite{jr:ECCE_EDT}.

For the projected statistical and systematic uncertainties, the following assumptions were made:
\begin{itemize}
    \item Integrated luminosity of 20~fb$^{-1}$ for the 5$\times$100~GeV measurement, as described above.
    \item Clean identification of exclusive $p(e,e^{\prime}\pi^+n)$ events by tagging the high energy, forward going neutron in the Zero Degree Calorimeter (ZDC), as determined by passing the DEMPgen events through a Geant4 simulation of the IP6 detector.
    \item Systematic uncertainties of 2.5\% point-to-point, and 12\% scale, similar to the ZEUS leading neutron measurement \cite{ZEUS:2007knd}.
    \item As the EIC cannot access the low $\epsilon<0.8$ needed to do a quality L/T-separation, a model is required to subtract the estimated $\sigma_T$ contribution from the measured cross sections to yield $\sigma_L$. For the error propagation, $R=\sigma_L/\sigma_T=0.013-0.14$ is assumed at the lowest $-t$, and $\delta R=R$ systematic uncertainty is assumed in the model \cite{VRPaper} subtraction to isolate $\sigma_L$.
    \item Pion pole channel dominance at small $-t$ over the measured $Q^2,W$ range will need to be confirmed in a separate measurement of exclusive $\pi^-/\pi^+$ ratios obtained from electron-deuteron collision data. If this check is not done, an additional systematic uncertainty in the pion pole dominance will be required.
    \end{itemize}

Under these conditions, we have concluded that $F_{\pi}$ measurements at the EIC are feasible up to $Q^2\approx 30$ GeV$^2$ (Fig.~\ref{fig:eic_fpi}). A consistent and robust EIC $F_{\pi}$ data set will probe deep into the region where $F_{\pi}(Q^2)$ exhibits strong sensitivity to both emergent mass generation via DCSB and the evolution of this effect with distance scale.

\clearpage
\subsection{Exclusive $p(e,e^{\prime}K^+\Lambda[\Sigma^{0}]$) Projections for the EIC}
\label{Sec:EIC_KPlus}

\begin{figure*}[hb]
    \centering
\includegraphics[width=0.8\linewidth]{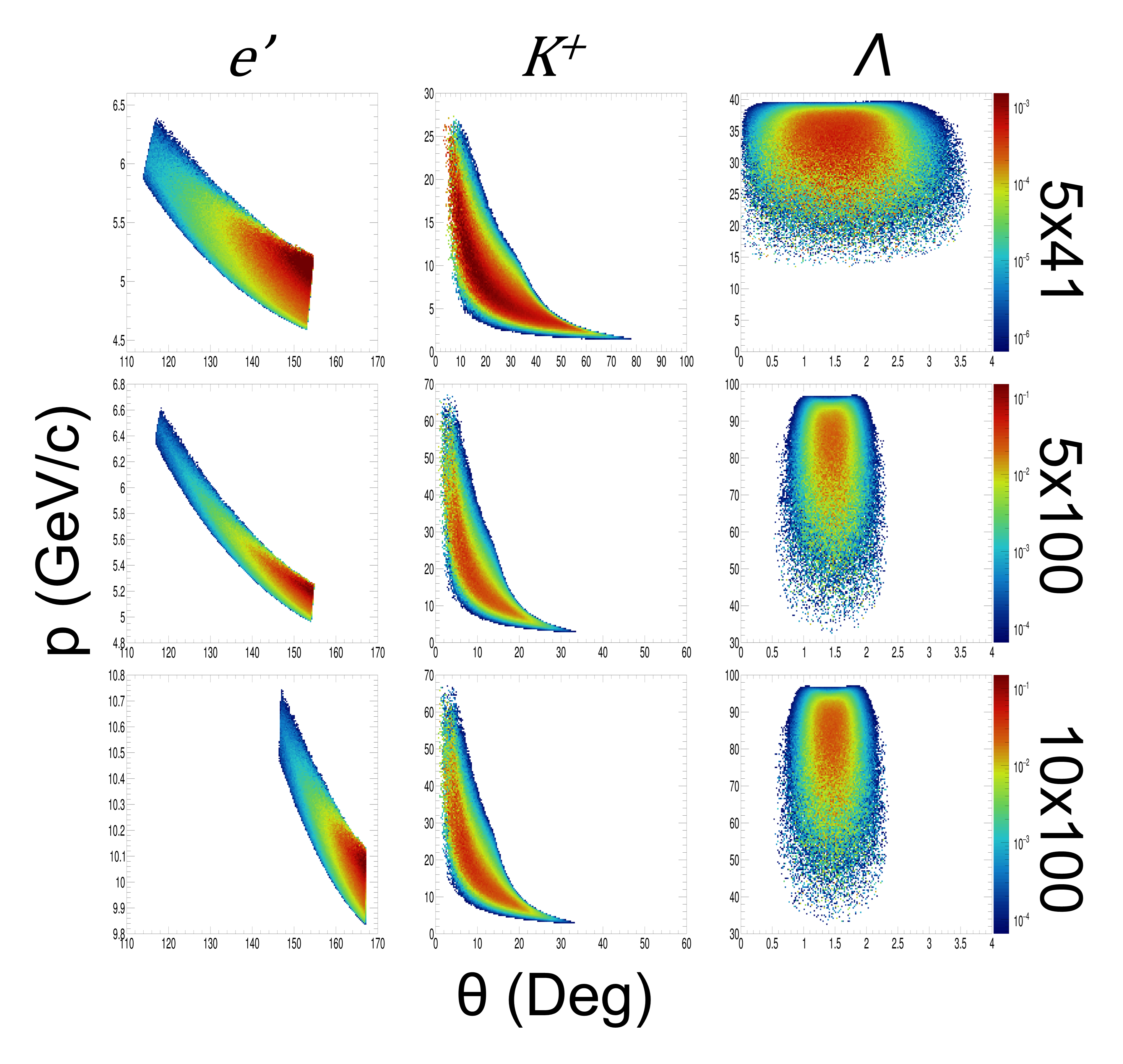}

    \caption{Exclusive $p(e,e^{\prime}K^+)\Lambda$ kinematic distributions for $e^{\prime}$ (left), $K^+$ (center), $\Lambda$ (right) at $5\times 41$ (top), $5\times 100$ (middle) and $10\times 100$ (bottom) GeV EIC beam energy combinations with $5<Q^{2}$ (GeV$^{2}$) $<35$. Due to the EIC beam crossing angle of 25~mrad, the $\Lambda$ event distributions are shifted from zero -- in reality, they are centered about a line tangent to the proton beam trajectory at the interaction point. The $z$-axis (color scale) is logarithmic and shows the rate (in Hz) per bin.
      }
    \label{fig:fKlambda_kin}
\end{figure*}

\begin{figure*}[hb]
    \centering
\includegraphics[width=0.8\linewidth]{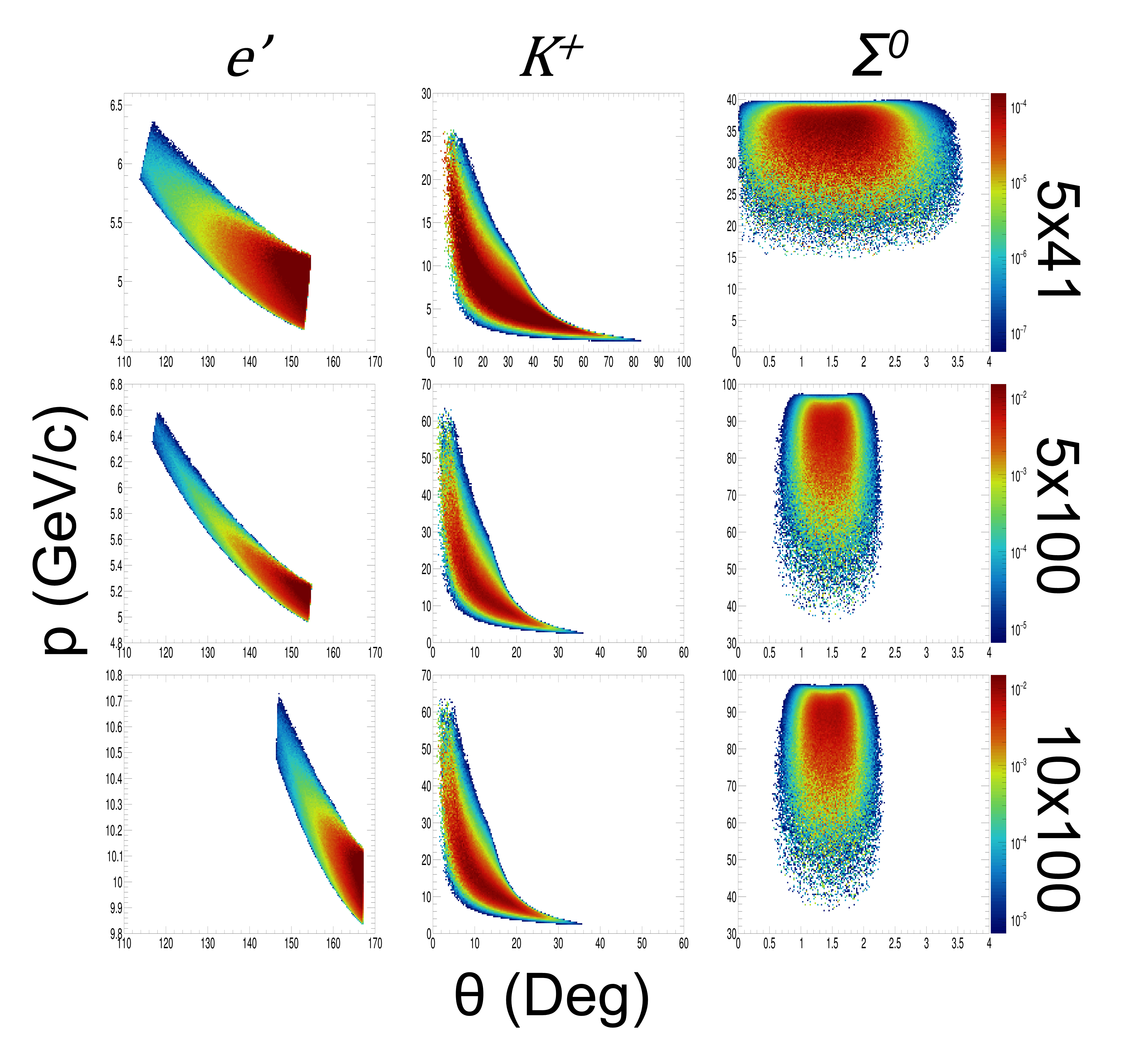}

    \caption{Exclusive $p(e,e^{\prime}K^+) \Sigma^{0}$ kinematic distributions for $e^{\prime}$ (left), $K^+$ (center), $\Sigma^{0}$ (right), analogous to Fig. \ref{fig:fKlambda_kin}.
      }
    \label{fig:fKsigma_kin}
\end{figure*}

\begin{figure}
    \centering
\includegraphics[width=\linewidth]{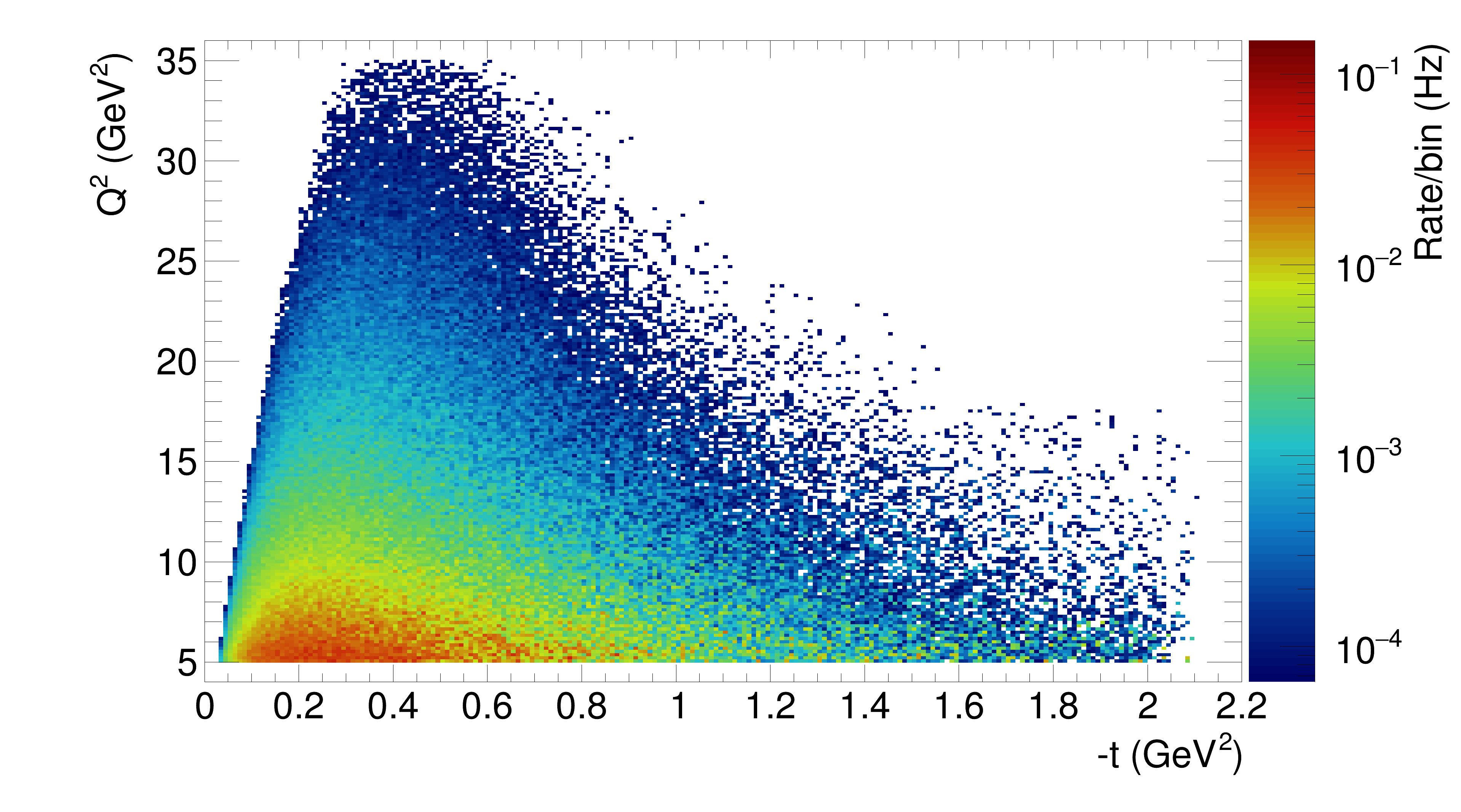}

    \caption{Weighted distribution of $p(e,e^{\prime}K^+)\Lambda$ reaction produced by DEMPgen for EIC kinematics ($5\times 100$ beam energy combination). The $z$-axis (color scale) is logarithmic and shows the rate (in Hz) per bin. }
    \label{fig:fKlambda_Q2_t}
\end{figure}

\begin{figure}
    \centering
\includegraphics[width=\linewidth]{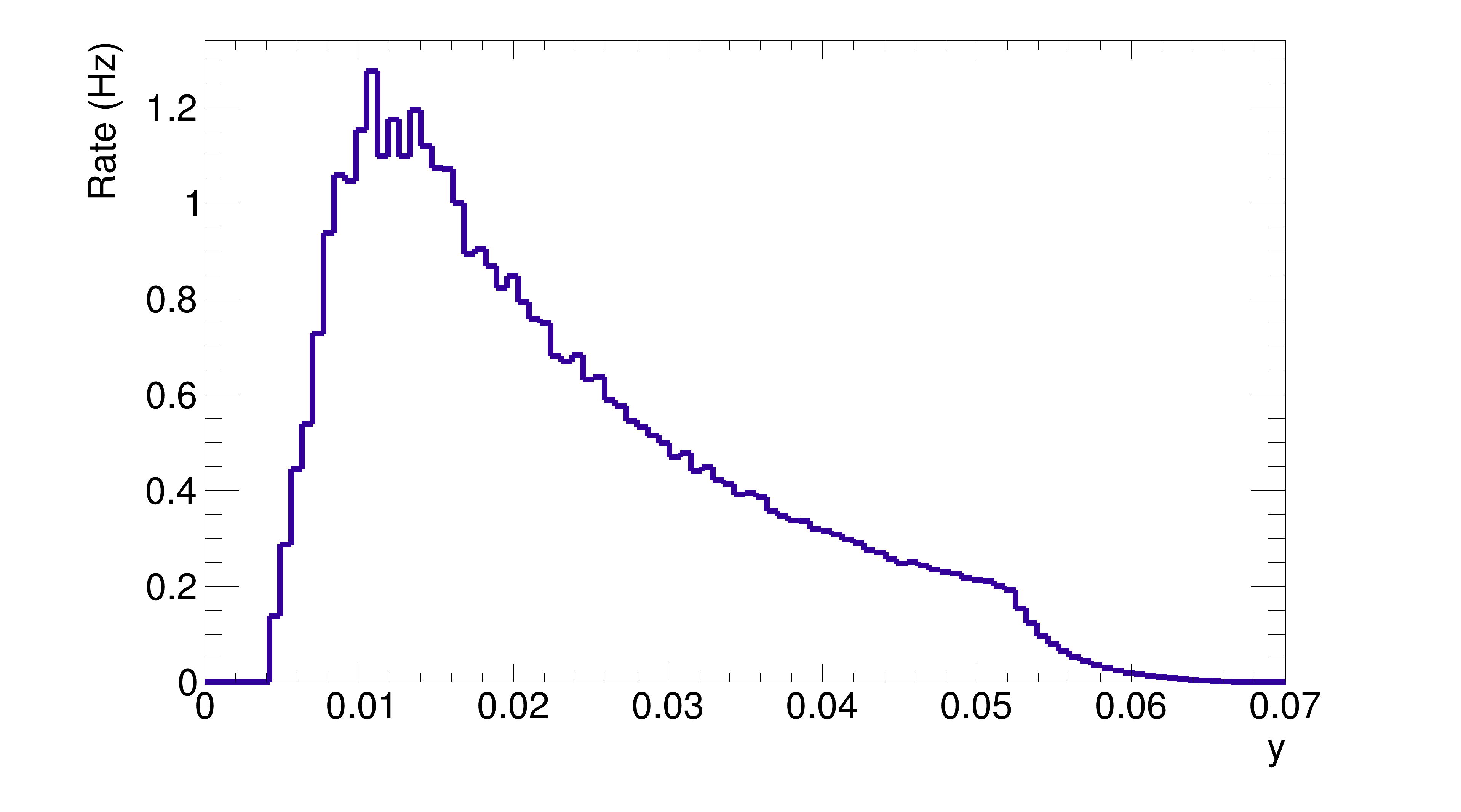}
    \caption{Weighted $y$ distribution of the $p(e,e^{\prime}K^+)\Lambda$ reaction produced by DEMPgen for EIC kinematics ($5\times 100$ beam energy combination) with $5<Q^{2}$ (GeV$^{2}$) $<35$.}
    \label{fig:fKlambda_Y}
\end{figure}

DEMP event kinematic distributions for the $\Lambda$ and $\Sigma^{0}$ channels are shown in the Figs.~\ref{fig:fKlambda_kin}, \ref{fig:fKsigma_kin}. The $K^+$ event distributions are highly similar to the $\pi^+$ distributions shown in Fig. \ref{fig:fpi_kin}, in that the electron is scattered (mostly) higher in energy in the electron endcap region, the $K^+$ has moderate momentum in the hadron endcap region, while the
ejectile hyperon takes most of the incident proton beam momentum at very forward angle.
Similarly, Fig.~\ref{fig:fKlambda_Q2_t} shows the $Q^{2}$ versus $-t$, and Fig.~\ref{fig:fKlambda_Y} presents the $y$ distribution of generated DEMP events for the $\Lambda$ channel.

A significant difference from the $\pi^+$ channel at the EIC is that the identification of the exclusive $K^+$ production reaction requires the efficient reconstruction of the $\Lambda [\Sigma^0]$ from its decay products in the far forward detectors. This is a non-trivial task for which the DEMPgen events are an essential prerequisite for the necessary detector reconstruction and acceptance studies. These simulation studies are in advanced progress, where we are investigating both the charged $\Lambda\rightarrow p\pi^-$ and neutral $\Lambda\rightarrow n\pi^0$ decay modes, as well as $\Sigma^0\rightarrow\Lambda\gamma$. When these studies are completed, we will have a significantly better understanding of the feasibility of $F_{K}$ measurements using the EIC far forward detectors, and will disseminate these results in a future publication.  In the meantime, DEMPgen has already been used by Zhoudunming Tu for proposed $\Lambda$ hyperon polarization studies at the EIC in Ref. \cite{Tu:2023few}.

%\clearpage
\subsection{Exclusive $\vec{^3He}(e,e^{\prime}\pi^-p)(pp)_{SP}$ Projections for 
SoLID at Jefferson Lab}

\begin{figure*}
  \centering
  \includegraphics[width=0.8\linewidth]{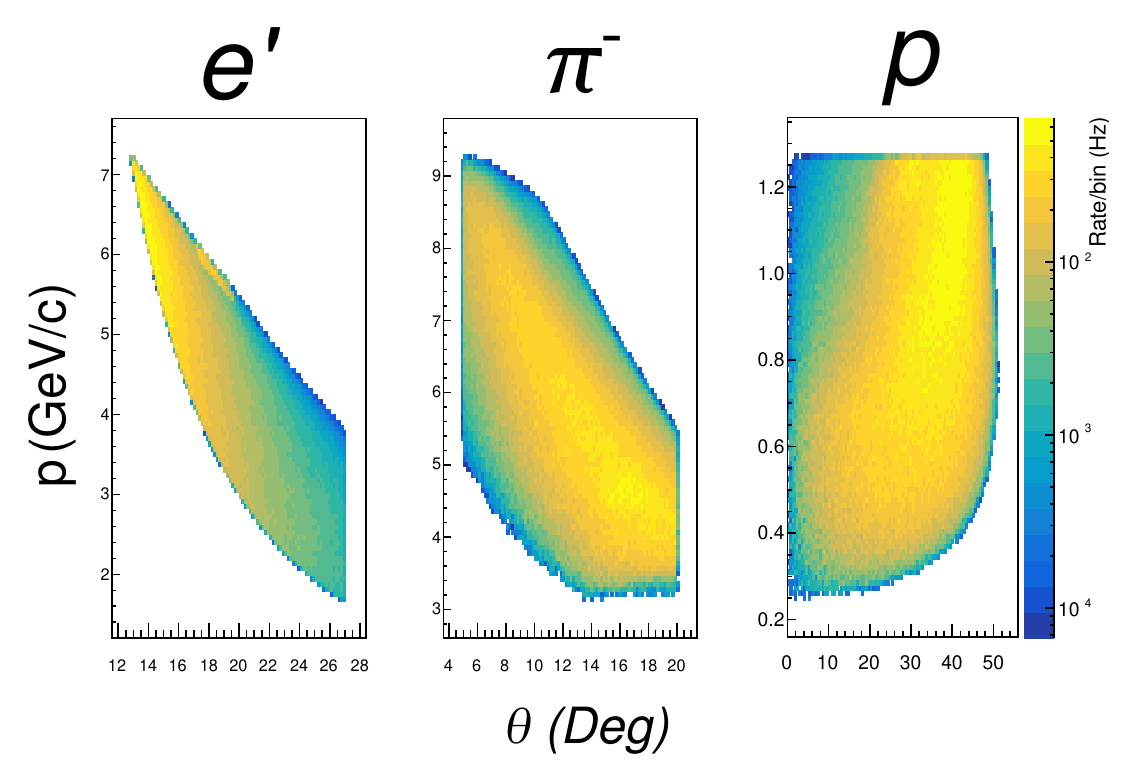}
  \caption{Weighted kinematic coverage of the three final state particles in the $^3$He$(e,e^{\prime}\pi^-p)(pp)_{SP}$ reaction produced by the DEMP event generator for SoLID experiment kinematics. The color axis represents the rate for each bin.}
  \label{fig:tripcov}
\end{figure*}

\begin{figure}
  \centering
  \includegraphics[width=\linewidth]{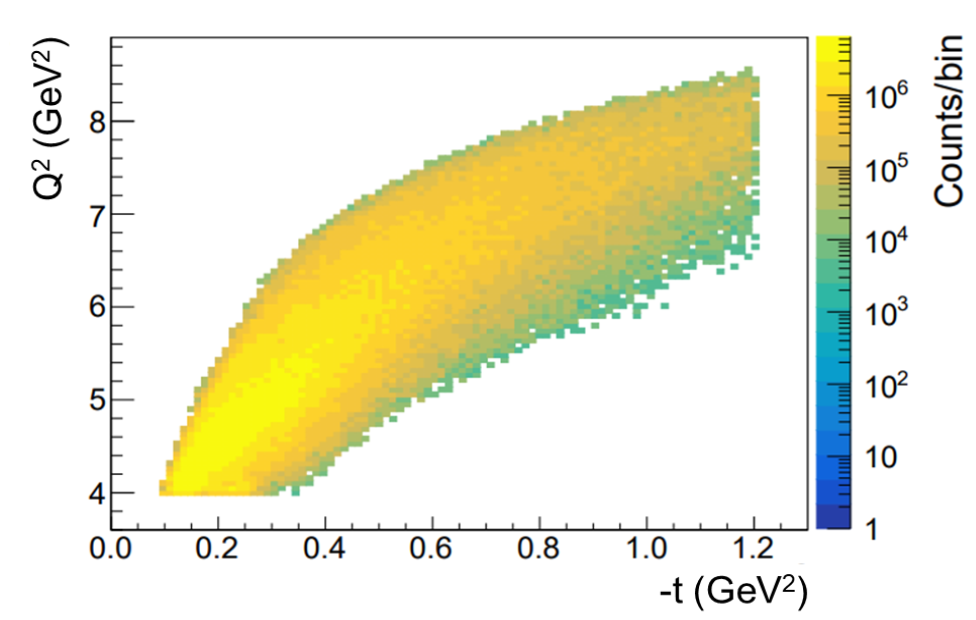}
  \caption{Weighted acceptance of 
    $^3$He$(e,e^{\prime}\pi^-p)(pp)_{SP}$ reaction produced by DEMPgen and measured in SoLID. The color axis represents the expected yield of DEMP events in the experiment.}
  \label{fig:acc_cov}
\end{figure}

\begin{figure}
\centering
\includegraphics[width=.95\linewidth]{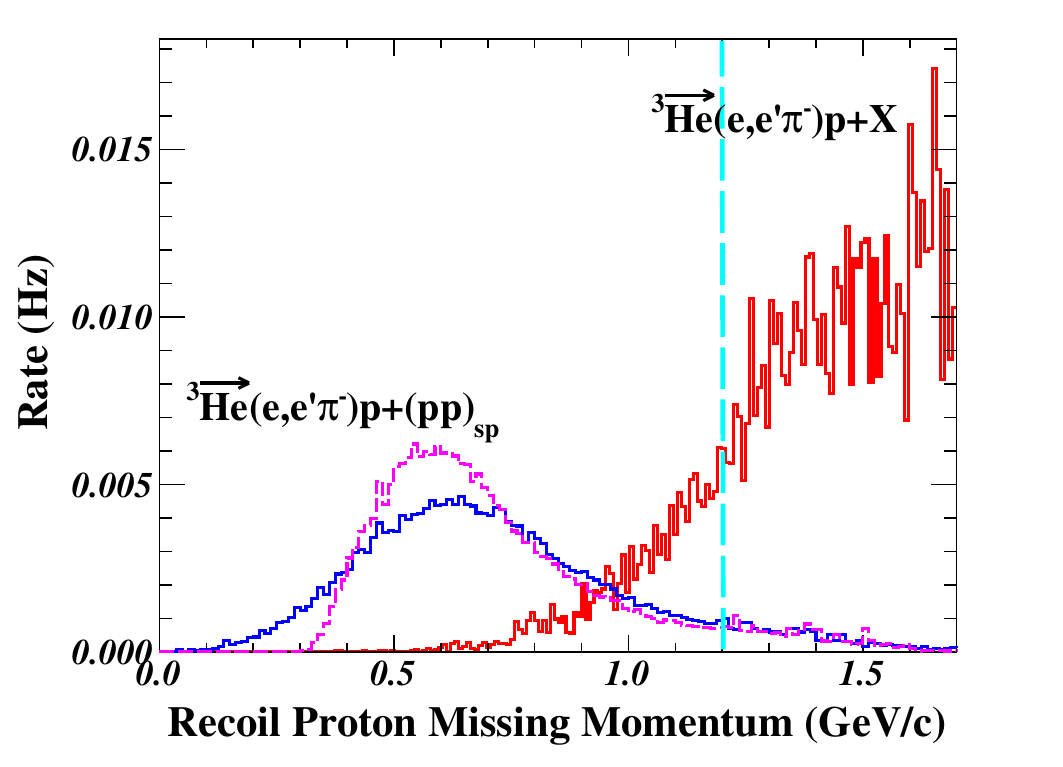}
\caption{Missing momentum spectra of recoil
    protons in DEMP (blue) and SIDIS (red) processes in the $^{3}$He$(e,e^{\prime}\pi^-p)(pp)_{SP}$ reaction. The dashed magenta curve is the DEMP missing mass only considering the Fermi motion, multiple scattering and the energy loss, while the blue and red curves have further taken into account the detector resolutions. The light-blue dashed line indicates $P_{miss}<1.2$~GeV/c, beyond which are mostly SIDIS events. Only generated events with $W>2$ GeV and $Q^2>4$ GeV$^2$ are shown. The normalization of the SIDIS background is approximate.
    }
  \label{fig:missing_mom}
\end{figure}

\begin{figure}
\centering
\includegraphics[width=0.95\linewidth]{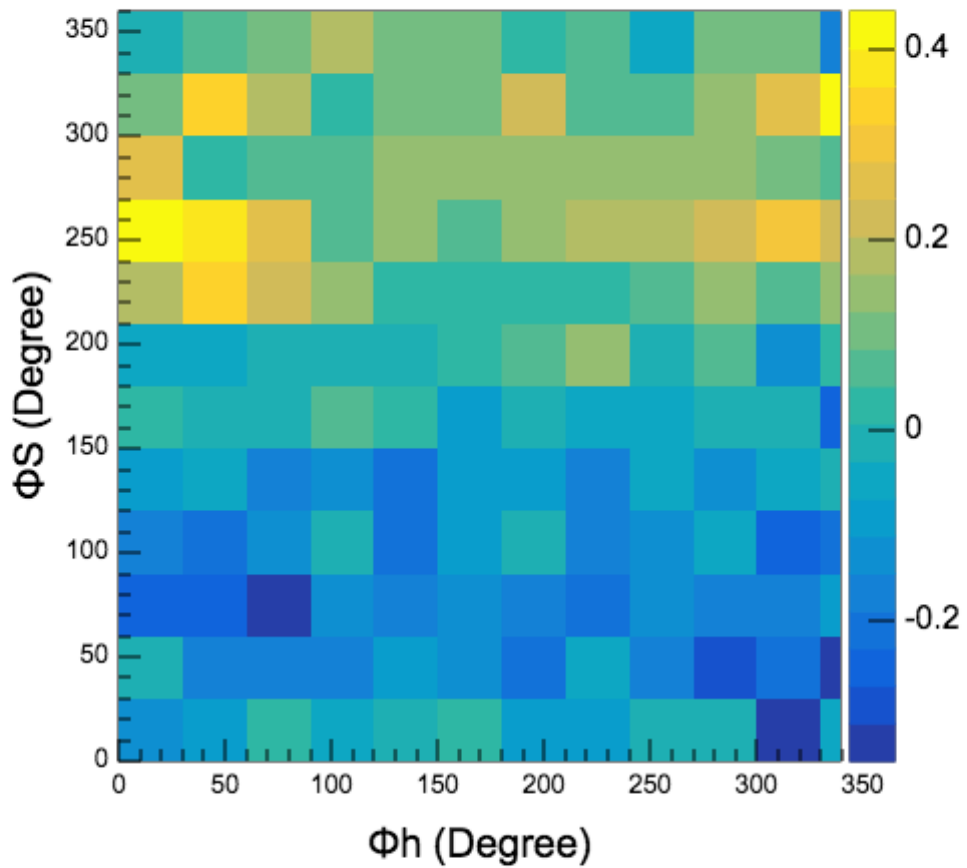}
\caption{\label{fig:asym4}
 $^3$He$(e,e^{\prime}\pi^-p)(pp)_{SP}$ $A_{UT}$ asymmetries of Eqn.~\ref{eqn:six_asym2} for a bin with central kinematics of $<-t>=0.45$ GeV$^2$, $<Q^2>=5.77$ GeV$^2$, $<x_B>=0.47$, binned as a 2-dimensional scatter plot for $12\times 12$ $(\phi,\phi_S)$ bins. Dark (bright) color indicates negative (positive) transverse target single-spin asymmetry for that bin.}
 \end{figure}

Like the EIC simulations discussed above, for the SoLID $^3$He$(e,e^{\prime}\pi^-p)(pp)_{SP}$ experiment we anticipate the necessity to detect the scattered electron, $\pi^-$ and recoil proton in coincidence to ensure exclusivity. The SoLID trigger will be configured so that all electron-$\pi^-$ coincidence events will be written to disk, and the event sample mined to select those events for which an additional proton is detected.

To study the feasibility of this approach, we generated events with DEMPgen using the following configuration:
\begin{itemize}
    \item $e^{-}$ Beam energy: 11 GeV
    \item $e^{\prime}$ energy: 1.1 to 9.9 GeV
    \item $e^{\prime}$ $\theta$: $5^{\circ}$ to $27^{\circ}$
    \item $\pi$ $\theta$: $6^{\circ}$ to $18^{\circ}$
\end{itemize}
In addition, any events meeting the following criteria are discarded due to being outside the accurate range for the cross section model:
\begin{itemize}
    \item $-t>1.2$~GeV$^2$
    \item $Q^2<4$~GeV$^2$
    \item $W<2$~GeV
\end{itemize}

Fig.~\ref{fig:tripcov} shows the weighted momentum and polar angle distribution of particles generated in this configuration. This is a fixed target experiment. The scattered electrons and $\pi^-$ are emitted at forward angles from 5 to 27$^{\circ}$ and take most of the beam momentum, up to 9 GeV/c. In contrast, the recoil proton has much lower momentum and is emitted over a much wider angular range (up to 50$^{\circ}$). The two spectator protons will essentially have only Fermi momentum and will be ranged out in the target and other detector elements. Fig.~\ref{fig:acc_cov} shows the $Q^2$ versus $-t$ coverage of the events in SoLID, and projected event statistics for a 48 day experimental run.

To further separate the DEMP events from competing non-exclusive reactions, such as SIDIS, missing mass and momentum cuts are planned to be used in the analysis of the experimental data. The missing momentum is also useful in identifying events which have undergone a final state interaction (Sec.\ref{Para:fsi}). The missing mass and momentum distributions produced by DEMPgen with Fermi momentum and various other corrective effects enabled are shown in Fig.~\ref{fig:missing_mom}. Similar to what was already indicated in Fig.~\ref{fig:eic_DEMP_SIDIS}, the DEMP events have unique kinematics on the edge of the SIDIS missing momentum distribution. With sufficiently good event reconstruction resolution, this makes it experimentally possible to cleanly separate the small DEMP cross section from the orders of magnitude larger SIDIS background.

DEMPgen can be used to study the experimental effects on the measured target asymmetries for different kinematics. For example, by turning on or off the FSI or Fermi momentum modules, their effects on the predicted asymmetries can be identified. Fig.~\ref{fig:asym4} shows the generated asymmetry from a transversely polarized $^3$He target for a bin with central kinematics of $<-t>=0.45$ GeV$^2$, $<Q^2>=5.77$ GeV$^2$, $<x_B>=0.47$. The axes are the two azimuthal angles in Fig.~\ref{fig:planes}, the angle with respect to the target polarization ($\phi_S$) on the vertical axis, and the angle between the scattering and reaction planes ($\phi$) on the horizontal axis. For these kinematics, the $A_{UT}^{sin(\phi_S)}$ modulation dominates, as evident by the trend from positive (yellow) asymmetry bins at top to negative (blue) asymmetry bins at bottom. The other modulations are significantly smaller, and are identified by a fitting of the full Eqn.~\ref{eqn:six_asym2} to the data. The SoLID projected data are expected to be a considerable advance over the only existing dataset from HERMES \cite{HERMES:2009gtv} in terms of both kinematic coverage and statistical precision. For more details on the SoLID experiment feasibility studies using DEMPgen, consult Ref.~ \cite{atpi_proposal}.

%%%%%%%%%%%%%%%%%%%%%%%%%%%%%%%%%%%%%%%%%%%%%%%%%%%%%%%%%%%%%%%%%%%%%%%%%%%%%%%
% Chapter 5
\section{Summary and Outlook
\label{sec:summary}}

We have developed DEMPgen, a powerful tool for future feasibility studies of proposed Deep Exclusive Meson Production (DEMP) measurements at Jefferson Lab (JLab) and the Electron-Ion Collider (EIC). Based on the results we have obtained, we expect exclusive $\pi^+$ form factor measurements at the EIC to be feasible up to about $Q^2=30$ GeV$^2$, which would be a considerable extension in kinematic range over what is possible with 11 GeV beam at JLab. 
DEMPgen was also essential for feasibility studies of proposed transverse target single-spin asymmetry measurements in exclusive $\pi^-$ production from the polarized neutron in $^3$He with SoLID. Both of these studies have established the reliability and utility of DEMPgen.

DEMPgen is modular in form so that additional reaction channels can be added to it over time. We have described a recent extension, namely $t$-channel $K^+$ production leading to the $\Lambda$ and $\Sigma^0$ final states. There is a lot of interest in our $K^+$ studies, to see whether it will be feasible to extract the $K^+$ form factor to high $Q^2$ at the EIC \cite{jr:EIC_YR}. We have used DEMPgen to generate predicted kinematic distributions and rates for the $e^{\prime}$, $K^+$, $\Lambda[\Sigma^0]$ final state. Detailed detector simulations to establish whether the reconstruction of the $\Lambda[\Sigma^0]$ from their detected decay products is sufficiently reliable are ongoing.

Further extensions of DEMPgen to a broader kinematic regime are envisioned. Particularly for the $K^+$ form factor feasibility studies, the lower beam energy combinations planned for the Electron-Ion Collider in China (EicC) \cite{jr:EicC} may prove to be vital for the reliable identification of the $\Lambda$ from its decay products, as the $\Lambda$ will decay closer to the interaction region than at the EIC, and the decay products correspondingly easier to detect.

Similarly, should the JLab 22 GeV upgrade \cite{accardi2023strong} come to fruition, it would be straightforward to extend the cross section and target asymmetry model parameterizations to cover the kinematic range enabled by the upgrade. There is also interest in our adding exclusive $\pi^{\pm}$ reactions from the deuteron to the generator \cite{jr:EIC_YR}, as the exclusive $\pi^{\pm}$ ratios versus $t$ are needed to determine the extent of non-pion pole contributions to the data from which the pion charge form factor are extracted (Sec.~\ref{Sec:EIC_PiPlus}).

Finally, we also plan to add $u$-channel exclusive $\pi^0$ production to DEMPgen, optimized for EIC studies.
If the reliability of EIC $u$-channel $\pi^0$ studies can be established, it would open up a novel kinematic range for the study of Transition Distribution Amplitudes (TDA) \cite{Pire_2021} in the backward colinear factorization regime.

%%%%%%%%%%%%%%%%%%%%%%%%%%%%%%%%%%%%%%%%%%%%%%%%%%%%%%%%%%%%%%%%%%%%%%%%%%%%%%%

\section*{Acknowledgements}

We would like to thank Weizhi Xiong and Zhihan Yu for their helpful comments about the DEMPgen code which led to several improvements, and 
Nathan Heinrich and Zhihong Ye for their assistance with Figs.~\ref{fig:fig_demp}, \ref{fig:missing_mom}, \ref{fig:asym4}.

This work was supported by the Natural Sciences and Engineering Research Council of Canada (NSERC) grants SAPIN-2021-00026 and SAPPJ-2023-00041, the UK Science and Technology Facilities Council (STFC) grants ST/V001035/1, ST/W004852/1, and Department of Energy grant DE-FG02-05ER41372 (WL). We gratefully acknowledge the support from the Stony Brook University Center for Frontiers of Nuclear Science Director Abhay Deshpande.

\clearpage
\bibliographystyle{plainurl,unsrt} 
%\input{DEMPgenBib.bbl}
%\bibliography{DEMPGenBib}

\begin{thebibliography}{10}

\bibitem{url:DEMPgen}
Z.~Ahmed, R.S. Evans, I.~Goel, G.M. Huber, S.J.D. Kay, W.B. Li, L.~Preet, and
  A.~Usman.
\newblock {DEMPgen - Latest release}.
\newblock \url{https://github.com/JeffersonLab/DEMPgen/releases}, 2024.

\bibitem{Fr00}
L.~L. Frankfurt, M.~V. Polyakov, M.~Strikman, and M.~Vanderhaeghen.
\newblock Hard exclusive electroproduction of decuplet baryons in the large
  ${N}_{c}$ limit.
\newblock {\em Phys. Rev. Lett.}, 84:2589--2592, Mar 2000.

\bibitem{NAP25171}
National~Academies of~Sciences~Engineering and Medicine.
\newblock {\em An Assessment of U.S.-Based Electron-Ion Collider Science}.
\newblock The National Academies Press, Washington, DC, 2018.

%\bibitem{eic}
%A.~Accardi et~al.
%\newblock Electron ion collider: The next qcd frontier - understanding the glue
%  that binds us all, 2014.

\bibitem{Horn_2016}
Tanja Horn and Craig~D Roberts.
\newblock The pion: an enigma within the standard model.
\newblock {\em Journal of Physics G: Nuclear and Particle Physics},
  43(7):073001, May 2016.

\bibitem{Carlson:1990zn}
C.~E. Carlson and Joseph Milana.
\newblock Difficulty in determining the pion form factor at high
  ${\mathit{q}}^{2}$.
\newblock {\em Phys. Rev. Lett.}, 65:1717--1720, Oct 1990.

\bibitem{gmhuber}
G.M. Huber et~al.
\newblock Charged pion form factor between ${Q}^{2}=0.60$ and 2.45
  GeV$^2$. ii. determination of, and results for, the pion form
  factor.
\newblock {\em Phys. Rev. C}, 78:045203, Oct 2008.

\bibitem{fpi-12gev}
G.M. Huber, D.~Gaskell, T.~Horn, et~al.
\newblock {Measurement of the Charged Pion Form Factor to High $Q^2$, JLab
  Experiment E12-06-101 (E12-19-006)}.
\newblock \url{https://www.jlab.org/exp_prog/proposals/19/E12-19-006.pdf},
  2006, 2010, 2019.

\bibitem{kaon-lt}
T.~Horn, G.M. Huber, P.~Markowitz, et~al.
\newblock {Studies of the L/T Separated Kaon Electroproduction Cross Sections
  from 5-11 GeV, Jefferson Lab 12 GeV Experiment E12-09-011}.
\newblock \url{https://www.jlab.org/exp_prog/proposals/09/PR12-09-011.pdf},
  2009.

\bibitem{Belitsky_2005}
A.V. Belitsky and A.V. Radyushkin.
\newblock Unraveling hadron structure with generalized parton distributions.
\newblock {\em Physics Reports}, 418(1-6):1--387, oct 2005.

\bibitem{Diehl_2003}
M.~Diehl.
\newblock Generalized parton distributions.
\newblock {\em Physics Reports}, 388(2-4):41--277, dec 2003.

\bibitem{Goeke_2001}
K.~Goeke, M.V. Polyakov, and M.~Vanderhaeghen.
\newblock Hard exclusive reactions and the structure of hadrons.
\newblock {\em Progress in Particle and Nuclear Physics}, 47(2):401--515, jan
  2001.

\bibitem{burkardt}
Matthias Burkardt.
\newblock Impact parameter dependent parton distributions and off-forward
  parton distributions for
  $\stackrel{\ensuremath{\rightarrow}}{\ensuremath{\zeta}}0$.
\newblock {\em Phys. Rev. D}, 62:071503, Sep 2000.

\bibitem{ji_1}
Xiangdong Ji.
\newblock Gauge-invariant decomposition of nucleon spin.
\newblock {\em Phys. Rev. Lett.}, 78:610--613, Jan 1997.

\bibitem{cuic}
Marija Cuic, Kresimir Kumericki, and Andreas Schafer.
\newblock Separation of quark flavors using dvcs data, 2020.

%\bibitem{gpv}
%K.~Goeke, M.V. Polyakov, and M.~Vanderhaeghen.
%\newblock Hard exclusive reactions and the structure of hadrons.
%\newblock {\em Progress in Particle and Nuclear Physics}, 47(2):401–515, Jan
%  2001.

\bibitem{frank}
L.~L. Frankfurt, P.~V. Pobylitsa, M.~V. Polyakov, and M.~Strikman.
\newblock Hard exclusive pseudoscalar meson electroproduction and spin
  structure of the nucleon.
\newblock {\em Phys. Rev. D}, 60:014010, Jun 1999.

\bibitem{Belitsky2004}
A.~V. Belitsky.
\newblock Renormalons in exclusive meson electroproduction.
\newblock {\em AIP Conference Proceedings}, 2004.

\bibitem{arrington2023solenoidal}
John Arrington et~al.
\newblock {The Solenoidal Large Intensity Device (SoLID) for JLab 12 GeV}.
\newblock {\em Journal of Physics G: Nuclear and Particle Physics}, 50:110501
  1--57, 2023.

\bibitem{spherepp}
{E.W. Wesstein}.
\newblock {Sphere Point Picking}.
\newblock \url{http://mathworld.wolfram.com/SpherePointPicking.html}.

\bibitem{Trento}
Alessandro Bacchetta, Umberto D'Alesio, Markus Diehl, and C.~Andy Miller.
\newblock Single-spin asymmetries: The trento conventions.
\newblock {\em Phys. Rev. D}, 70:117504, Dec 2004.

\bibitem{hand}
L.~N. Hand.
\newblock Experimental investigation of pion electroproduction.
\newblock {\em Phys. Rev.}, 129:1834--1846, Feb 1963.

\bibitem{VRPaper}
Tom Vrancx and Jan Ryckebusch.
\newblock Charged-pion electroproduction above the resonance region.
\newblock {\em Phys. Rev. C}, 89:025203, Feb 2014.

\bibitem{Choi:2015yia}
Tae~Keun Choi, Kook~Jin Kong, and Byung~Geel Yu.
\newblock Pion and proton form factors in the regge description of
  electroproduction p(e, e'$\pi$+)n.
\newblock {\em Journal of the Korean Physical Society}, 67(7):1089--1094, Oct
  2015.

\bibitem{Basnet:2019cpg}
S.~Basnet et~al.
\newblock Exclusive ${\ensuremath{\pi}}^{+}$ electroproduction off the proton
  from low to high $\ensuremath{-}t$.
\newblock {\em Phys. Rev. C}, 100:065204, Dec 2019.

\bibitem{ROOT}
{Brun, R. and Rademakers, E.}
\newblock {ROOT - An Object Oriented Data Analysis Framework}.
\newblock \url{http://root.cern.ch/}.

\bibitem{VGL_paper1}
M.~Guidal, J.-M. Laget, and M.~Vanderhaeghen.
\newblock Pion and kaon photoproduction at high energies: forward and
  intermediate angles.
\newblock {\em Nuclear Physics A}, 627(4):645--678, 1997.

\bibitem{VGL_paper2}
M.~Vanderhaeghen, M.~Guidal, and J.-M. Laget.
\newblock Regge description of charged pseudoscalar meson electroproduction
  above the resonance region.
\newblock {\em Phys. Rev. C}, 57:1454--1457, Mar 1998.

\bibitem{VGL_paper3}
M.~Guidal, J.-M. Laget, and M.~Vanderhaeghen.
\newblock Electroproduction of strangeness above the resonance region.
\newblock {\em Phys. Rev. C}, 61:025204, Jan 2000.

\bibitem{VGL_paper4}
M.~Guidal, J.-M. Laget, and M.~Vanderhaeghen.
\newblock Exclusive electromagnetic production of strangeness on the nucleon:
  Regge analysis of recent data.
\newblock {\em Phys. Rev. C}, 68:058201, Nov 2003.

\bibitem{VRPaper2}
Tom Vrancx, Jan Ryckebusch, and Jannes Nys.
\newblock ${K}^{+}\ensuremath{\Lambda}$ electroproduction above the resonance
  region.
\newblock {\em Phys. Rev. C}, 89:065202, Jun 2014.

\bibitem{Lovepreet_thesis}
Love Preet.
\newblock Understanding hadronic mass through light meson structure at EIC,
  University of Regina M.Sc. thesis, 2023.

\bibitem{strangecalc}
{Nys, J. and Ryckebusch, J.}
\newblock {StrangeCalc}.
\newblock \url{http://rprmodel.ugent.be/calc/}.

\bibitem{Blok08}
H.~P. Blok et~al.
\newblock Charged pion form factor between ${Q}^{2}=0.60$ and 2.45
  ${\mathrm{gev}}^{2}$. i. measurements of the cross section for the
  ${}^{1}\mathrm{H}(e,{e}^{'}{\ensuremath{\pi}}^{+})n$ reaction.
\newblock {\em Phys. Rev. C}, 78:045202, Oct 2008.

\bibitem{GKPriv}
{Goloskokov, S.V. and Kroll, P.}
\newblock {Private Communications}, {2009--2017}.

\bibitem{Diehl_2005}
M.~Diehl and S.~Sapeta.
\newblock On the analysis of lepton scattering on longitudinally or
  transversely polarized protons.
\newblock {\em The European Physical Journal C}, 41(4):515–533, June 2005.

\bibitem{atpi_proposal}
{Huber, G. M. and Ahmed, Z. and Ye, Z. and others}.
\newblock {Measurement of Deep Exclusive $\pi^-$ Production using a
  Transversely polarized $^3$He Target and the SoLID Spectrometer, Jefferson
  Lab Experiment E12-10-006B}.
\newblock
  \url{https://misportal.jlab.org/pacProposals/proposals/1317/attachments/99852/Proposal.pdf}.

\bibitem{HERMES:2009gtv}
A.~Airapetian et~al.
\newblock {Single-spin azimuthal asymmetry in exclusive electroproduction of
  $\pi^+$ mesons on transversel y polarized protons}.
\newblock {\em Phys. Lett. B}, 682:345--350, 2010.

\bibitem{Goloskokov:2009ia}
S.~V. Goloskokov and P.~Kroll.
\newblock {An Attempt to understand exclusive $\pi^+$ electroproduction}.
\newblock {\em Eur. Phys. J. C}, 65:137--151, 2010.

\bibitem{SCHIAVILLA1986219}
R.~Schiavilla, V.R. Pandharipande, and R.B. Wiringa.
\newblock Momentum distributions in a = 3 and 4 nuclei.
\newblock {\em Nuclear Physics A}, 449(2):219 -- 242, 1986.

\bibitem{pcdr}
{The SoLID Collaboration}.
\newblock {SoLID (Solenoidal Large Intensity Device) Preliminary Conceptual
  Design Report}.
\newblock \url{https://hallaweb.jlab.org/12GeV/SoLID/files/solid_precdr.pdf}.

\bibitem{PhysRevC.42.2310}
J.~L. Friar, B.~F. Gibson, G.~L. Payne, A.~M. Bernstein, and T.~E. Chupp.
\newblock Neutron polarization in polarized $^{3}\mathrm{He}$ targets.
\newblock {\em Phys. Rev. C}, 42:2310--2314, Dec 1990.

\bibitem{PhysRevC.64.024004}
F.~Bissey, A.~W. Thomas, and I.~R. Afnan.
\newblock Structure functions for the three-nucleon system.
\newblock {\em Phys. Rev. C}, 64:024004, Jul 2001.

\bibitem{phase-shifts}
Glenn Rowe, Martin Salomon, and Rubin~H. Landau.
\newblock Energy-dependent phase shift analysis of pion-nucleon scattering
  below 400 mev.
\newblock {\em Phys. Rev. C}, 18:584--589, Jul 1978.

\bibitem{Shinozaki}
A.~Shinozaki.
\newblock {\em {Total Cross-Section for the $(\gamma,\pi^+\pi^-)$ process on
  $^2$H and $^{12}$C from 550-1105 MeV}}.
\newblock PhD thesis, University of Regina, 2002.

\bibitem{williamsweight}
{Williams, W. S. C.}
\newblock {\em An introduction to elementary particles}.
\newblock Academic Press, 1971.

\bibitem{dedrickweight}
K.~G. Dedrick.
\newblock Kinematics of high-energy particles.
\newblock {\em Rev. Mod. Phys.}, 34:429--442, Jul 1962.

\bibitem{Catchen}
Gary~L. Catchen, Javed Husain, and Richard~N. Zare.
\newblock Scattering kinematics: Transformation of differential cross sections
  between two moving frames.
\newblock {\em The Journal of Chemical Physics}, 69(4):1737--1741, 1978.

\bibitem{sidis}
Duane Byer et~al.
\newblock Sidis-rc evgen: a monte-carlo event generator of semi-inclusive deep
  inelastic scattering with the lowest-order qed radiative corrections, 2023.

\bibitem{Amendolia:1984nz}
S.~R. Amendolia et~al.
\newblock {A Measurement of the Pion Charge Radius}.
\newblock {\em Phys. Lett. B}, 146:116--120, 1984.

\bibitem{Amendolia:1986wj}
S.~R. Amendolia et~al.
\newblock {A Measurement of the Space - Like Pion Electromagnetic Form-Factor}.
\newblock {\em Nucl. Phys. B}, 277:168, 1986.

\bibitem{Ackermann:1977rp}
H.~Ackermann, T.~Azemoon, W.~Gabriel, H.~D. Mertiens, H.~D. Reich, G.~Specht,
  F.~Janata, and D.~Schmidt.
\newblock {Determination of the Longitudinal and the Transverse Part in pi+
  Electroproduction}.
\newblock {\em Nucl. Phys. B}, 137:294--300, 1978.

\bibitem{Brauel:1979zk}
P.~Brauel, T.~Canzler, D.~Cords, R.~Felst, Guenter Grindhammer, M.~Helm, W.~D.
  Kollmann, H.~Krehbiel, and M.~Schadlich.
\newblock {Electroproduction of $\pi^+ n$, $\pi^- p$ and $K^+ \Lambda$, $K^+
  \Sigma^0$ Final States Above the Resonance Region}.
\newblock {\em Z. Phys. C}, 3:101, 1979.

\bibitem{Volmer:2000ek}
J.~Volmer et~al.
\newblock {Measurement of the Charged Pion Electromagnetic Form-Factor}.
\newblock {\em Phys. Rev. Lett.}, 86:1713--1716, 2001.

\bibitem{Horn:2006tm}
T.~Horn et~al.
\newblock {Determination of the Pion Charge Form Factor at ${Q}^{2}=1.60$ and
  2.45 GeV/c$^2$}.
\newblock {\em Phys. Rev. Lett.}, 97:192001, 2006.

\bibitem{Nesterenko:1982gc}
V.~A. Nesterenko and A.~V. Radyushkin.
\newblock {Sum Rules and Pion Form-Factor in QCD}.
\newblock {\em Phys. Lett. B}, 115:410, 1982.

\bibitem{Chang:2013nia}
L.~Chang, I.C. Cloet, C.D. Roberts, S.M. Schmidt, and P.C. Tandy.
\newblock {Pion electromagnetic form factor at spacelike momenta}.
\newblock {\em Phys. Rev. Lett.}, 111(14):141802, 2013.

\bibitem{Simula:2023ujs}
Silvano Simula and Ludovico Vittorio.
\newblock {Dispersive analysis of the experimental data on the electromagnetic
  form factor of charged pions at spacelike momenta}.
\newblock {\em Phys. Rev. D}, 108(9):094013, 2023.

\bibitem{Albino_2022}
L.~Albino, I.M. Higuera-Angulo, K.~Raya, and A.~Bashir.
\newblock Pseudoscalar mesons: Light front wave functions, gpds, and pdfs.
\newblock {\em Physical Review D}, 106(3), August 2022.

\bibitem{Bakulev:2004cu}
A.~P. Bakulev, K.~Passek-Kumericki, W.~Schroers, and N.~G. Stefanis.
\newblock {Pion form-factor in QCD: From nonlocal condensates to NLO analytic
  perturbation theory}.
\newblock {\em Phys. Rev. D}, 70:033014, 2004.
\newblock [Erratum: Phys.Rev.D 70, 079906 (2004)].

\bibitem{jr:ECCE_EDT}
A.~Bylinkin et~al.
\newblock Detector requirements and simulation results for the EIC exclusive,
  diffractive and tagging physics program using the ecce detector concept.
\newblock {\em Nuclear Instruments and Methods in Physics Research Section A:
  Accelerators, Spectrometers, Detectors and Associated Equipment},
  1052:168238, 2023.

\bibitem{jr:EIC_YR}
R.~Abdul~Khalek et~al.
\newblock Science requirements and detector concepts for the electron-ion
  collider: Eic yellow report.
\newblock {\em Nuclear Physics A}, 1026:122447, 2022.

\bibitem{ZEUS:2007knd}
S.~Chekanov et~al.
\newblock {Leading neutron energy and $p_T$ distributions in deep inelastic
  scattering and photoproduction at HERA}.
\newblock {\em Nucl. Phys. B}, 776:1--37, 2007.

\bibitem{Tu:2023few}
{Tu, Zhoudunming}
\newblock{Deep exclusive meson production as a probe to the puzzle of \ensuremath{\Lambda} hyperon polarization].
\newblock  Phys. Rev. C, 109:055205, 2024.
\newblock  arXiv:2308.09127.
%\newblock    doi = "10.1103/PhysRevC.109.055205",
}

\bibitem{jr:EicC}
D.P. Anderle et~al.
\newblock Electron-ion collider in china.
\newblock {\em Frontiers of Physics}, 16(6), Jun 2021.

\bibitem{accardi2023strong}
A.~Accardi et~al.
\newblock Strong interaction physics at the luminosity frontier with 22 GeV
  electrons at Jefferson Lab, 2023.

\bibitem{Pire_2021}
B.~Pire, K.~Semenov-Tian-Shansky, and L.~Szymanowski.
\newblock Transition distribution amplitudes and hard exclusive reactions with
  baryon number transfer.
\newblock {\em Physics Reports}, 940:1–121, December 2021.

%\bibitem{simc}
%{Jefferson Lab Hall C Collaboration}.
%\newblock {SIMC}.
%\newblock \url{https://hallcweb.jlab.org/wiki/index.php/SIMC\_Monte\_Carlo}.

\bibitem{EIC_Accel_Dev_CAP22}
{Seryi, A.}
\newblock {Electron-Ion Collider accelerator development}.
\newblock
  \url{https://indico.cern.ch/event/1072579/contributions/4796856/attachments/2456676/4210776/CAP-EIC-June-7-2022-Seryi-r2.pdf}.

\end{thebibliography}

%%%%%%%%%%%%%%%%%%%%%%%%%%%%%%%%%%%%%%%%%%%%%%%%%%%%%%%%%%%%%%%%%%%%%%%%%%%%%%%

% SJDK 30/11/22 - Force a clear page and then begin the appendix
\clearpage
\appendix

%%%%%%%%%%%%%%%%%%%%%%%%%%%%%%%%%%%%%%%%%%%%%%%%%%%%%%%%%%%%%%%%%%%%%%%%%%%%%%%

% 04/10/22 - New appendix section with JSON control cards example
\section{\label{sec:Appendix_json} json Control Cards}

DEMPgen utilizes .json control cards to customise various parameters when running the event generator. Some parameters are only applicable to EIC or SoLID simulations and others are applicable in both cases. Table \ref{tab:Common_json_Input} outlines the options that are common in both cases, Table \ref{tab:EIC_json_Input} specifies the EIC module options and the SoLID module options are listed in Table \ref{tab:SoLID_json_Input}.

Note that some quantities, such as the energy/angular ranges over which the scattered electron and ejectile are generated, have additional checks within the generator. These checks are needed due to their interconnection with cuts on kinematic quantities ($Q^{2}$, $W$, $t$) within the event generator. This is discussed further in \ref{sec:Appendix_PSF}.

\begin{table*}[h!]
\centering
\caption{Common input .json file options}
    \label{tab:Common_json_Input}
    \begin{tabular}{>{\centering\arraybackslash}m{3.5cm} | >{\centering\arraybackslash}m{12.5cm}}
             \begin{center} Parameter \end{center} & \begin{center} Description \end{center}\\ \hline
             Experiment &``eic'' or ``solid'', controls the type of event generated subsequently. \\ \hline
             file\_name & Name of the output file created by the generator.\\ \hline
             \begin{center} n\_events \end{center} & Number of events to be thrown by the generator, this is \emph{not} the number of total events that will be saved in the output. This is equivalent to $N_{Requested}$ as described eslewhere in the text.\\ \hline
             generator\_seed & Random number generator seed used by the event generator.\\ \hline
            Kinematics\_type & 1 for FF (EIC) or 2 for TSSA (SoLID). May include TSSA for EIC in the future. \\ \hline 
            particle & Choices are omega, pi+, pi0 or K+. This is the produced ejectile (meson) in the reaction.\\
    \end{tabular}
\end{table*}

\begin{table*}[h!]
\centering
\caption{EIC module only input .json file options}
    \label{tab:EIC_json_Input}
    \begin{tabular}{>{\centering\arraybackslash}m{3.5cm} | >{\centering\arraybackslash}m{12.5cm}}
           \begin{center} Parameter \end{center}  & \begin{center} Description \end{center}\\ \hline
           hadron & Lambda or Sigma0, only used in EIC kaon DEMP event generation.\\ \hline
             ebeam & Incident electron beam energy. Typically 5, 10 or 18, but can be set arbitrarily.\\ \hline
              hbeam & Incident hadron (ion) beam energy, typically, 41, 100 or 275, but can be set arbitrarily.\\ \hline
             hbeam\_part & Hadron (ion) beam particle, proton, deut or helium3. Work in progress.\\ \hline
            \begin{center} det\_location \end{center} & Detector location, ``ip6'' or ``ip8''. This option sets the beam crossing angle accordingly for each potential EIC interaction point.\\ \hline
             \begin{center}OutputType\end{center} & Fixes the output file type options are LUND, Pythia6 or HEPMC3. SoLID output is LUND only and this flag is not used. \\ \hline \begin{center}ROOTOut\end{center} & True or false flag to enable/disable the generation of a .root output file in addition to event by event file (in the format specified by OutputType parameter).\\ \hline
            \begin{center}Ee\_Low\end{center} & Minimum scattered electron energy generated, as a multiple of the electron beam energy. Default is 0.5. \\ \hline
            \begin{center}Ee\_High\end{center} & Maximum scattered electron energy generated, as a multiple of the electron beam energy. Default is 2.5. \\ \hline
            e\_Theta\_Low & Minimum scattered electron $\theta$ that will be generated, in degrees. Default is 60.\\ \hline
            e\_Theta\_High & Maximum scattered electron $\theta$ that will be generated, in degrees. Default is 175.\\ \hline
            EjectileX\_Theta\_Low & Minimum ejectile (meson) $\theta$ that will be generated, in degrees. Default is 0.\\ \hline
            EjectileX\_Theta\_High & Maximum ejectile (meson) $\theta$ that will be generated, in degrees. Default is 50.\\
    \end{tabular}
\end{table*}

\begin{table*}[h!]
\centering
\caption{SoLID module only input .json file options}
    \label{tab:SoLID_json_Input}
    \begin{tabular}{>{\centering\arraybackslash}m{3.5cm} | >{\centering\arraybackslash}m{12.5cm}}
           \begin{center} Parameter \end{center}  & \begin{center} Description \end{center}\\ \hline
             beam\_energy & Incoming electron beam energy\\ \hline
             Targ\_dir & Target direction, 1 for up or 2 for down.\\ \hline
             targ\_pol\_x & Target polarisation in the $x$ direction.\\ \hline
             targ\_pol\_y & Target polarisation in the $y$ direction.\\ \hline
             targ\_pol\_z & Target polarisation in the $z$ direction.\\ \hline
             scat\_elec\_Emin & Minimum scattered electron energy generated, in MeV.\\ \hline
             scat\_elec\_Emax & Maximum scattered electron energy generated, in MeV.\\ \hline
             scat\_elec\_thetamin & Minimum scattered electron $\theta$ generated, in degrees.\\ \hline
             scat\_elec\_thetamax & Maximum scattered electron $\theta$ generated, in degrees.\\ \hline
             prod\_pion\_thetamin & Minimum $\pi$ $\theta$ generated, in degrees. \\ \hline
             prod\_pion\_thetamax & Maximum $\pi$ $\theta$ generated, in degrees. \\ \hline
             multiple\_scattering & ``true'' or ``false'' to enable or disable multiple scattering effects.\\ \hline 
             ionisation & ``true'' or ``false'' to enable or disable ionisation effects.\\ \hline
             bremmstrahlung & ``true'' or ``false'' to enable or disable bremmstrahlung effects.\\ \hline
             final\_state\_interaction & ``true'' or ``false'' to enable or disable final state interaction effects.\\ \hline
             fermi\_momentum & ``true'' or ``false'' to enable or disable Fermi momentum effects.\\ \hline
             weight\_cut & ``true'' or ``false'' to enable or disable a cut on the event weight\\ \hline
             w\_cut &``true'' or ``false'' to enable or disable a cut on the $W$ value of generated events.\\ \hline
             w\_min & Minimum value of $W$ retained by the generator (if cut is enabled).\\ \hline
             Qsq\_cut & ``true'' or ``false'' to enable or disable a cut on the $Q^{2}$ value of generated events.\\ \hline
             Qsq\_min & Minimum value of $Q^{2}$ retained by the generator (if cut is enabled).\\ \hline
             t\_cut & ``true'' or ``false'' to enable or disable a cut on the $t$ value of generated events.\\ \hline
             t\_min & Minimum value of $t$ retained by the generator (if cut is enabled).\\
    \end{tabular}
\end{table*}

%%%%%%%%%%%%%%%%%%%%%%%%%%%%%%%%%%%%%%%%%%%%%%%%%%%%%%%%%%%%%%%%%%%%%%%%%%%%%%%

% 12/05/23 - New appendix section with EIC luminosity values.
\section{\label{sec:Appendix_EIC_Lumi} EIC Luminosity}

In processing events, DEMPgen utilizes the luminosity to determine the event weight. For EIC event generation, the luminosity is set depending upon the beam energy combination specified. The luminosity values for each beam energy combination are specified in Table \ref{tab:EIC_Lumi}. Note that some sample beam energy combinations for the Electron-Ion Collider China (EicC) have been included too\cite{jr:EicC}.

\begin{table}[!htb]
    \centering
\begin{tabular}{|>{\centering\arraybackslash}m{1.8cm}  |>{\centering\arraybackslash}m{1.8cm} | >{\centering\arraybackslash}m{2.8cm} | >{\centering\arraybackslash}m{2cm}|}
  \hline
  $E_{e}$ (GeV) & $E_{p}$ (GeV) & $\mathcal{L}$ ($10^{33}\,cm^{-2}s^{-1}$) & Note \\
  \hline
  2.8 & 13 & 0.7 & EicC\\
  \hline
  3.5 & 20 & 2 & EicC \\
  \hline
  5 & 41 & 0.44 & -\\
  \hline
  5 & 100 & 3.68 & -\\
  \hline
  10 & 100 & 4.48 & -\\
  \hline
  18 & 275 & 1.54 & -\\
  \hline
   - & - & 1 & Default \\
  \hline
  \end{tabular}
    \caption{Luminosity values used by DEMPgen for different electron/proton beam energy combinations. Values for the EIC were taken from \cite{EIC_Accel_Dev_CAP22}.} 
    \label{tab:EIC_Lumi}
\end{table}

% https://indico.cern.ch/event/1072579/contributions/4796856/attachments/2456676/4210776/CAP-EIC-June-7-2022-Seryi-r2.pdf

% 06/1/24 - New appendix section for Quadratic Equation in Analytical Solution
\section{\label{sec:Appendix_quad} Quadratic Equation in Analytical Solution}

While solving for the energy of the ejectile using Eqn~.\ref{eqn:kinsolve} in the analytical solution, a quadratic equation can be generated. This is given by:
\begin{equation}
  \label{eqn:kinsolve22}
  a [E^{2}_{Ej}] + b [E_{Ej}] + c=0,
\end{equation}
where $a$, $b$ and $c$ are constants and depends on the four momentum of initial particles as well as the direction of ejectile as:

\begin{equation}
    \label{eqn:a}
    a = 4 [\vec{P}^{2}_{i}(\hat{P}_{i}.\hat{P}_{Ej})^{2} - E^{2}_{i}]
\end{equation}
\begin{equation}
    \label{eqn:b}
    b = 4 [E_{Ej}(E^{2}_{Ej}-\vec{P}^{2}_{i}+M^{2}_{Ej}-M^{2}_{Rec})]
\end{equation}
\begin{equation}
    \label{eqn:c}
    c = - [4 (\vec{P}^{2}_{i}(\hat{P}_{i}.\hat{P}_{Ej})^{2})M^{2}_{Ej} + (E^{2}_{Ej}-\vec{P}^{2}_{i}+M^{2}_{Ej}-M^{2}_{Rec})^{2}]
\end{equation}

Here, $\vec{P}_{i}$ is the net initial three momenta, $\hat{P}_{in}$ and $\hat{P}_{Ej}$ are the net initial and ejectile unit vectors and $E_{i}$ is the net initial energy.

%%%%%%%%%%%%%%%%%%%%%%%%%%%%%%%%%%%%%%%%%%%%%%%%%%%%%%%%%%%%%%%%%%%%%%%%%%%%%%%

% 04/10/22 - New appendix section with info on the phase space factor
\section{\label{sec:Appendix_PSF} Phase Space Factor}

The phase-space-factor is the fraction of the total kinematically accessible phase space that is covered by the event generator. This factor is a function of the incoming electron beam energy and the angles over which the scattered electron and produced ejectile are generated. This is calculated as -

\begin{equation}
    PSF = \left( \left(E_{e^{\prime}Max} - E_{e^{\prime}Min} \right) d\Omega_{e^{\prime}}\left(\theta, \phi\right) d\Omega_{Ej}\left(\theta, \phi\right)\right),
\end{equation}
where $E_{e^{\prime}Max}$ and $E_{e^{\prime}Min}$ are the maximum and minimum energy that the generated scattered electron can have. $d\Omega_{e^{\prime}}$ and $d\Omega_{Ej}$ are the solid angles over which the scattered electron and ejectile are generated respectively. The resulting value of the PSF is a quantity in units of $MeVsr^{2}$ (SoLID) or $GeVsr^{2}$ (EIC).

$E_{e^{\prime}Max}$ and $E_{e^{\prime}Min}$ can be specified by the user in the .json input file (see \ref{sec:Appendix_json} for more details). The $\theta$ range over which scattered electrons and ejectiles are generated is also specified by the user in the same manner, however, the generator is hard coded to generate both of these particles across $2\pi$ in $\phi$.

Due to other cuts that are applied by the generator on quantities such as $Q^{2}$, $W$ and $t$, the user specified ranges may be ``too large''. After the initialisation step, DEMPgen conducts a check of the phase space factor, represented by the ``PSF Check'' box in Fig.~\ref{fig:EIC_Flowchart}. In this step, DEMPgen determines the real, maximum ranges over which $E_{e^{\prime}Max}$, $E_{e^{\prime}Min}$, $\theta_{e\prime}$ and $\theta_{Ej}$ can be generated within the $Q^{2}$, $W$ and $t$ limits for the user specified beam energy combination. If the user specified values for $E_{e^{\prime}Max}$, $E_{e^{\prime}Min}$, $\theta_{e\prime}$ or $\theta_{Ej}$ exceed these limits, DEMPgen will adjust the generation range to fit within the calculated limit, printing a warning to the user as it does so. $PSF$ will be recalculated using these new limits. Without this adjustment, the phase space factor would be artificially inflated in some cases. The values will not be adjusted if the specified generation ranges are smaller than the ``maximum'' limits imposed by the cuts on kinematic quantities.

After all events are processed, DEMPgen also calculates the PSF based upon the range of scattered electron and ejectile angles that were actually generated and retained. This is the ``True PSF Calculation'' box in Fig.~\ref{fig:EIC_Flowchart}. This value, $PSF_{Gen}$ is retained and printed to the cut summary file for user verification. For large numbers of $N_{Requested}$, $PSF_{Gen}\approx PSF$. If this is not the case, the user can scale the generated weights by the ratio of $PSF_{Gen}$ to $PSF$. The user specified, recalculated and actual generated ranges over which the scattered electron and ejectile were thrown are all printed to the cut summary output file.
%\end{linenumbers}

\end{document}